\documentclass[12pt,a4paper,parskip=half-,twoside,BCOR=8mm,DIV=13
				]{scrreprt}

\usepackage[ngerman, english]{babel}
\usepackage[utf8]{inputenc}
\usepackage[T1]{fontenc}
\usepackage{csquotes}
\usepackage{graphicx}
\usepackage{color}
\usepackage{amsmath}
\usepackage{amssymb}

\usepackage{tabularx}
\newcolumntype{L}[1]{>{\raggedright\arraybackslash} m{#1}} 
\newcolumntype{C}[1]{>{\centering\arraybackslash}   m{#1}} 
\newcolumntype{R}[1]{>{\raggedleft\arraybackslash}  m{#1}} 

\usepackage{booktabs} 
\usepackage{pxfonts}
\usepackage{user}

\graphicspath{{img/}}

\usepackage[backend=biber,url=true,sorting=none,giveninits=true,maxnames=2,isbn=true,arxiv=abs]{biblatex}

\newcommand{\dcauthorpre}{} 
\newcommand{\dcauthorsurname}{Schmeißer}
\newcommand{\dcauthorname}{Martin} 
\newcommand{\dcauthoradd}{geboren am 21.01.1989 in T\"ubingen}

\newcommand{\dctitle}{In-situ measurements of the intrinsic emittance of photocathodes for high brightness electron beams} 
\newcommand{\dcsubtitle}{~}

\newcommand{\dcconsulta}{Dr. Thorsten Kamps}

\newcommand{\dcapprovala}{Prof. Dr. Andreas Jankowiak} 
\newcommand{\dcapprovalb}{Prof. Dr. Kurt Aulenbacher}

\newcommand{\dcdegree}{Master of Science (M. Sc.)} 
\newcommand{\dcsubject}{Physik} 
\newcommand{\dcfaculty}{Mathematisch-Naturwissenschaftlichen Fakult\"at I}
\newcommand{\dcinstitute}{Institut für Physik}
\newcommand{\dcuniversity}{Humboldt-Universit\"at zu Berlin}

\newcommand{\dcdatesubmitted}{\today} 

\newcommand{\dckeydea}{Schlagwort 1}
\newcommand{\dckeydeb}{Schlagwort 2}
\newcommand{\dckeydec}{Schlagwort 3}
\newcommand{\dckeyded}{Schlagwort 4}

\newcommand{\dcpdfsubject}{\dctitle}
\usepackage{ifpdf}

\ifpdf

\usepackage[
	unicode=true,
    colorlinks=true,
	linkcolor=black,
	filecolor=black,
	urlcolor=black,
	citecolor=black,
	plainpages=false,
	hypertexnames=false,
	pdfpagelabels=true,
	hyperindex=true,
    bookmarksnumbered=true]{hyperref}
	
\hypersetup{pdftitle={\dctitle},
	pdfsubject={\dcpdfsubject},
	pdfkeywords={\dckeydea, \dckeydeb, \dckeydec, \dckeyded},
	pdfpagemode=UseOutlines,
	pdfauthor={\dcauthorsurname\ \dcauthorname}	}
	
\else
	
\fi

\AtEveryBibitem{
  \ifentrytype{misc}{  
  }{%
  \ifentrytype{book}{  
  \clearfield{url}
  \clearfield{urldate}
  \clearfield{day}
  \clearfield{month}
  \clearfield{endday}
  \clearfield{endmonth}
  \clearfield{issue}
  \clearfield{number}
  \clearfield{language}
  }{%
  \clearfield{url}
  \clearfield{urldate}
  \clearfield{day}
  \clearfield{month}
  \clearfield{endday}
  \clearfield{endmonth}
  \clearfield{issue}
  \clearfield{number}
  \clearfield{language}
  \clearfield{isbn}
  \clearfield{issn}
  }%
  }
}

\begin{document}
\pagenumbering{roman}
\setcounter{page}{1}
\pagestyle{plain}


\author{\small von \\ \small \dcauthorpre\ \dcauthorname\ \dcauthorsurname\ \\ \small  \dcauthoradd  }

\title{\vspace{-1.5cm}\dctitle \\ 
\vspace{0.5cm}
\large{\dcsubtitle} \\ 
\vspace{0.5cm} {\Large{MASTERARBEIT}}\\ 
\vspace{0.5cm} \small{zur Erlangung des akademischen Grades \\ 
\dcdegree\\ im Fach \dcsubject \\\vspace{1.5cm}
\parbox{6cm}{\centering \includegraphics[width=5.5cm]{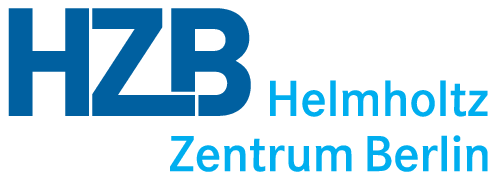}  } 
\hfill
\parbox{6cm}{\centering \includegraphics[width=4cm]{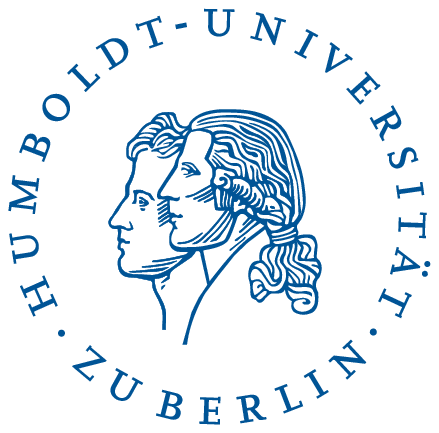}}\\
\hfill\\
\vspace{1cm} eingereicht an der \\ 
\dcfaculty \\ 
\dcinstitute\\
\dcuniversity \\ \vspace{-1cm} }}

\date{\small{\vspace{1cm}
%
%
%
%
%
%
\vspace{-0.5cm}
\raggedright{
\begin{tabbing}
Eins \= 2.2.  \= Drei \kill
Betreuung: \vspace{0.4cm}\\
\> \>\dcconsulta \vspace{0.8cm}\\
Begutachtung: \vspace{0.4cm}\\
\>1. \>\dcapprovala \vspace{0.4cm}\\
\>2. \>\dcapprovalb
\end{tabbing}
}
}}
\maketitle
\selectlanguage{english}
\setcounter{page}{2}
\tableofcontents
\listoffigures
\cleardoublepage

\pagenumbering{arabic}
\setcounter{page}{1}

\catcode`_=\active
\newcommand_[1]{\ensuremath{\sb{\mathrm{#1}}}}
\catcode`^=\active
\newcommand^[1]{\ensuremath{\sp{\mathrm{#1}}}}

\noindent

\chapter{Introduction} \label{chap:intro}

For novel accelerator applications of electron beams it is essential that the next generation of electron sources provides a high brightness beam with full control over the beam parameters. The normalized brightness $B_n$ of a beam is an important figure of merit to characterise the beam of an accelerator. In a light source, the brightness of the photon beam depends on the peak brightness of the electron beam. For electron beams, brightness is defined as the beam current $I$ divided by its normalized trace space volume \cite{Reiser2008}:

\begin{equation}
B = \frac{2}{\pi^2} \frac{I}{\varepsilon_{x,n} \varepsilon_{y,n}} \, ,
\end{equation}

where $\varepsilon_n$ are the normalized emittances. Thus, the goals for the source are to provide a beam with low emittance, high peak current in short bunches and low energy spread. Photoemission from materials with a high quantum efficiency allows to achieve such characteristics where control over the drive laser pulse shape gives control over the electron beam parameters. In order to mitigate space charge effects, the beam is accelerated immediately after its generation in a high gradient cavity. The operation in continuous wave mode at high gradients is possible with superconducting radio frequency (SRF) cavities.


The fundamental limit of the emittance achievable from a photoinjector is set by the initial emittance of the photocathode. At the same time the injector must present an ideal environment for acceleration, so as not to disturb the beam quality. Therefore, this work covers both aspects, the measurement of the beam parameters from a full injector setup as well as the resolution of the initial transverse momentum distributions of photocathode materials.

At Helmholtz-Zentrum Berlin, a high-brightness electron source which will use the SRF photoinjector concept is under development for the prospected energy recovery linac bERLinPro \cite{bERLinPro-CDR}. Energy recovery linacs (ERLs) will deliver high brightness beams where the fundamental limit is set by the beam source instead of equilibrium conditions as in storage rings. At the same time, they allow cw operation at high currents where management (dumping) of the spent high energy beam and operational cost forbid such modes in conventional linacs. Thus, ERLs hold the promise to combine the advantages of linear accelerators and storage rings.
bERLinPro is intended as 
a demonstration facility for high current, cw operation of an ERL. It will accelerate a 100\,mA electron beam to 50\,MeV with a 1.3\,GHz repetition rate.


The beam parameters of complete prototype of the bERLinPro injector that consisted of a lead cathode in an 1.6 cell superconducting cavity were measured. The measurements are discussed in chapter \ref{chap:measuring-emittance} and indicate that the beam emittance is governed by a multitude of contributions from the cathode and the injector.
%
Cathodes for practical use must have a high enough QE to deliver the required current at moderate illumination intensities and be able to sustain the operation for several days. Alkali antimonides are good candidates for this purpose, as discussed in chapter \ref{sec:alkali-antimonides}. The preparation of such cathodes with QEs of a few percent in the visible range has been performed routinely \cite{Cultrera2013,mySmedley2009} but the details of the growth and the influence of process parameters on the cathode's properties are not well understood yet.
A theoretical treatment of photoemission is presented, and expressions for quantum efficiency and initial emittance of photocathodes are reviewed in section \ref{sec:theory}.
In order to study Potassium Caesium Antimonide (\PCA) cathodes and their fabrication a preparation and characterization system has been developed at HZB.
In section \ref{sec:alkali-antimonides} this experimental setup is explained, details of the alkali antimonide cathodes are discussed, and the preparation of a caesium antimonide cathode is presented.
Finally, chapter \ref{chap:momentatron} focuses on the physics and engineering design of spectrometer that will resolve the initial transverse momentum of the photoelectrons released from cathode samples, the momentatron. It is installed in the analysis chamber of the cathode preparation system and allows \textit{in-situ} characterization of the samples. 
\chapter{Measurement of the Emittance from an Injector}
\label{chap:measuring-emittance}

\begin{figure}[!ht]
\centering
\includegraphics[width=0.85\textwidth]{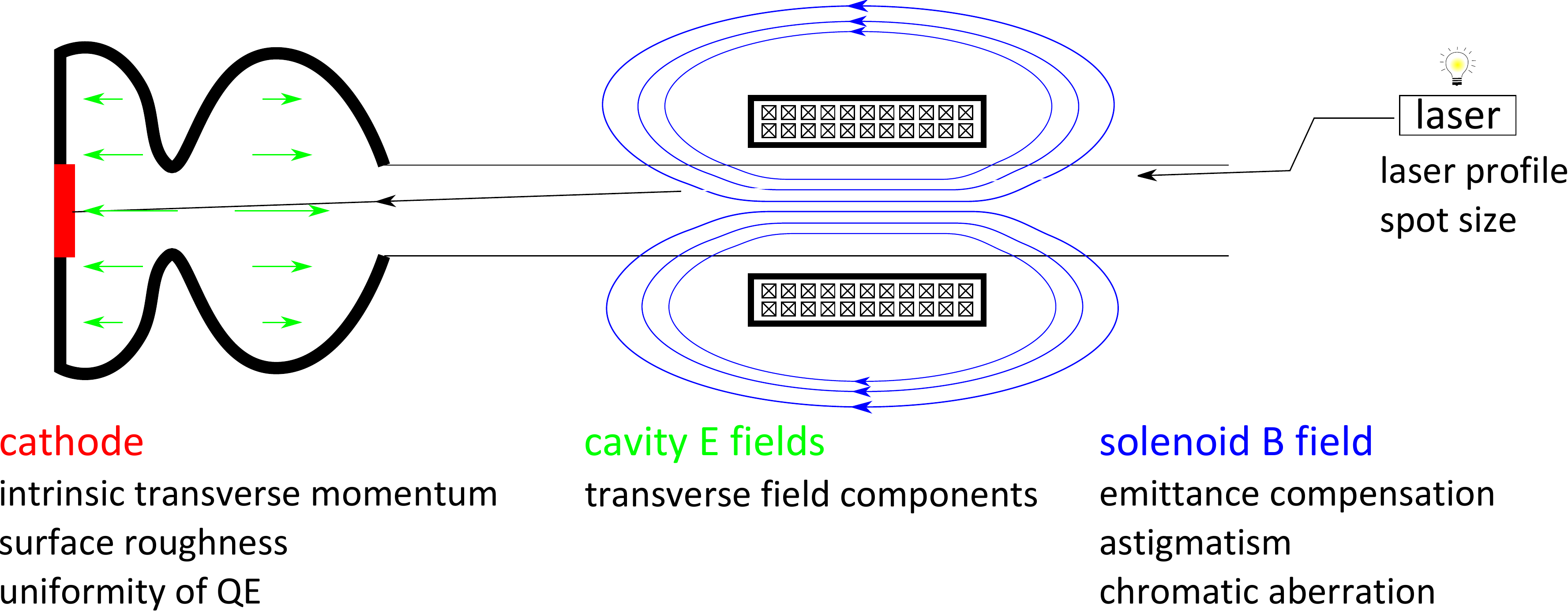}
\caption{Sources of emittance in an RF photoinjector.}
\label{fig:gun-emittance}
\end{figure}

The emittance gives a good measure of a particle beams quality and directly influences the brightness that can be achieved at user stations. It is defined as the phase space volume populated by the particles in one bunch. Because the beam quality often degrades by filamentation while the populated phase space area remains constant due to Liouville's theorem it is feasible to define the (projected) geometric rms-emittance through the second moments of the beam's particle density distribution as \cite{Reiser2008}
\begin{equation}
\label{eq:eps-definition}\varepsilon_{rms,\, g} = \sqrt{<u^2>\,<u^{\prime \, 2}> - <u \, u^{\prime}>^2} \, ,
\end{equation}
where $u$ and $u^\prime$ are position and divergence in one transverse direction. The normalized emittance
\begin{equation}
\label{eq:normeps-definition}\varepsilon_{n} = \beta \gamma \varepsilon_g
\end{equation}
remains constant during acceleration and allows comparison between beams of different energies. $\beta$ and $\gamma$ are the relativistic factors. Figure \ref{fig:gun-emittance} illustrates contributions to the total emittance of the beam from a photoinjector. During the measurements, the influences of the laser spot size and solenoid field errors were studied.\footnote{The measurements covered in this chapter have been published previously in \cite{Schmeisser2013a, Schmeisser2013}.} The effects of the cathode's properties are addressed in chapters \ref{chap:photocathodes} and \ref{chap:momentatron}.
Two techniques to measure the emittance are discussed: both yield the aforementioned rms emittance, one by summing up second moments of the particle distribution, the other by fitting parameters of a linear beam optics model that depend on the rms distribution.

\section*{Slit Based Technique}
\label{chap:slit}

\begin{figure}[ht]
\centering
\includegraphics[width=.75\textwidth]{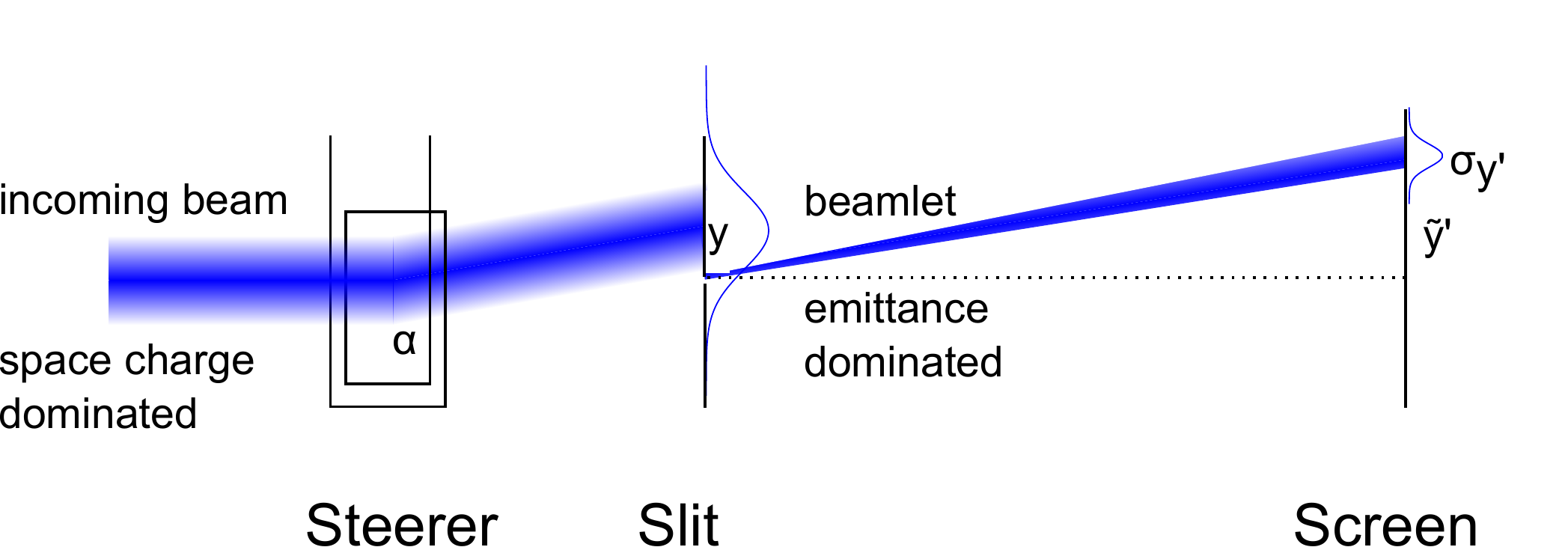}
\caption{Scheme of the measurement setup for a slit mask measurement.}
\label{fig:scheme-slit-mask}
\end{figure}

The slit mask measurement allows to reconstruct the projected phase space distribution in one transverse direction at the location of the slit, as illustrated in figure \ref{fig:scheme-slit-mask}. The slit serves two purposes. For one, the slit selects a narrow band of the particle distribution, which allows to scan the beam's diameter and correlate transverse position and divergence. The width and mean value of the divergence of the beamlet are calculated from the transverse distance travelled in a drift path. Furthermore, if the beam dynamics upstream of the slit is space charge dominated, the aperture reduces the bunch charge so one may assume emittance dominated dynamics in the downstream beamlet. At large bunch charges ($\sim10$\,pC) the emittance is overestimated due to space charge effects in the beamlets \cite{Bazarov2008}. Such effects can be neglected in this work because the bunch charges are typically 0.1\,pC.


\begin{figure}[ht]
\centering
\includegraphics[width=.75\textwidth]{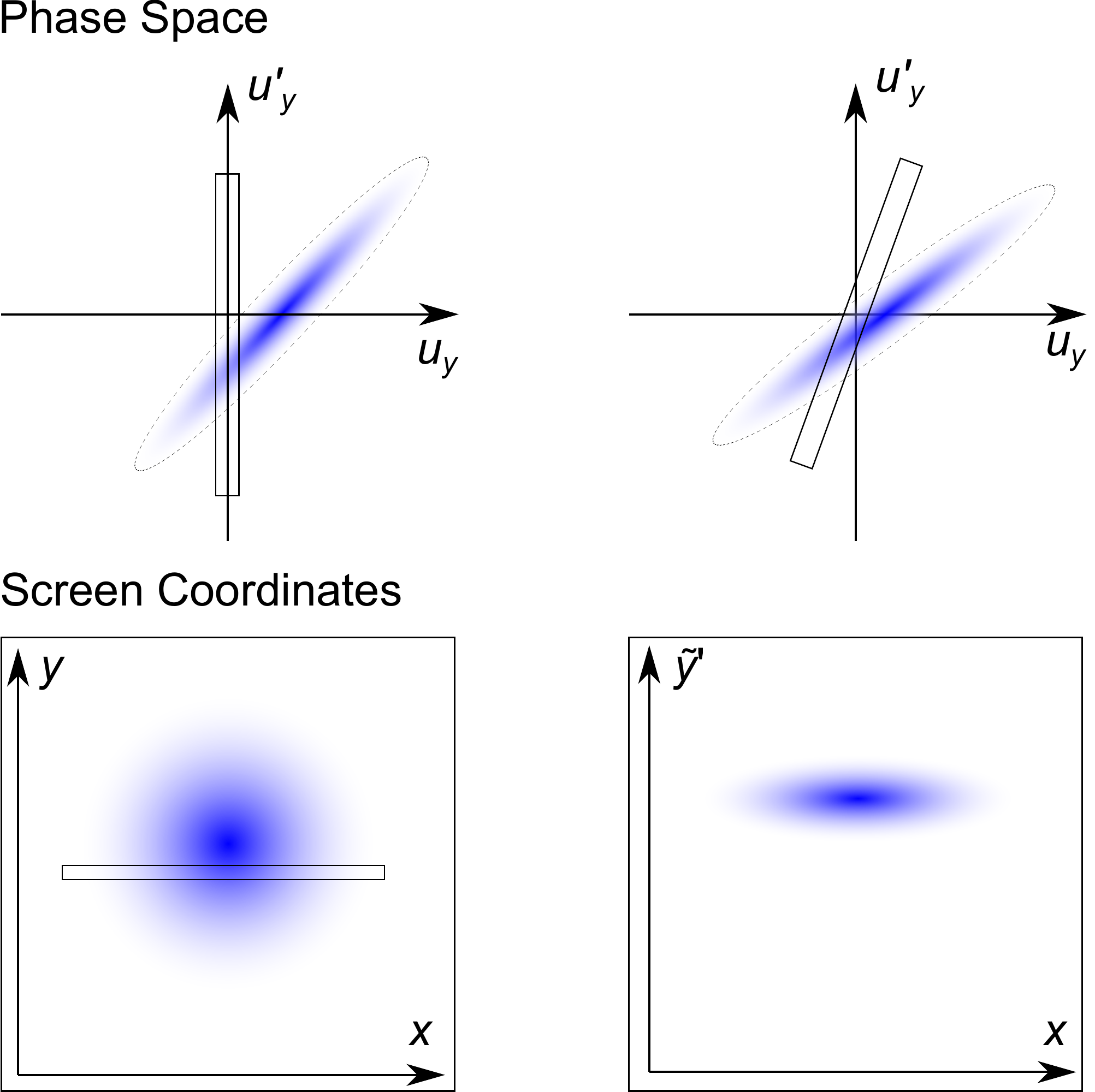}
\caption[Particle distributions during a slit mask measurement]{Illustration of real and phase space distributions during a slit mask measurement.}
\label{fig:real-and-phase-space}
\end{figure}

The emittance can be evaluated using the second moments as noted hereafter.
$y$ and $y'$ denote spatial positions of particles on a screen, $u_y$ and $u_y^\prime$ denote phase space coordinates for vertical position and divergence as indicated in figure \ref{fig:real-and-phase-space}:

\begin{equation}
\langle u_y^2 \rangle_{Beam} = \frac{\sum\limits_i^{Particles} y_{i}^2}{\sum\limits_i^{Particles} 1} = \frac{\sum\limits_j^{Beamlets} I_j \langle y^2 \rangle_{j}}{\sum\limits_j^{Beamlets} I_j}
\end{equation}

\begin{equation}
\langle u_y^{\prime 2} \rangle_{Beam} = \frac{\sum\limits_i^{Particles} y_i^{\prime 2}}{\sum\limits_i^{Particles} 1} =
 \frac{\sum\limits_j^{Beamlets} I_j  \langle y^{\prime 2} \rangle_j}{\sum\limits_j^{Beamlets} I_j} =
 \frac{\sum\limits_j^{Beamlets} I_j ( \mu_j^2 + \sigma_j^2 )}{\sum\limits_j^{Beamlets} I_j} 
\end{equation}

\begin{equation}
\langle u_y u_y^{\prime} \rangle_{Beam} = \frac{\sum\limits_j^{Beamlets} I_j \langle y\rangle_j \langle y^{\prime}\rangle_j}{\sum\limits_j^{Beamlets} I_j}
\end{equation}

here, $I_j$ is the measured intensity of the $j$-th beamlet on the screen, $\mu_j$ and $\sigma_j$ are center and standard deviation of an assumed normal distribution of the individual divergences $y^{\prime}_i$ in one beamlet.\\
Note, that in case of steering $y$ and $\mu_j$ need to be corrected for the collective divergence due to beam steering:

\begin{equation}
y^{\prime} = \frac{\tilde{y}^{\prime} - \alpha L}{l_{beamlet}}
\end{equation}

\begin{equation}
\mu_j = \langle y^{\prime} \rangle = \frac{\langle \tilde{y}^{\prime} \rangle - \alpha L}{l_{beamlet}}
\end{equation}

where $\alpha$ is the steering angle, $l_{beamlet}$ is the drift length of the beamlet between slit mask and screen and $L$ is the length between steerer and screen. $\tilde{y}^{\prime}$ denotes the measured position on a screen, that is shifted by the steerer angle.

\clearpage
\pagebreak
\section*{Solenoid Scanning}

\begin{figure}[ht]
\centering
\includegraphics[width=.75\textwidth]{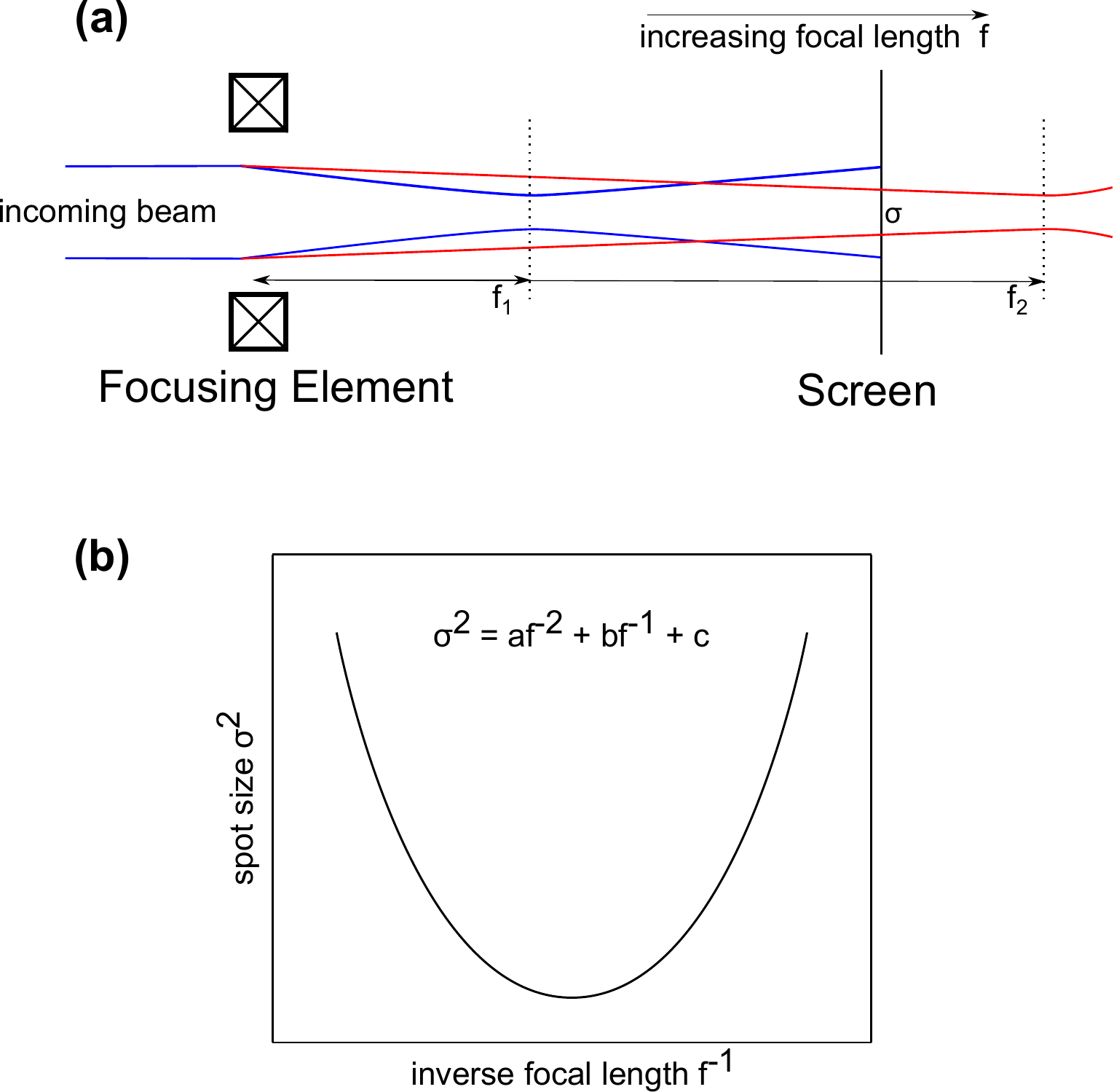}
\caption[Measurement setup for a solenoid scan]{Scheme of the measurement setup for a solenoid scan (a) and illustration of the quadratic fit to obtain parameters for the emittance calculation (b).}
\label{fig:scheme-solenoid-scan}
\end{figure}

Using the solenoid to scan the beam waist through a screen yields information about the initial beam size, divergence and their correlation, as illustrated in figure \ref{fig:scheme-solenoid-scan}b.
Photoinjectors are equipped with an emittance compensating solenoid that is used for these measurements.

In a linear beam optics model, the beam radius $r_{f,L}$ at the screen position $L$ depends on the initial beam radius $r_i$ and the divergence after the solenoid $r_f^\prime$. $r$ and $r'$ are the radial position and divergence of a particle at initial (before solenoid) and final (screen) positions, indicated by an i or f subscript, respectively:
\begin{equation}
\langle r_{f,L}^2 \rangle = \langle ( r_i + r_f^\prime \cdot L)^2 \rangle \, .
\end{equation}
Expanding the equation above with the solenoid's linear effect on the divergence
\begin{equation}
r_f^\prime = r_i^\prime - \frac{r_i}{f}
\end{equation}
yields a quadratic dependence of the measured beam radius on the solenoid's focal length $f$ \cite{myVolker2012}
\begin{equation}
\langle r_{f,L}^2 \rangle = \frac{1}{f^2} \underbrace{(\langle r_i^2 \rangle L^2)}_a + \frac{1}{f} \underbrace{(-2L(L\langle r_i r_i^\prime \rangle + \langle r_i^2 \rangle))}_b + \underbrace{\langle r_i^{\prime 2} \rangle L^2 + 2L\langle r_i r_i^\prime \rangle + \langle r_i^2 \rangle}_c
\end{equation}
The parameters $a, b$ and $c$ of the parabola can be obtained from a quadratic fit of the beam size, plotted against the inverse focal lengths of the solenoid and the rms emittance is given by \cite{myVolker2012}
\begin{equation}
\varepsilon_{rms} = \frac{\sqrt{ac-\frac{b^2}{4}}}{L^2}
\end{equation}



%

\section{Experimental Setup}
\label{chap:Setup}

An SRF injector is driven by a laser source that extracts photoelectrons from a cathode. The cathode is located in the back wall of an RF cavity where the electron bunches are accelerated by a strong electric field gradient, timed to match the correct phase of the RF wave.
The beam is fed into a diagnostics beam line, where fundamental beam parameters can be measured.

\begin{figure}[ht]
\centering
\includegraphics[width=.75\textwidth]{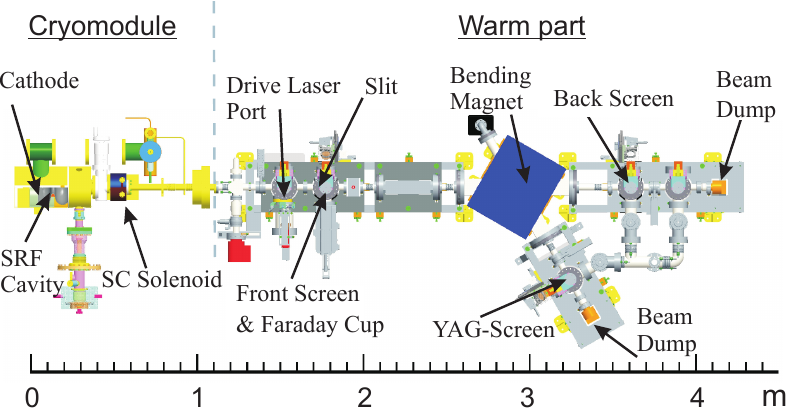}
\caption[Cold part in the HoBiCaT cryomodule and the diagnostic beamline]{Overview of the cold part in the HoBiCaT cryomodule and the diagnostic beamline.}
\end{figure}

\subsection{Electron Gun}

The electron gun was the second prototype ("Gun 0.2") in a step-by-step approach towards the \berlinpro~injector.
It consisted of a 1.6 cell superconducting niobium cavity that was located in the HoBiCaT cryostat, at an operational temperature of 1.8\,K, and a cathode plug inserted into the back wall of the cavity.
The cavity was fed with microwave power from an inductive output tube (IOT) or a solid state amplifier at 1.3\,GHz and was operated at field gradients up to 28\,MV/m, limited by field emission followed by slow quench. Stable operation was possible up to 27\,MV/m.
The cathode plug was made of niobium covered with a thin lead film. Lead has an approximately one order of magnitude higher quantum efficiency than niobium and is also superconducting below 7.2\,K.
Table \ref{tab:gun-parameters} lists typical operational parameters of the gun.

\begin{table}[!ht]
\centering
\setlength{\arrayrulewidth}{0.25pt}
\begin{tabular}{l c c}
\textbf{Parameter} & \textbf{Value} & \textbf{Unit}    \\ \midrule

Average Current    & $<$ 1          & nA               \\
Bunch Charge       & 0.187          & pC               \\
QE                 & $10^{-5}$      &                  \\
Beam Energy        & 1-2.5          & MeV              \\
Laser Power        & $<$ 0.5        & mW               \\
Laser wavelength   & 258            & nm               \\
Pulse Length       & 2.5 \ldots 3   & ps fwhm, gaussian\\
Rep rate           &  8             & kHz              \\
$\mathrm{E_{max}}$ & \parbox{5cm}{\centering 10 - 12.5 w solid amp\\ 22 with IOT\\ 27 peak field} & MV/m  \\ 
\end{tabular}
\caption{Typical operational parameters of Gun 0.2 with a lead coated cathode. The pulse length refers to the emission time of the bunch. }
\label{tab:gun-parameters}
\end{table} 

\subsection{Diagnostic Beam Line}

Several screens were employed to image beam profiles and to aid beam positioning. 
A superconducting solenoid with an effective magnetic length of 41.5\,mm and a field amplitude of $44.1\,\frac{\mathrm{mT}}{\mathrm{A}}$ \cite{myVolker2012} was located inside the cryomodule at a distance of 439\,mm to the cathode.
The solenoid can be used to focus the beam on the front or back screens in the straight beam line,
where a solenoid scan yields emittance information as described in the beginning of section \ref{chap:measuring-emittance}.
Three beam stops made of copper served as faraday cups for current measurements. They were located at the first screen station and at the two ends of the beamline.

The slit at the first screen station was used to select a narrow beamlet which was imaged on the back screen.
The slit mask was made of 1.5\,mm thick tungsten with an $100\,\mu$m aperture and located directly in front of the first screen and faraday cup. A thickness of 1.5\,mm gives a good compromise between efficient suppression of the background and sufficient transmission.

In the dispersive section, after the beam has passed the dipole, it was possible to evaluate the momentum of the particles. Because the dipole could not be driven into saturation it was calibrated by cycling between +8\,A and -8\,A \cite{myMatveenko}. The beam momentum can be calculated as

\begin{equation}
pc = 0.88 \cdot \left( \frac{I_{Dipole}}{[A]} + 0.12 \right) [MeV].
\end{equation}

\subsection{Procedures}

Three individual measurements were conducted to characterize the phase space of one beam setting using the slit mask. First, the beam was imaged on the front screen without slit to calibrate the steerer angle and vertical offset. In a second sweep the front screen was used again, but with the slit to measure the beamlet intensity. Finally, the beamlets were allowed to drift towards the back screen in order to measure the divergence of each beamlet.
All measured values were taken from ten polls per beamlet on the back screen and five polls per beamlet on the front screen. Mean and standard deviation were recorded.
The phase space distributions were corrected for the steerer angle and summed up to obtain second moments and the emittance as described in \cite{Anderson2002}.

In order to conduct a solenoid scan, the current in the solenoid was varied over a specified range \cite{myVolker2012}.
Screen images were saved for a single poll per solenoid setting,
while beam positions and diameters were averaged over six polls. 
Beam parameters and the emittance are obtained from a quadratic fit of the beam width against the inverse focal length of the SC solenoid.

\section{Results}
\label{chap:results}

All measurements were conducted in November 2012 during the last run of the "Gun 0.2" setup. At that time, the cavity was operated with full RF power from the IOT only for test measurements and energy calibration. Due to a failure in the power supply of the IOT only a solid state amplifier was available for the subsequent beam measurements with a field gradient of up to 12.5\,MV/m.


\subsection{Slit Mask Measurements}

\begin{figure}[ht]
\centering
\includegraphics[width=.75\textwidth]{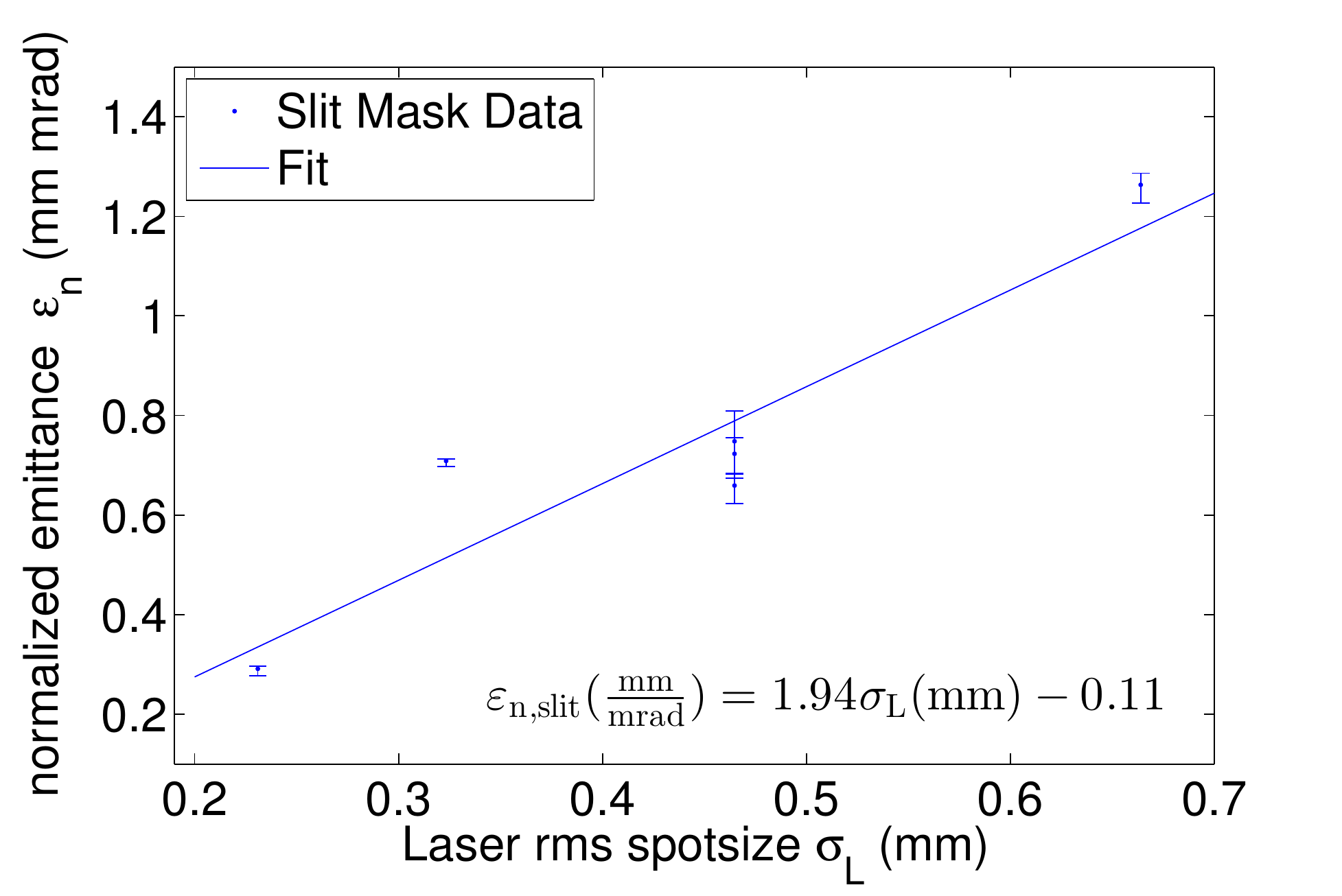}
\caption[Results from slit mask measurements]{Results of emittance measurements using the slit mask technique. Bunch charges are not equal.}
\label{fig:slit-mask-results}
\end{figure}

The vertical phase space was characterized at different laser spot sizes with a constant cavity gradient of 10\,MV/m at an emission phase of approximately 15\,deg using the slit mask as described in section \ref{chap:slit}. An aperture in the laser beam line was used to manipulate the spot size, however, other laser settings remained constant so the bunch charges were not equalized among the measurements (but all below the space charge regime). At a spot size of 0.46\,mm\,rms three solenoid settings were studied. The resulting emittance values are displayed in figure \ref{fig:slit-mask-results}.

\begin{figure}[ht]
\centering
\mbox{\includegraphics[width=.5\textwidth]{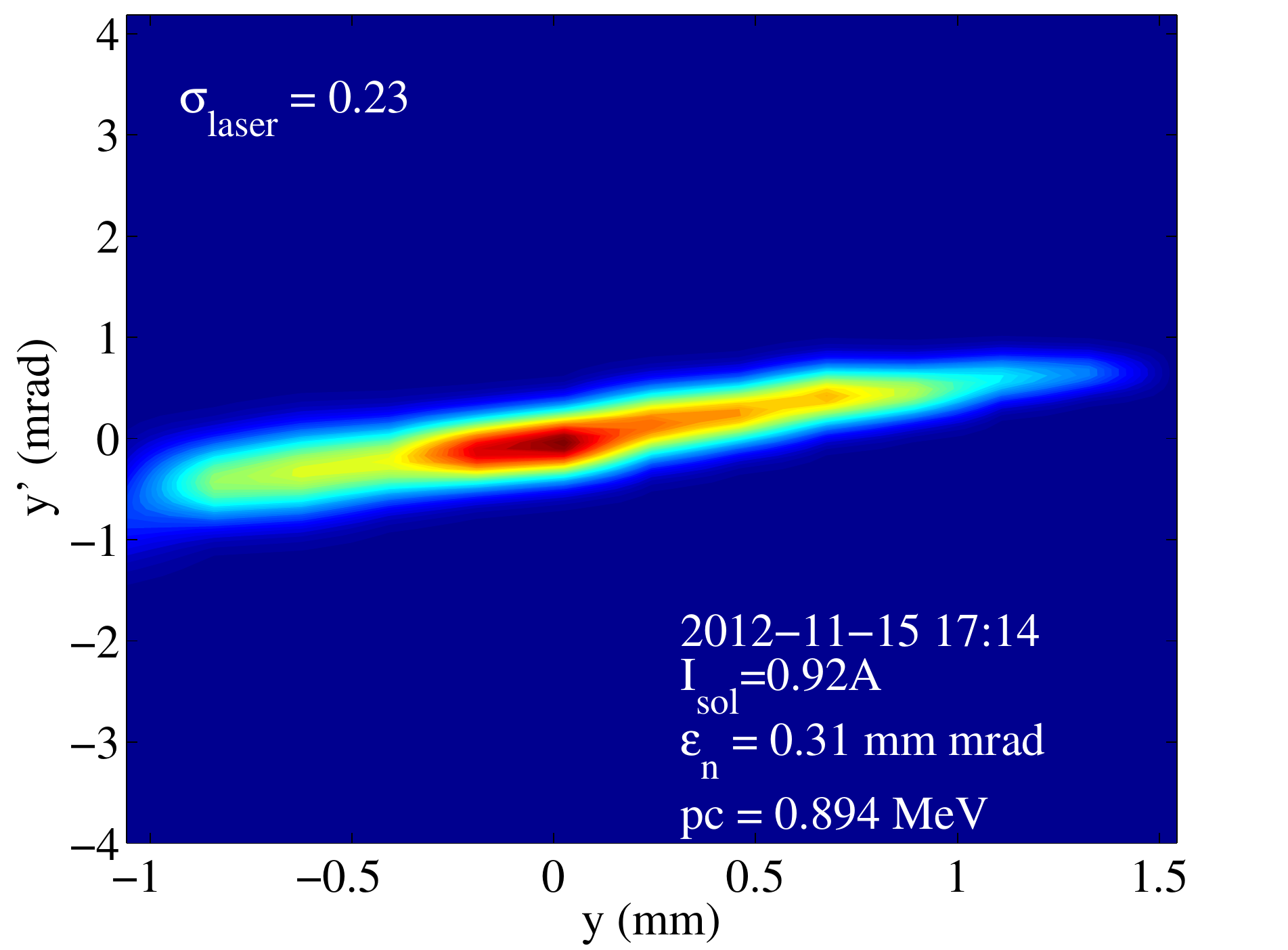}
\includegraphics[width=.5\textwidth]{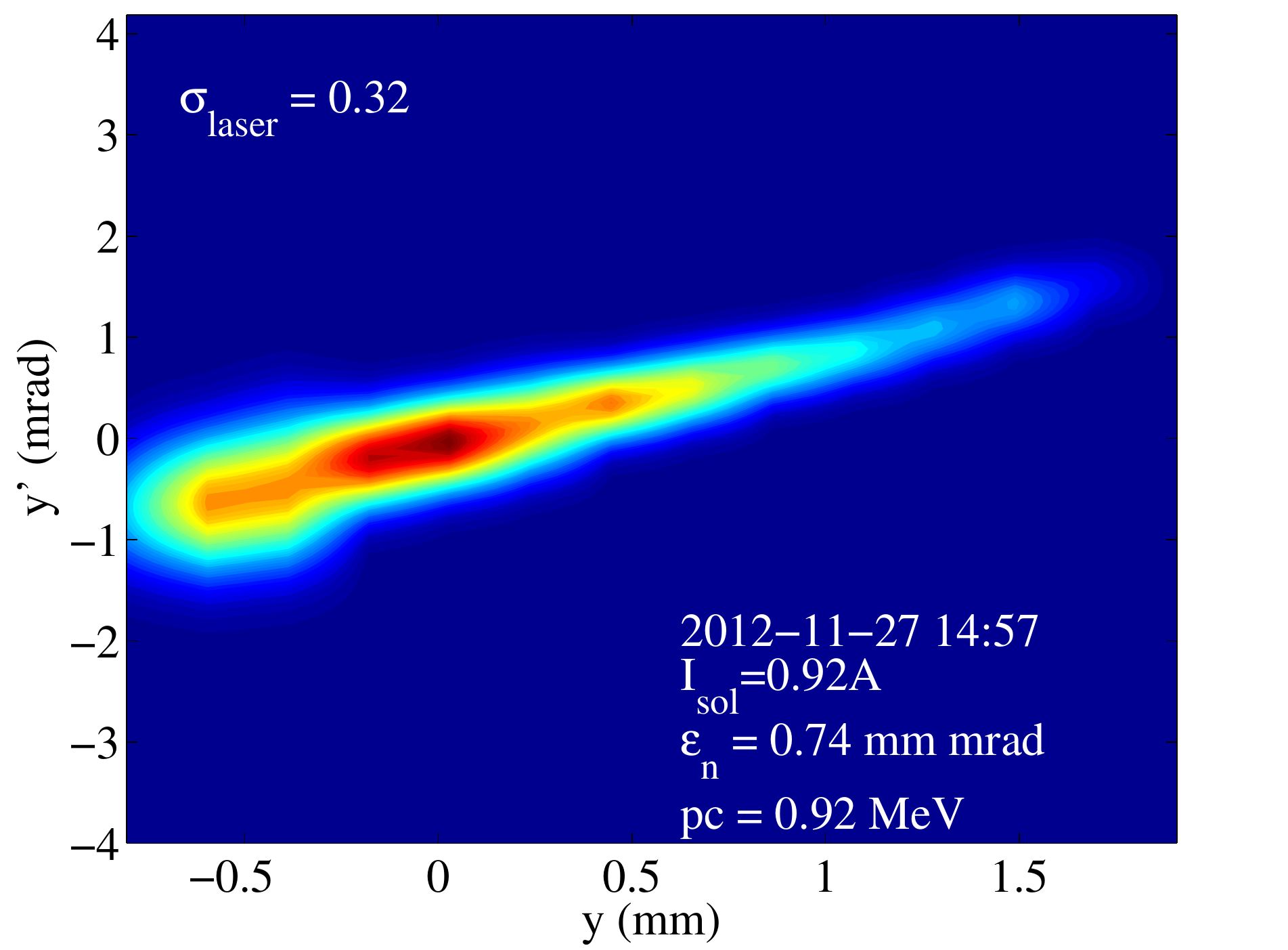}}
\mbox{\includegraphics[width=.5\textwidth]{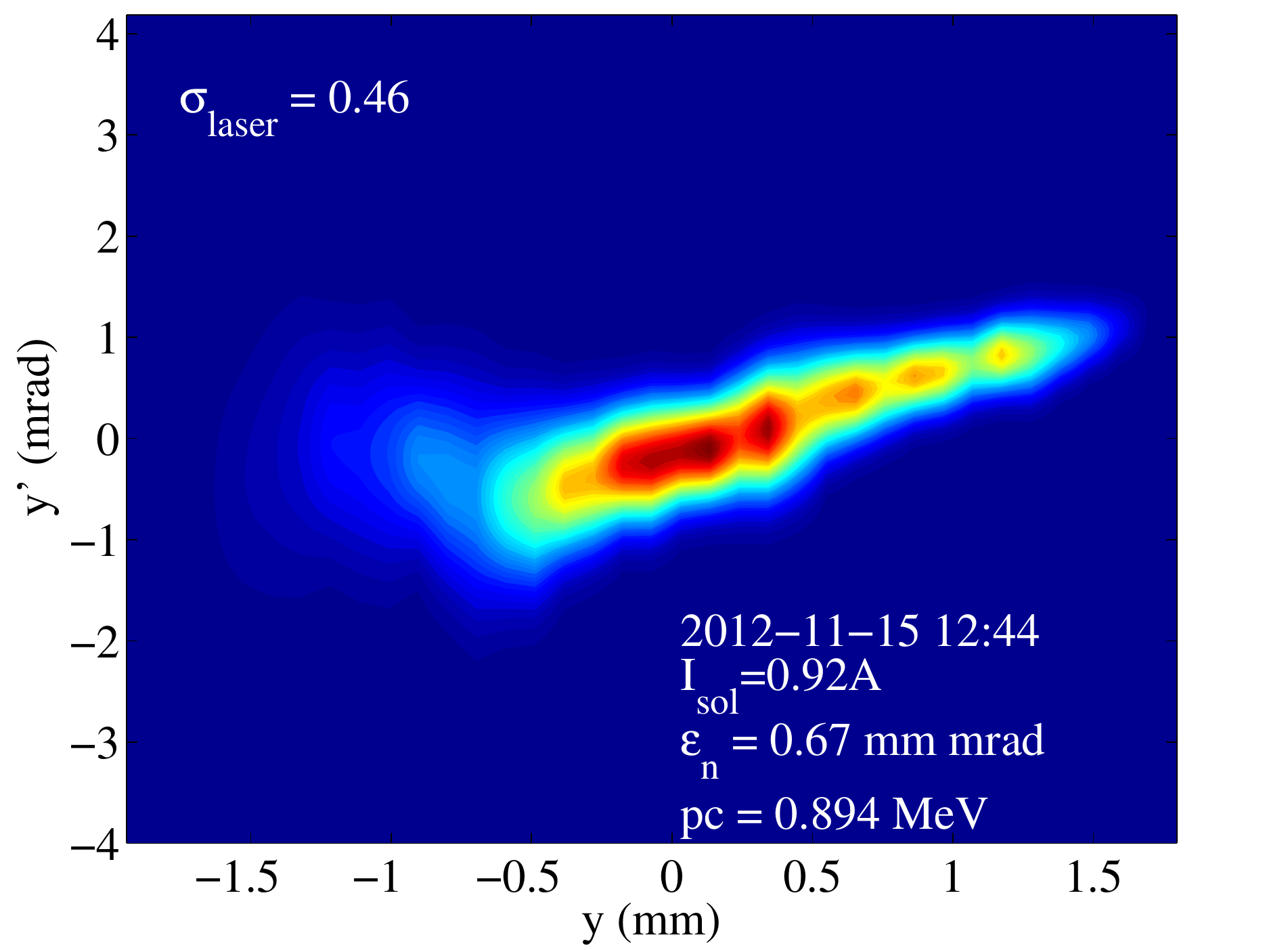}
\includegraphics[width=.5\textwidth]{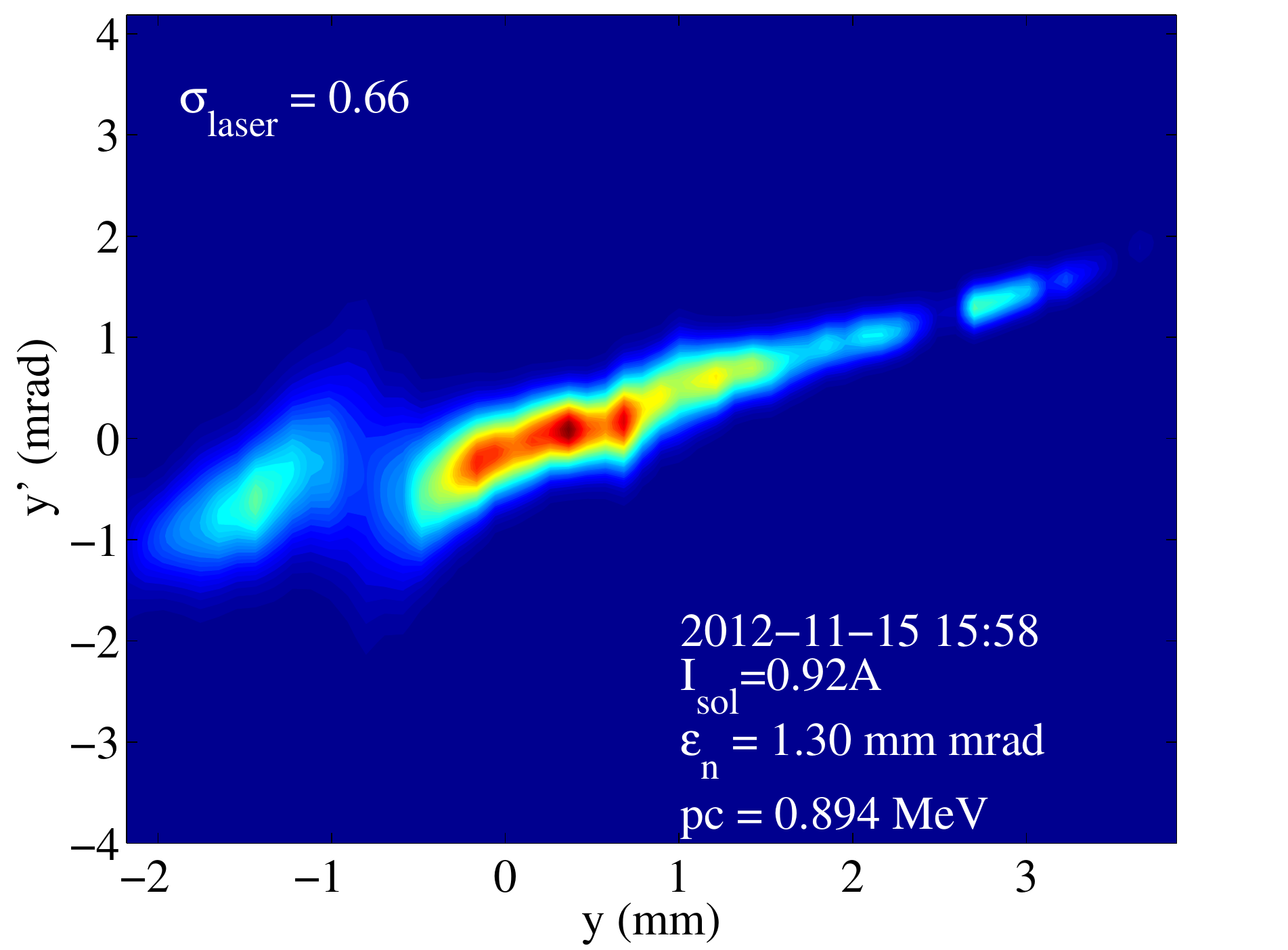}}
\caption[Phase space reconstructions at different laser spot sizes]{Reconstructions of vertical phase spaces, laser spot size increases in reading direction and is indicated in the figures.}
\label{fig:phase-spaces-laser}
\end{figure}

A linear dependency of the normalized emittance with respect to the rms laser spot size can be recognized, despite the scattering of the data. The corresponding phase spaces are displayed in figure \ref{fig:phase-spaces-laser}, where at large spot sizes a structured phase space was obtained. This hints at structured emission from hot spots on the cathode surface which may be covered with protrusions and droplets.

Uncertainties due to temporal deviations and other statistical errors are estimated to amount to less than 5\,\%. The finite thickness of the YAG screen together with a 45\,deg viewing angle may introduce an overestimation of the measured beam size of up to 30\,$\mu$m \cite{myBardayIBIC2012}. From a numerical estimate this will introduce a systematic error of about +3\,\% in the emittance. Space charge effects are negligible at the observed bunch charges below 1\,pC. Only these effects are taken into account by the error bars in figure \ref{fig:slit-mask-results}.

The accuracy of the evaluated emittance is largely defined by the dynamic range of the entire measurement, which differs between measurements because beamlets at the tail of the distribution might move off the screen due to geometrical constraints.
The range differs between 2.5 and 6.4\,dB, which implies that differences as large as
20\,\% in the number of considered particles occur.
For reference, the dependence of the emittance on the amount of particles imaged is shown in a typical charge cut curve of a round, gaussian beam that was tracked through the photoinjector using ASTRA.

\begin{figure}[ht]
\vspace{-0.1cm}
\centering
\includegraphics[width=.45\textwidth]{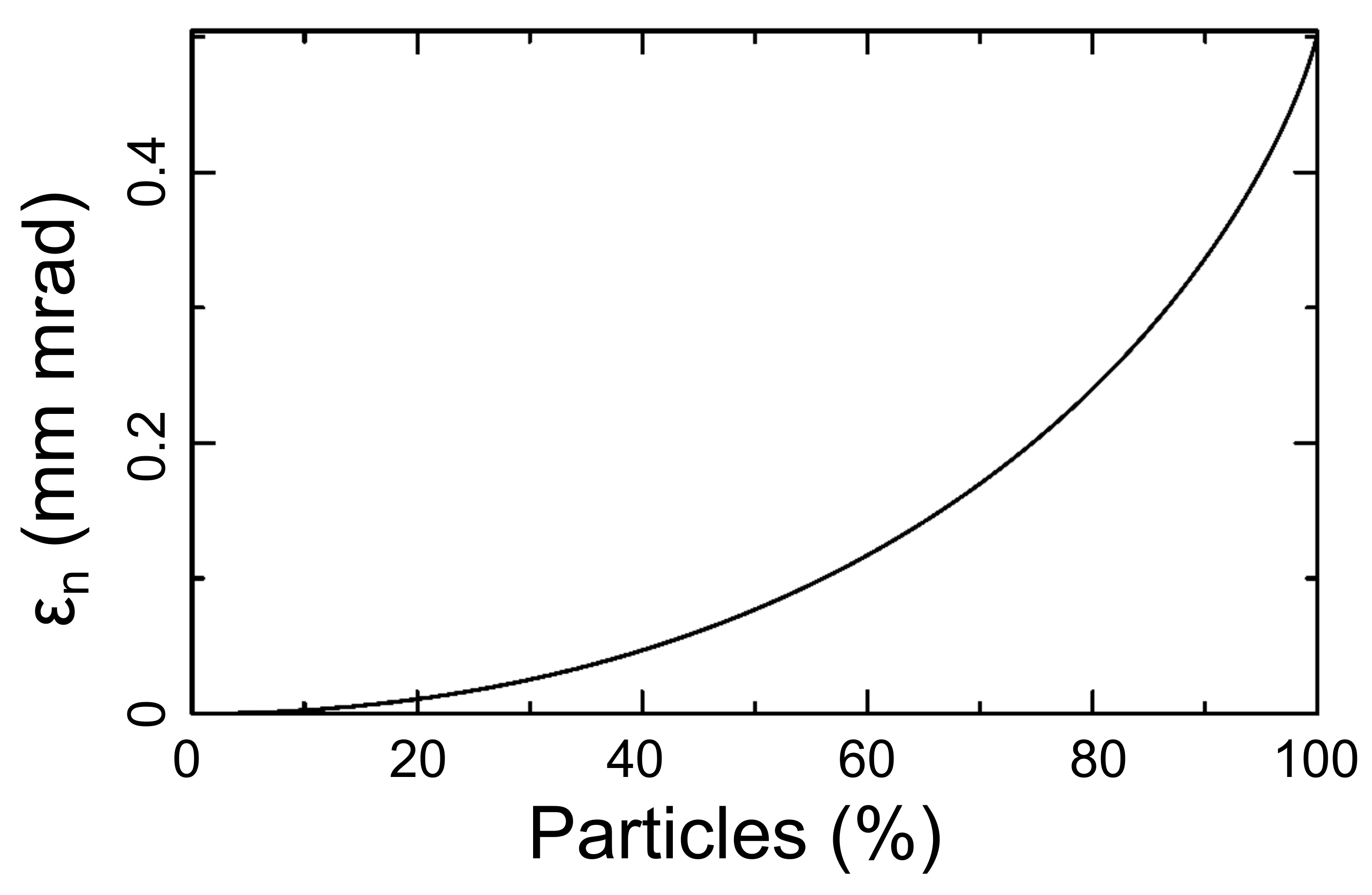}
\caption[Charge cut curve]{Typical charge cut curve of a round, gaussian beam from the photoinjector.}
\label{fig:charge-cut}
\end{figure}

\begin{figure}[ht]
\centering
\mbox{\includegraphics[width=.5\textwidth]{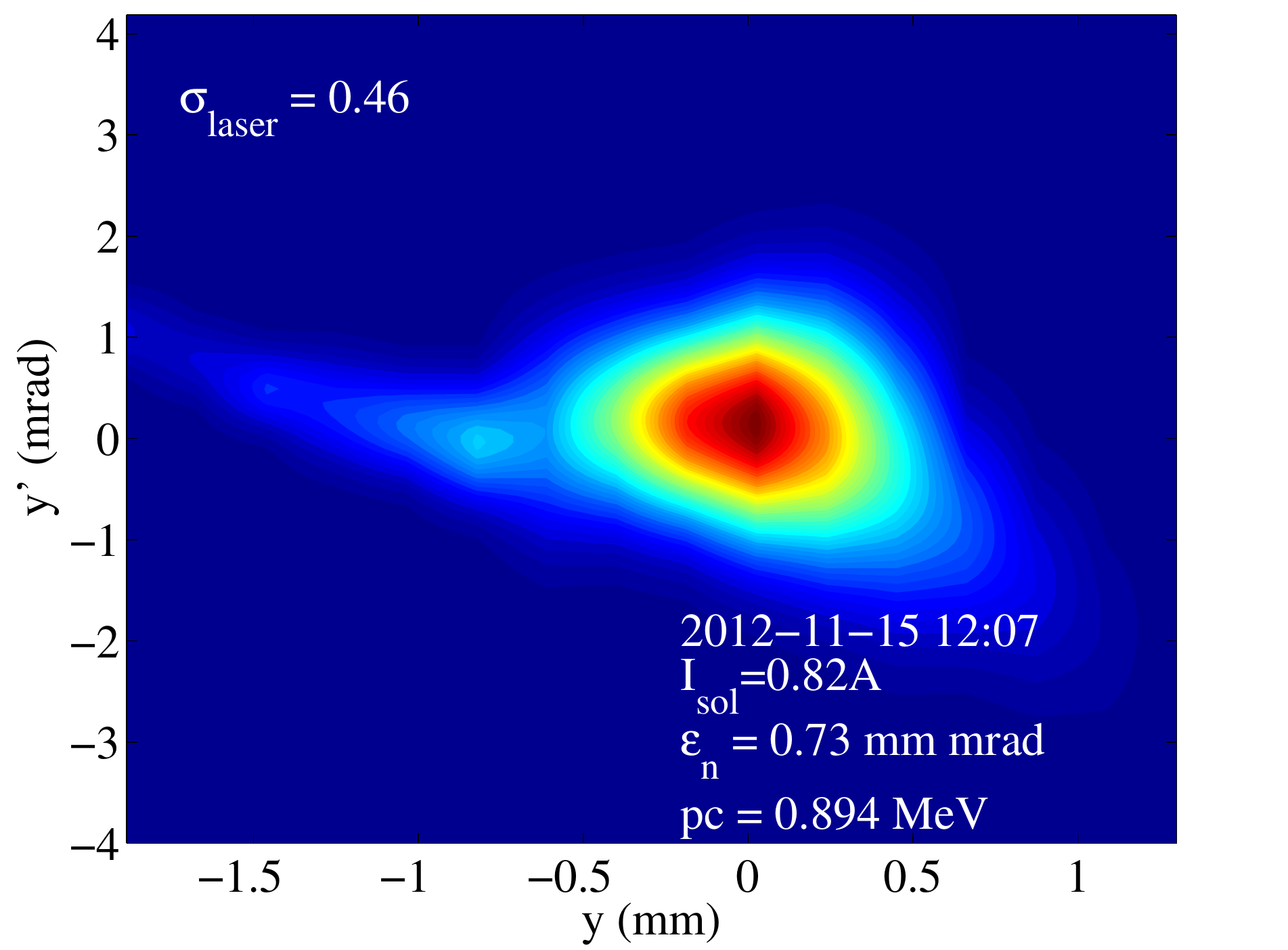}
\includegraphics[width=.5\textwidth]{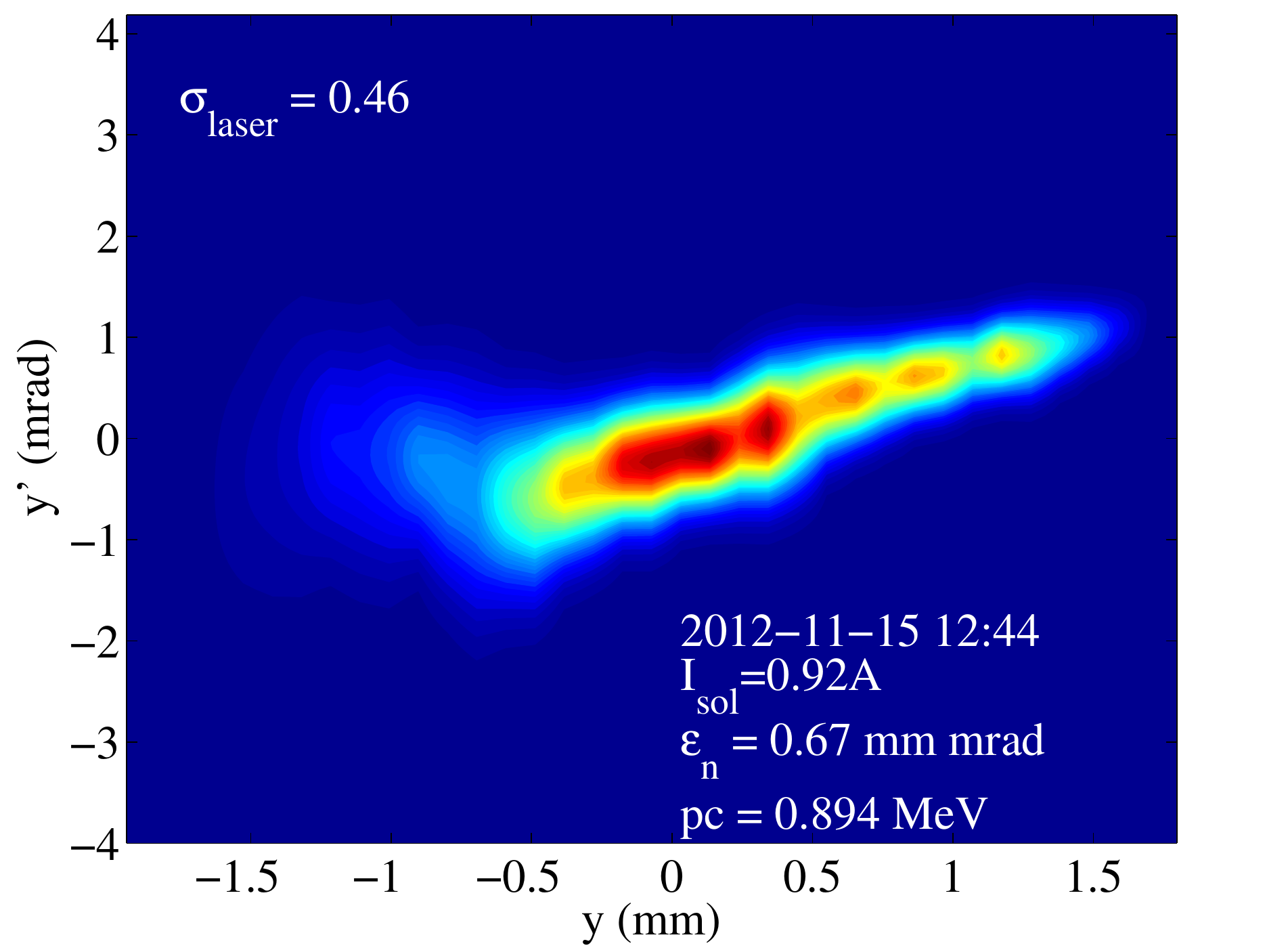}}
\includegraphics[width=.5\textwidth]{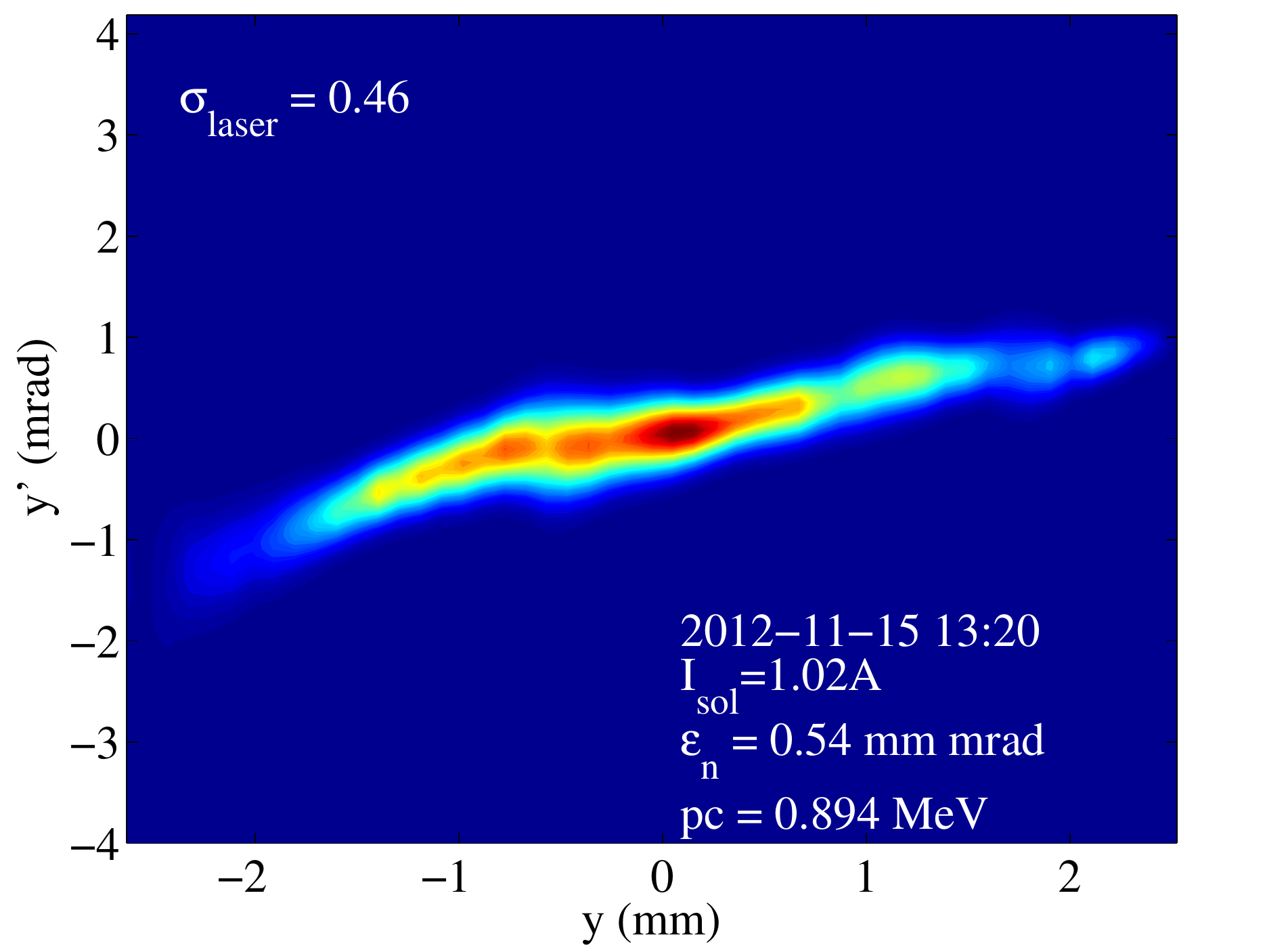}
\caption[Phase space reconstructions at different solenoid currents]{Reconstructions of vertical phase spaces at increasing solenoid currents, as indicated in the figures.}
\label{fig:phase-spaces-sol}
\end{figure}

\clearpage
Three measurements with identical beam settings but increasing solenoid current were performed. The phase spaces are printed in figure \ref{fig:phase-spaces-sol}. The correlation between the vertical position and divergence is clearly increasing with the solenoid current as the focal point moves closer to the solenoid.

\subsection{Solenoid Scans}

\begin{figure}[!ht]
\centering
\includegraphics[width=.75\textwidth]{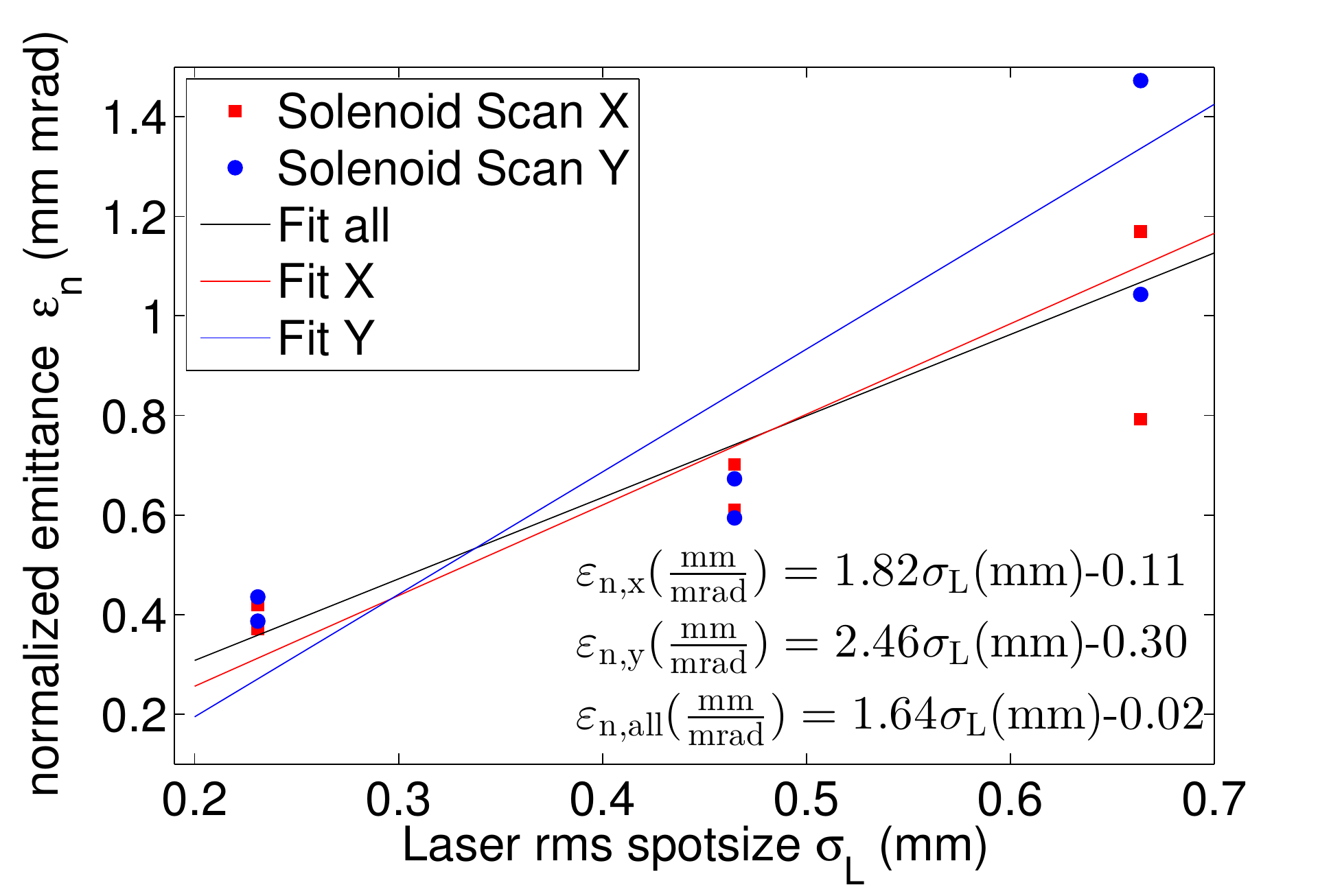}
\caption[Results from solenoid scan measurements]{Emittance measurements using the solenoid scan technique at different laser diameters. Bunch charges are not equal. The lower values are from measurements at 15\,deg launch phase, the higher ones at 25\,deg. Horizontal and vertical fits are for the 15\,deg measurement only.}
\label{fig:sol-scan-laser}
\end{figure}

Emittance measurements with different laser spot sizes were also carried out using the solenoid scan, see figure \ref{fig:sol-scan-laser}. The data was taken with a field amplitude of 10\,MV/m at launch phases of 15 and 25\,deg, which corresponds to beam energies of 0.94 and 0.90\,MeV, respectively. The beam was imaged on the front screen, at a distance of 1.121\,m to the center of the solenoid coil. As was observed with the slit mask measurements, the data shows a linear dependency, however, the values are slightly larger in this case.

\begin{figure}[ht]
\centering
\includegraphics[width=.75\textwidth]{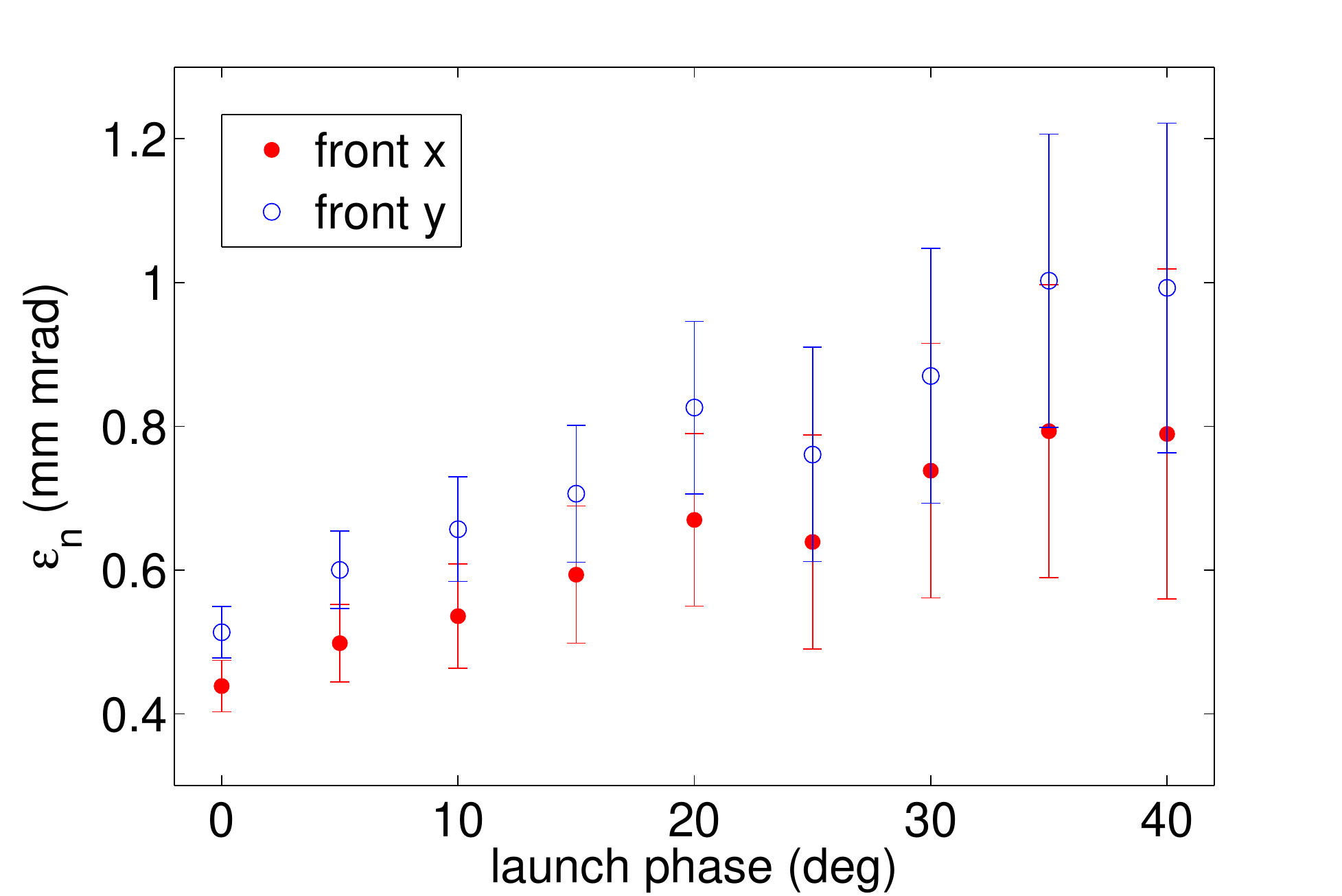}
\caption[Emittance measurements for different emission phases - front]{Emittance measurements for different emission phases on the front screen. Error bars account for chromatic aberration and astigmatism.}
\label{fig:PhaseScan-emittances-front}
\end{figure}

\begin{figure}[ht]
\centering
\includegraphics[width=.75\textwidth]{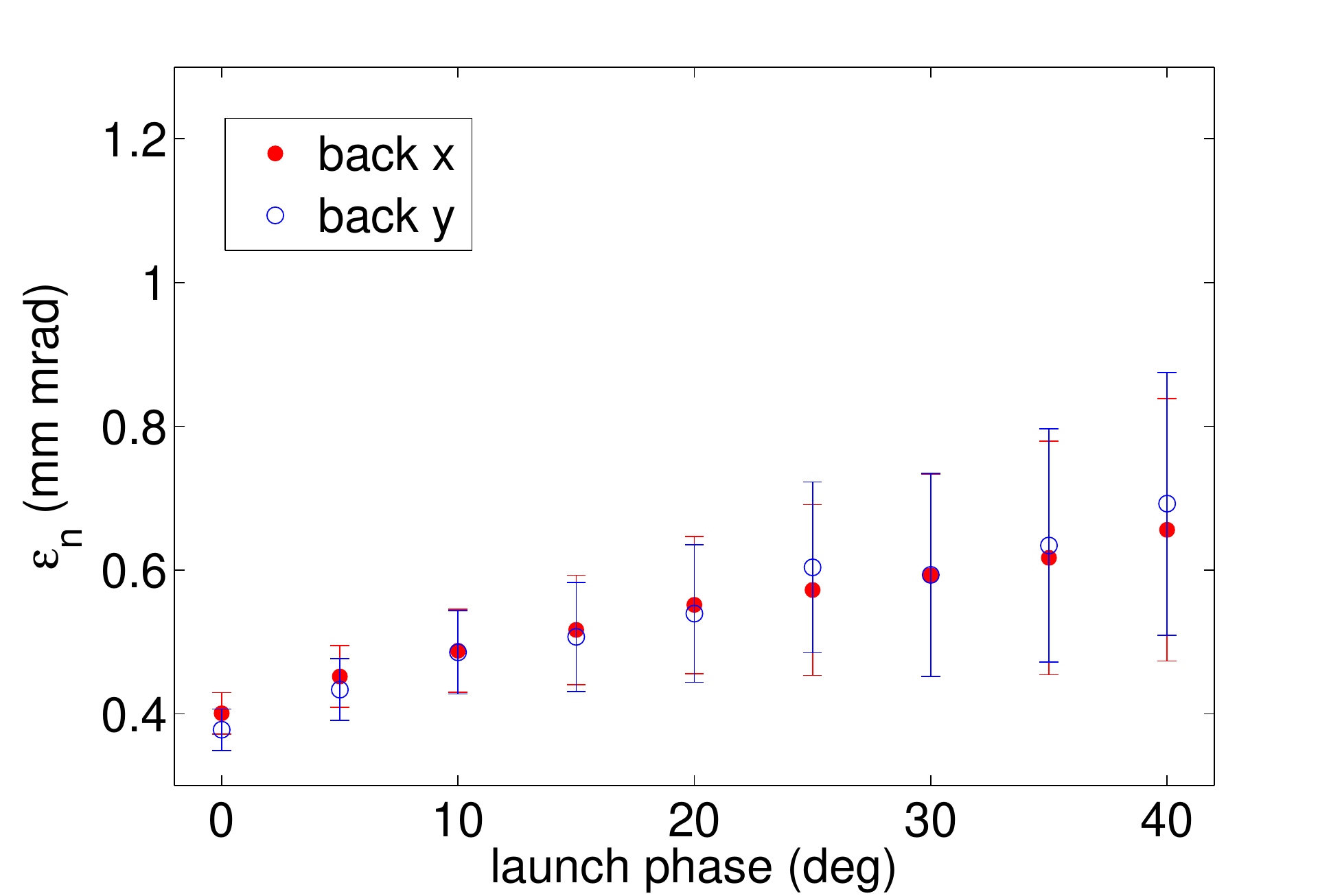}
\caption[Emittance measurements for different emission phases - back]{Emittance measurements for different emission phases on the back screen. Error bars account for chromatic aberration and astigmatism.}
\label{fig:PhaseScan-emittances-back}
\end{figure}

\begin{figure}[ht]
\centering
\includegraphics[width=.75\textwidth]{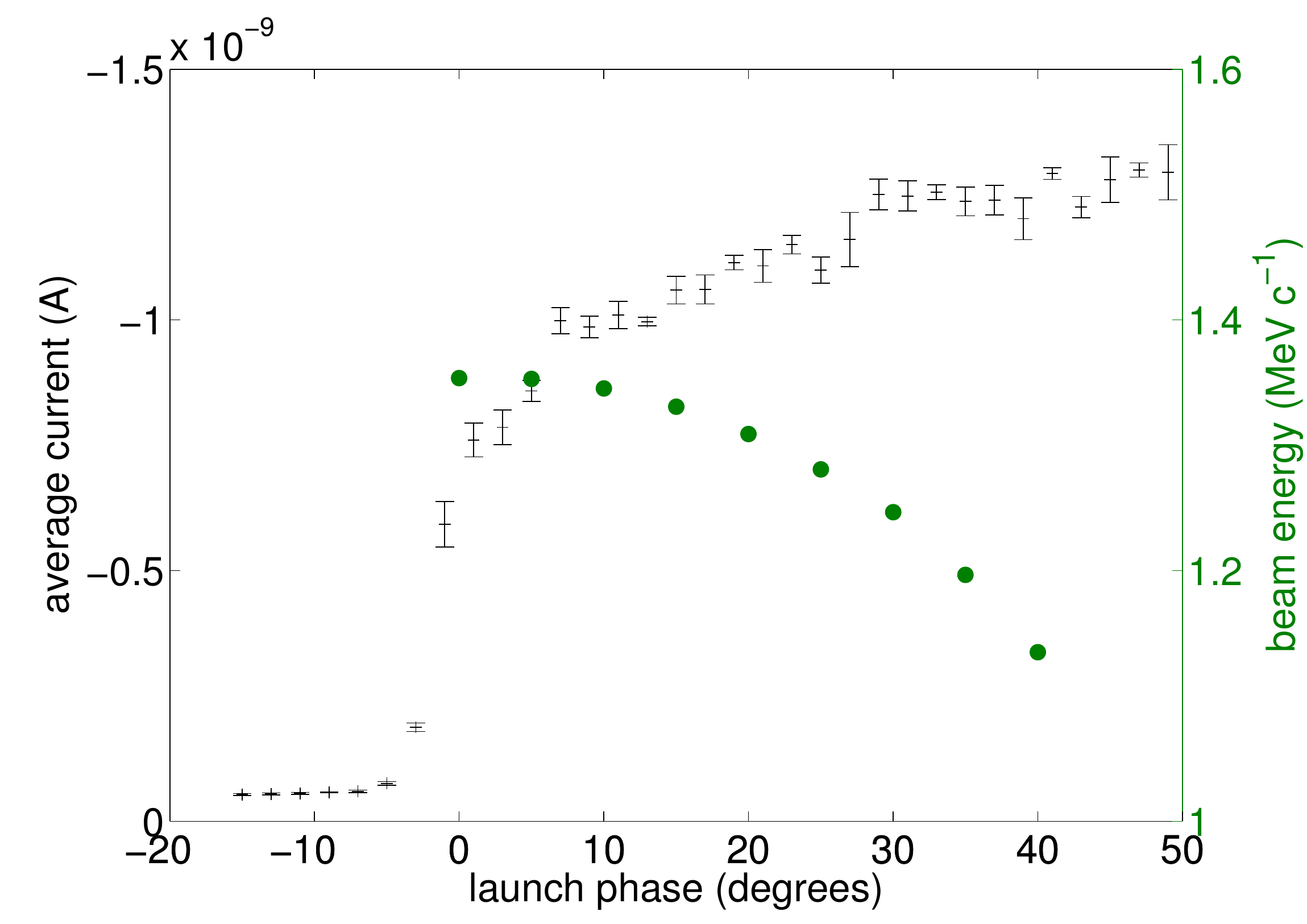}
\caption{Phase scan of the beam current and energy.}
\label{fig:PhaseScan-current-energies}
\end{figure}

Additionally, emittance measurements for different emission phases were conducted at a gradient of 12.5\,MV/m.
The resulting emittances are displayed in figures \ref{fig:PhaseScan-emittances-front} and \ref{fig:PhaseScan-emittances-back}.
For reference, the corresponding average currents and beam energies can be found in figure \ref{fig:PhaseScan-current-energies}.
The emittance rises linearly with the launch phase up to 30\,deg and seems to flatten between 30 and 40\,deg.

Values measured using the front screen are clearly larger than the ones measured on the back screen.
Starting from 5\,\% at low launch phases the difference increases also linearly with the phase to up to 20\,\% at phases of 30 and 40\,deg.
Higher emittance values on the front screen, where the beam has to be focused stronger, suggest that strong focussing introduces aberrations that lead to an increase of the emittance.
The effect of chromatic and spherical aberrations of the solenoid was investigated with several simulations in \cite{myVolker2012}.
For similar beam parameters and the same solenoid it was shown that the spherical aberration coefficient $C_s$ increases from about 630~m$^{-3}$ to about 960~m$^{-3}$ when moving the focus from the back to the front screen.
The emittance increase due to spherical aberrations depends on higher-order moments of the particle distribution \cite{myVolker2012}:
\begin{equation}
\varepsilon^2 = \varepsilon^2_{(lin)} + 2 \, C_s\left(\langle r^3_i r_i^\prime \rangle \langle r_i^2 \rangle - 3 \langle r_i r_i^\prime \rangle\langle r_i^2 \rangle^2 \right) + 6 \, C_s \, \langle r_i^2 \rangle^4
\end{equation}
which could not be measured, however an estimate with values from simulations and the slit mask measurements yields an emittance increase that is several orders of magnitude lower than the measured difference.

\begin{figure}[ht]
\centering
\includegraphics[width=.75\textwidth]{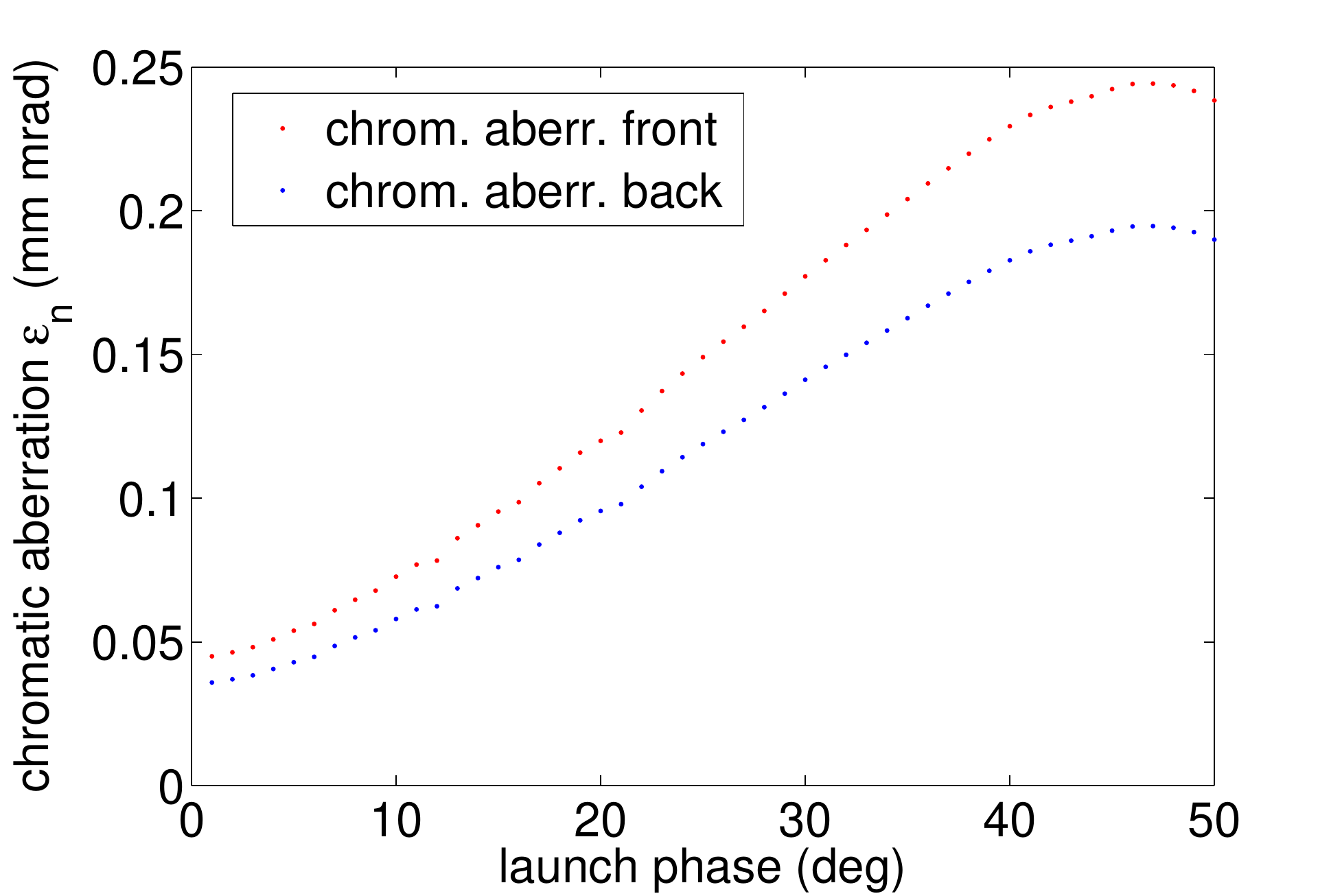}
\caption[Influence of the chromatic aberration of the solenoid on the beam emittance]{Influence of the chromatic aberration of the solenoid on the beam emittance for a beam at 12.5\,MV/m accelerating gradient. Front and back refers to solenoid settings where the beam is focused on the front and back screen, respectively.}
\label{fig:chromaticity}
\end{figure}

The chromatic aberration follows \cite{Dowell}
\begin{equation}
\varepsilon_{n,chrom} = \sigma^2_{x,sol} \, \frac{\sigma_p}{mc} \, K |\sin KL + KL \cos KL|
\end{equation}
where $\sigma_{x,sol}$ is the rms beam width at the solenoid, $\sigma_p$ is the rms momentum spread, $K=\frac{eB_0}{2p}$, and $L$ is the solenoid's effective length. Because no vertical slit was installed in front of the dipole, a precise measurement of the momentum spread was not possible. The influence of the chromatic aberration was estimated using beam size and momentum spread values from Astra simulations. As can be seen in figure \ref{fig:chromaticity}, the influence is much stronger when focusing the beam on the front screen. Overall, the chromatic aberration may be responsible for the increase of the measured emittance with the launch phase as it also increases nearly linearly below 30\,deg and flattens above 30\,deg. Additionally, it explains the discrepancy between the measurements on the front and back screen, because focusing on the front screen introduces higher aberrations due to the stronger solenoid field.

A third source of uncertainty is caused by the astigmatism of the solenoid. It causes two separated foci for the two transverse directions. When the beam is focused in $x$ direction, it may still have twice its minimal size in $y$ direction and vice versa. Because the beam is rotated by the larmor angle $\Theta_L$ with respect to the lab frame, this may result in a vast overestimation of the beam diameter on the screen and, subsequently, of the reported emittance. This effect is especially prominent in the measurements on the front screen. It is assumed that the difference between the horizontal and vertical emittance in the measurements on the front screen is due to the astigmatism. The lower uncertainty level was estimated to
\begin{equation}
\Delta \varepsilon = C \, \sigma_y\bigg|_{\sigma_x = min} \, \sin(\Theta_L)
\end{equation}
where $\sigma_y$ is the beam diameter in $y$ direction at the location of the beam waist in $x$ direction. C is a constant and value of 0.25 fits the present results. It is somewhat arbitrary because the influence of the larmor rotation on the projected intensity is not known for the specific distribution.
Chromatic aberration and astigmatism effects are accounted for in the error bars of figure \ref{fig:PhaseScan-emittances-front} and  \ref{fig:PhaseScan-emittances-back}.


\chapter{Photocathodes} \label{chap:photocathodes}

\section{Photoemission Theory} \label{sec:theory}

In 1905, Albert Einstein published his paper "Über einen die Erzeugung und Verwandlung des Lichtes betreffenden heuristischen Gesichtspunkt" \cite{Einstein1905} on the quantum nature of light and photoemission for which he received the 1921 Nobel Price in physics. It was over 50 years later that photoemission was confirmed as a bulk process instead of a surface phenomenon: in 1958, Spicer proposed the three step model of photoemission which he used to interpret his and Sommer's results \cite{Spicer1958}. The model is commonly employed to derive expressions for quantum yield, time response, and emittance of various photoemissive materials.

The derivation of theoretical expressions for the quantum yield and response time of photoemissive materials is provided by Spicer \& Herrera-Gómez \cite{Spicer1993}. More recent works propose models for the intrinsic emittance of metals (Jensen \cite{Jensen2007} and Dowell et. al. \cite{Dowell2009, Dowell2010}) and semiconductors (Flöttmann \cite{Flottmann1997}). 
In the following sections, basic features of the three step model model are reviewed and expressions for quantum yield and emittance are derived.


\subsection*{Three Step Model}

The three step model assumes that the photoemission process can be separated into three independent steps \cite{Spicer1958} :

\textbf{1. Excitation} : An electron is excited by absorption of incident light from the initial state to an excited state.

\textbf{2. Transport} : The excited electron moves through the crystal towards the surface (or away from it). Scattering effects during this process need to be discussed.

\textbf{3. Emission} : Electrons close to the surface may leave the crystal if their momentum component normal to the surface is high enough to overcome the surface barrier.

\begin{figure}[!hb]
\begin{center}
	\includegraphics[width=0.85\textwidth]{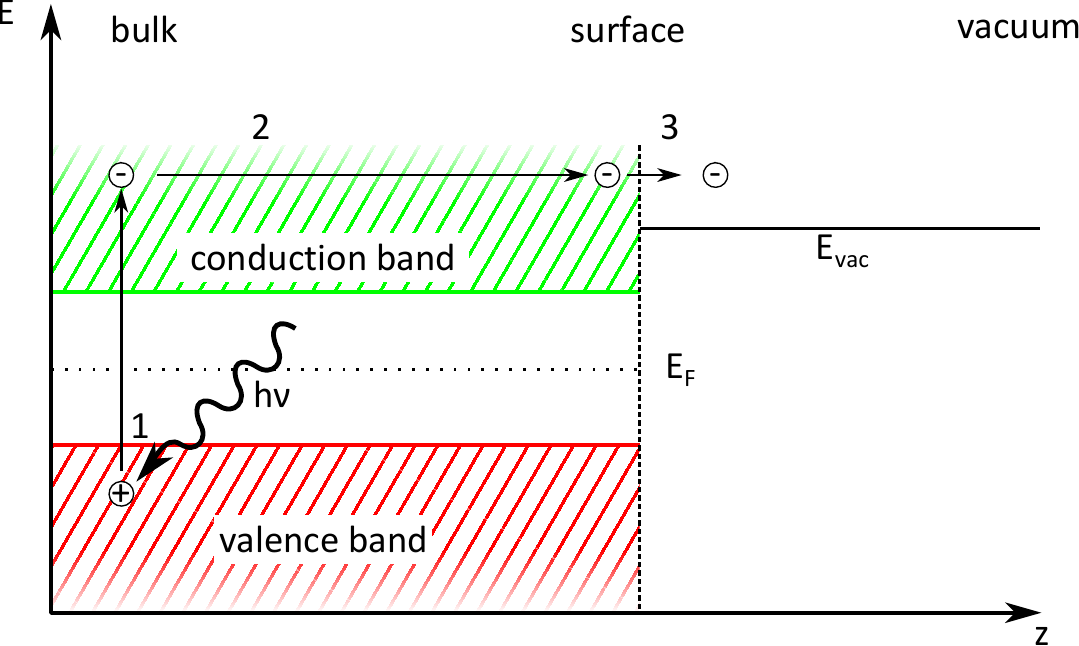}
\end{center}
\caption[Illustration of the three step model of photoemission.]{Illustration of the three step model of photoemission. The steps are indicated as: 1 -- excitation, 2 -- transport to the surface and 3 -- emission through the surface.}
\label{fig:three-step-model}
\end{figure}

The three step model itself represents an approximation to the photoemission process. It neglects the true quantum mechanic nature of the many electron ensemble by treating the excited electron as a single particle. Still, only the application of additional, strong approximations leads to a model where analytical results can be obtained: The band structure and densities of states of the crystal are reduced to a rectangular model as illustrated in figure \ref{fig:three-step-model}. Furthermore, the transition probability is assumed to depend solely on the population density predicted by the Fermi-Dirac-Distribution. Thus, the transition matrix element between the initial and final states is set to unity and the true electronic structure of the initial and final orbitals is neglected. Finally, all remaining orbitals are assumed to be frozen, that means, correlation effects between the excited electron and the ensemble are neglected.

In the transport step, one assumes electron-phonon dominated scattering in semiconductors and both electron-phonon and electron-electron scattering in metals.

\clearpage
\subsection*{Quantum Efficiency}

The respective properties of metallic and semiconducting emitters are reviewed with a focus on their influence on the quantum efficiency (QE). It is discussed how a favourable interplay of the physical quantities related to electron excitation and transport can lead to high yield cathodes. To a large extent, the argumentation from Spicer and Herrera-Gómez is applied \cite{Spicer1993}.

Consider the three steps of Spicer's emission model with respect to quantum efficiency. In the first step, light shines on the cathode and excites electrons.
The intensity $I$ inside the material at a distance $x$ from the surface is
\begin{equation}
I(x,h \nu) = (1-R) I_0 e^{-\alpha x}
\end{equation}
and the absorbed intensity within a thin sheet of thickness $\dd x$ is
\begin{equation}
\dd I(x) = (1-R) I_0 e^{-\alpha x} \alpha \Dd x \, .
\end{equation}
Here, $I_0$ is the incident intensity, $R$ is the reflectivity of the surface, $\alpha$ is the bulk absorption coefficient and $x$ is the distance from the surface. Note, that $I, I_0, R$ and $\alpha$ depend on the light's energy.

Once an electron is excited, in the second step it may travel to the surface, and third, escape the crystal. The probability of escape consists of
\begin{equation}
P_{esc} = P_{0\alpha} \, P_{e-e} \, P_{dir} \, .
\end{equation}
$P_{0\alpha}$ is the probability of excitation above the vacuum level
\begin{equation}
P_{0\alpha} = \alpha_{PE} \, I_0 \, (1-R) \, e^{-\alpha x} \Dd x \, ,
\end{equation}
where $\alpha_{PE}$ is the fraction of excited electrons that have an energy above the vacuum level. $P_{e-e}$ is the probability that an electron will travel the distance $x$ without suffering electron-electron scattering which would reduce its energy below the vacuum level:
\begin{equation}
P_{e-e} = e^{-\frac{x}{L(h \nu)}} \, ,
\end{equation}
where $L(h \nu)$ is the average electron-electron scattering length for electrons excited by photons of energy $h \nu$. $P_{dir}$ is the probability that the electron's momentum vector points in the right direction, that is, has a component normal to the surface that is large enough to overcome the surface barrier.

The total electron yield $i$ and QE are thus
\begin{equation}
i(h \nu) = \int\limits_0^\infty P_{esc} \Dd x = \int\limits_0^\infty I_0 (1-R) \alpha_{PE} e^{-\alpha x} e^{-\frac{x}{L}} P_{dir} \Dd x = I_0 (1-R) \frac{\alpha_{PE}}{\alpha + \frac{1}{L}} P_{dir}
\end{equation}
\begin{equation}
QE(h\nu) = \dfrac{i(h\nu)}{I_0} = (1-R)\dfrac{\dfrac{\alpha_{PE}}{\alpha} P_{dir}}{1+ \dfrac{1}{\alpha L}} \label{eq:QE1}
\end{equation}

By definition of the absorption length $l_a(h\nu) = 1 / \alpha(h \nu)$ one can rewrite equation \ref{eq:QE1} as

\begin{equation}
QE(h\nu) = (1-R)\dfrac{\dfrac{\alpha_{PE}}{\alpha} P_{dir}}{1+ \dfrac{l_a}{ L}} \label{eq:QE2} \, .
\end{equation}


Two terms in this expression will greatly influence the QE of a material. The ratio $\alpha_{PE} / \alpha$ is the ratio of electrons that get excited above the vacuum level. Clearly, it rises with the excitation energy $\hbar\omega$ and with lower electron affinity. Materials that exhibit a high QE have a $\alpha_{PE}/\alpha$ value between 0.1 and 1.0 where emitters with negative electron affinity may have values close to 1.0.

In the denominator, the ratio of absorption length to scattering length, $l_a/L$, defines the spatial region, from which photoexcited electrons may contribute to the yield. 
The optical absorption length at photon energies where photoemission takes place is in the range of a few 100\,\AA. For metals, the electron scattering length can be as low as 10\,\AA~ due to pronounced electron-electron scattering. It is approximately 40\,\AA~ for copper when $h\nu$ is 1\,eV above the threshold \cite{Spicer1993}. Considering equation \ref{eq:QE2}, this means that the quantum yield for copper is reduced by about one order of magnitude by scattering of excited electrons.

\begin{figure}[!hb]
\label{fig:e-bands-compare-sh}
\begin{center}
	\includegraphics[width=0.85\textwidth]{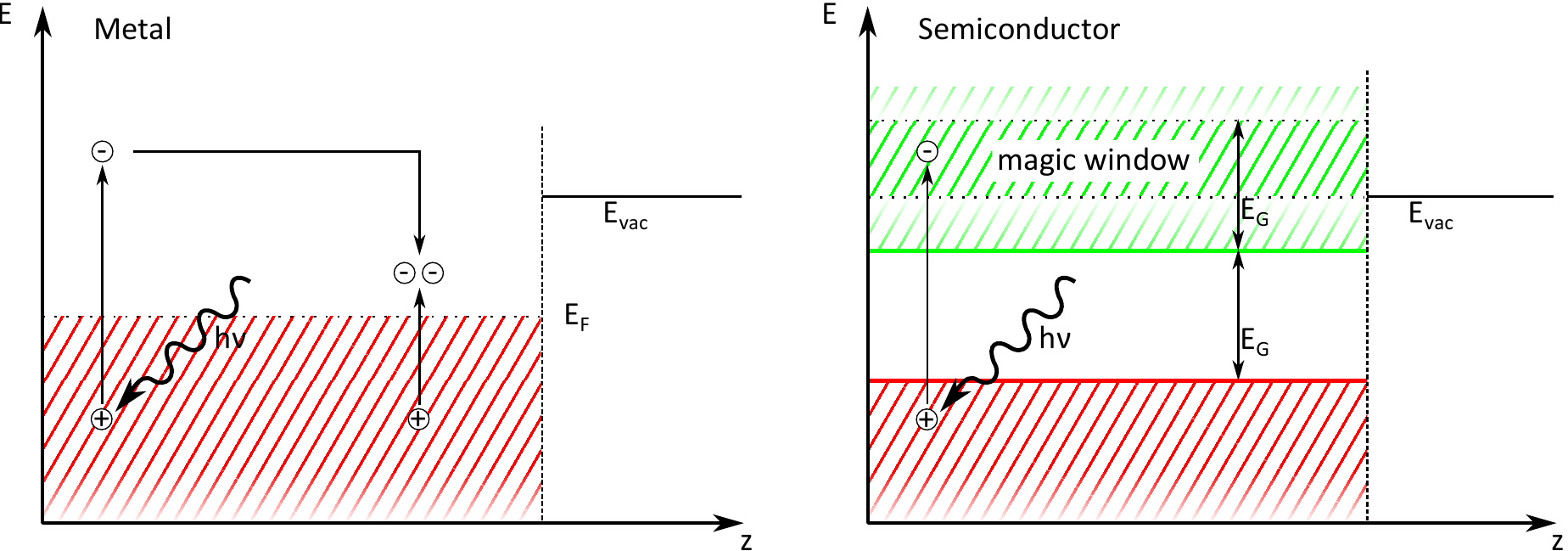}
\end{center}
\caption[Electron-electron scattering in metals and semiconductors.]{Electron-electron scattering in metals and semiconductors. Pair production in metals allows electrons to loose the energy needed to escape while in semiconductors a "magic window" can be present where electrons cannot produce pairs but have enough energy to escape.}
\end{figure}

In semiconductors electron-electron scattering is suppressed in a "magic" energy window $E_{MW}$ above the vacuum level where $E_{VL} < E_{MW} < E_{VBM} + 2E_G $ if the electron affinity is less than the band gap (thus, $E_{VL} < E_{VBM} + 2E_G$). This is because 
there are no final states in the band gap available and the energy of the electrons is too low to raise a scattering partner from the valence band to the conduction band. The various energy levels are illustrated in figure \ref{fig:e-bands-compare-sh}.  In efficient emitters, the scattering length of electrons with energy just above the vacuum level can be in the order of the optical absorption length, so the fraction $\frac{l_a}{L}$ is close to unity.

The direction of the momentum vector of the excited electron plays an important role when electron-electron scattering is pronounced. Because a single scattering event will reduce the energy of an excited electron below the vacuum level only those electrons that reach the surface without scattering can be considered for emission. Thus, a narrow escape cone is defined by the maximum angle $\vartheta_{max}$ between the electron's momentum vector and the surface normal that will allow the electron to escape.
Assuming clean and even surfaces, $\vartheta_{max}$ is defined by the conservation law of transverse momentum $p_x$ at the crystal-vacuum interface and the requirement, that the longitudinal component $p_z$ be larger than the surface barrier \cite{Dowell2006}.
\begin{eqnarray}
p_z = \sqrt{2m(E+\hbar\omega)} \, \cos \vartheta \geq \sqrt{2m(E_F+\phi)} \, ,\\
\cos \vartheta_{max} = \sqrt{\dfrac{E_F + \phi}{E+\hbar\omega}} \, .
\end{eqnarray}
In order to obtain a more quantitative QE expression (for metals), one needs to consider the dependence of both $\vartheta_{max}$ and $\alpha_{PE}$ on the energy $E$ of the electron before the excitation. The term $\frac{\alpha_{PE}}{\alpha}$ is obtained by integrating over the energy range in the valence band that can contribute to the photoemission yield. $P_{dir}$ is taken into account by restricting the integration to angles below $\vartheta_{max}$. The derivation, which is shown in appendix \ref{appendix:QE} yields
\begin{equation}
QE_{metal} = \dfrac{1-R(\omega)}{1+\frac{l_a}{ L}} \dfrac{E_F+\hbar\omega}{2\hbar\omega} \left[ 1- \sqrt{\dfrac{E_F+\phi}{E_F+\hbar\omega}} \right]^2
\label{eq:qe-metals}
\end{equation}

\begin{figure}[!ht]
\label{fig:QE-compare}
\begin{center}
\includegraphics[width=0.9\textwidth]{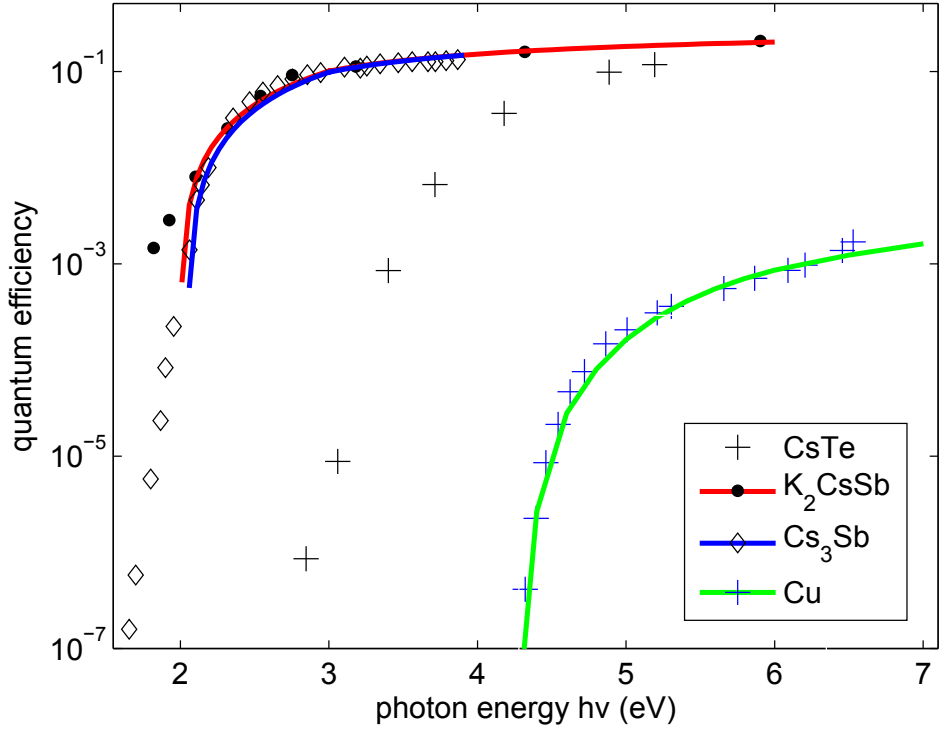}
\end{center}
\caption[Quantum efficiencies of common photoemitters, measured and theoretical values]{Quantum efficiencies of common photoemitters. Experimental values are compiled from \cite{mySmedley2009, Spicer1958, Schreiber2012, Dowell2006} for \PCA, Cs_3Sb, CsTe and copper.}
\end{figure}

Finally, the QE for semiconducting emitters is discussed. Above, electron-phonon scattering needs not be considered because electron-electron scattering is dominant and scattered electrons have a negligible chance to escape. If, however, electron-electron scattering is suppressed in a magic window, then electron-phonon scattering may dramatically increase $P_{dir}$. Consider electron-phonon scattering as a random-walk process with a sink at the crystal surface \cite{Spicer1993}. The scattering process is nearly inelastic, so the direction of the electron is randomized but it looses only a small fraction of its energy. Several scattering events are possible before an electron looses too much energy to overcome the surface barrier and in materials with low or negative effective electron affinity all excited electrons will eventually escape. Thus, $P_{dir}$ may be approximated by a step function for efficient semiconducting emitters. Equation \ref{eq:qe-metals} reduces to
\begin{equation}
QE_{sc} = \dfrac{1-R(\omega)}{1+\dfrac{l_a}{ L}} \dfrac{\hbar\omega-\phi}{\hbar\omega} \, .
\label{eq:qe-SCs}
\end{equation}
The models presented here are in good agreement with experimental data, as can be seen in figure \ref{fig:QE-compare}. For copper, the following parameters were used \cite{Dowell2006} : $E_F$ = 7\,eV, $\phi$ = 4.31\,eV and scattering length $L$ = 2.2\,nm. Optical constants from \cite{Palik1985} were obtained through \cite{Polyanskiy}. For \PCA, a work function of $\phi$ = 2\,eV and $L$ = 5.5\,nm were assumed. The absorption length was obtained from \cite{myKalarasse2010} and approximated by a trapezoidal form. Similarly, for Cs_3Sb, $\phi$=2.05\,eV and $L$=11\,nm were assumed and the absorption length was obtained from \cite{Spicer1958}.
%
%

\subsection*{Emittance}
In order to obtain the normalized rms emittance
\begin{equation}
\varepsilon_n :=  \beta \gamma \sqrt{<x^2><x^{\prime2}>-<xx'>^2}
\end{equation}
it is assumed that the emission occurs uniformly from a circular illuminated spot on the cathode surface with radius $r$, thus $<x^2> = r/2$.
Because the emittance is considered just outside the cathode surface where electron flow is laminar the correlation term can be neglected.
$\beta$ and $\gamma$ are the relativistic factors that normalize the emittance with respect to the average kinetic energy.
The $<p_x^2>$ term can be obtained from emission models.

Again, one has to distinguish between different types of emitters. For metals, only electrons that reach the surface without scattering are considered and the transverse momentum is conserved during passage of the surface barrier.
Thus, the normalized emittance is conserved \cite{Flottmann1997}.
One can move the integration over the transverse momentum distribution from outside (vacuum) to the inside of the cathode crystal.

If transverse momentum conservation at the surface holds, there is a maximum angle $\vartheta_{max}$ of the electron's momentum vector with respect to the surface normal up to which escape is possible
\begin{equation}
\vartheta_{max} = \sqrt{\frac{E_F+\varphi}{E+\hbar\omega}} \, ,
\end{equation}
because the longitudinal component needs to be higher than the surface barrier.
Here, $E_F$ is the Fermi Energy, $\varphi$ is the work function, $E$ is the energy of the electron prior to excitation and $\hbar\omega$ is the incident light energy.
$\vartheta_{max}$ introduces an energy dependency of the angular distribution.

The mean transverse energy can be computed by integration of the momentum vectors of the excited electrons over the angle, restricted by $\vartheta_{max}$, and energies that are above the vacuum level and can be reached by excitation with energy $\hbar\omega$:
\begin{equation}
<k_r^2> \propto \int\limits_{E(k) = \mu-\hbar\omega+\phi}^\mu \dd k \, \int\limits_0^{\cos^{-1} \vartheta_{max}} \dd \vartheta \, k^4 (\sin \vartheta)^{n+1} f_\lambda (\cos \vartheta, E(k))
\end{equation}
The distribution of momentum vectors of the excited electron is assumed to be isotropic.
Solving the integral, as shown in the appendix, yields an expression for the normalized divergence of photoemitted electrons
\begin{equation}
\beta \gamma \sigma_{x\prime} = \sqrt{\frac{\hbar\omega-\varphi}{3mc^2}} \, ,
\end{equation}
and thus the normalized emittance
\begin{equation}
\varepsilon_{n, metal} = \sigma_x \sqrt{\frac{\hbar\omega - \varphi}{3mc^2}} \, .
\label{eq:metal-emittance}
\end{equation}

Several assumptions made above need to be revised when considering semiconducting emitters. Experimental data does not support the premise that transverse momentum is conserved at the boundary \cite{Karkare2013}.
Also, a significant contribution to the QE comes from scattered electrons that leave the cathode with a delay.
However, they are not necessarily fully thermalized, so they have to be treated separately and no analytical representation of their momentum distribution is known.

Monte Carlo simulations of the scattering processes in GaAs cathodes have been performed and yield good agreement with experimental data \cite{Karkare2013}.
Such treatment would be beneficial to the understanding of alkali antimonide cathodes as well but is beyond the scope of this work.

For the time being, a simplistic model is employed where one assumes that the energy distribution of the emitted electrons is uniform and isotropic
\begin{eqnarray}
<E> = <E_x> +  <E_y> +  <E_z> \, ,\\
<E_x> = <E_y> = <E_z> \, ,
\end{eqnarray}
so the mean transverse energy (MTE) depends only on the excess energy
\begin{equation}
MTE = \frac{2<E>}{3} \, .
\end{equation}
If the kinetic energy of the photoelectrons is distributed uniformly between the excess energy $\hbar\omega - E_G - E_A$ and 0, then $<E> = (\hbar\omega - E_G - E_A)/2$,
thus, the intrinsic emittance for photoemission from a semiconducting (sc) cathode is
\begin{equation}
\varepsilon_{n,sc} = \sigma_x \sqrt{\frac{\hbar\omega - E_G - E_A}{3mc^2}} \, .
\label{eq:sc-emittance-E-xcs}
\end{equation}
Here, the excess energy term $\hbar\omega - E_G - E_A$ corresponds to $\hbar\omega - \varphi$ in equation \ref{eq:metal-emittance}. $E_G$ is the band gap and $E_A$ the electron affinity.
\section{Alkali Antimonides} \label{sec:alkali-antimonides}

Several types of materials have been employed as photocathodes in the past. Metals are used because they are easily available and highly polished surfaces can be obtained. Their fast pulse response in the fs range allows pulse shaping and the generation of ultra-short bunches. Additionally, metals are insensitive to residual gases and ion back bombardment, so their operational lifetime can be virtually infinite. Especially copper is a common photocathode because of its relatively high QE, good thermal and electrical conductivity, and the commercial availability of polished high purity copper structures.
Recently, an SRF plug-gun has been commissioned at HZB with a lead cathode that was plasma arc deposited on the cathode plug \cite{Neumann2012, Schmeisser2013, BardayIPAC2013}. The cathode showed a QE of up to $8\cdot 10^{-3}$\,\% after UV-laser cleaning.
Generally, however, metals have too low quantum efficiencies to generate high average currents. Furthermore, the laser power required to generate one mA of current is in the kW range which is prohibitive, especially at UV wavelengths.

Semiconductors may have a high QE because electron-electron scattering is greatly reduced.
%
Table \ref{tab:cathodes} lists properties of commonly used photoemissive materials. Alkali antimonides, especially K$_2$CsSb, are good candidates for demanding photoinjector applications. Their high QE and sensitivity to visible wavelengths relax the requirements on the drive laser by orders of magnitude. The cathode is exposed to less radiation intensity which reduces the load on the cooling system. Nominally, 100\,mA of beam current are produced when the laser power at the cathode is
\begin{equation}
P_{laser} = \frac{124\,\mathrm{W\,nm}}{\lambda \cdot QE} \,
\end{equation} 
for the laser powers in table \ref{tab:cathodes}, the harmonic wavelengths of a Nd:YAG laser (1064, 532, and 266\,nm) and 1\% QE for \PCA~were assumed.

\begin{table}[!hb]
\hspace{-8pt}
\begin{tabular}{L{3.5cm} C{2cm} | C{2.8cm} C{2.6cm} | C{3.3cm} }
 & \textbf{Metals} & \multicolumn{2}{c|}{\textbf{Semiconductors}} & \textbf{NEA} \\
 &            Cu, Pb         & K$_2$CsSb & CsTe  & GaAs     \\
 \midrule
Sensitivity & UV             & VIS       & UV      & VIS-IR \\
QE (\%)     & $10^{-2}$      & >1        & 10      & 10     \\
Laser Power at Cathode (W)   &
              5000           & 23        & 5       & 1      \\
Robustness  & unreactive     & fairly reactive (<$10^{-10}$ mbar) & reactive (<$10^{-9}$ mbar) & very reactive (<$10^{-11}$ mbar) \\
Dominating Mode of Scattering
            & e^- \,-\, e^-  & \multicolumn{2}{c|}{e^- \,-\, phonon} & e^-\, - \,phonon + thermal diffusion \\
Response Time & fs & ps & fs & ps + tail
\end{tabular}   
\caption{Properties of common photoemitters. Data from \cite{BardayIPAC2013, Kirsch2014, Jensen2006}.}
\label{tab:cathodes}
\end{table}

A comprehensive handbook on preparation, use, and properties of a vast amount of materials is provided by Sommer \cite{Sommer1968}. Especially alkali antimonides are covered in Sommer's work.
The most stable alkali antimonide compounds all have a stoichiometry of M$_3$Sb where M is one or more alkali metal. Interestingly, the materials can be divided in two groups. One group typically crystallizes in a cubic structure and exhibits high QEs, the other crystallizes in a hexagonal lattice and has lower QEs. Members of the first group, for example \PCA~ and Cs$_3$Sb, are p-type semiconductors while the latter group consists of n-type semiconductors.

Alkali antimonides are classically prepared by exposing an antimony specimen to vapour of the respective alkali metal. Antimony can be prepared as a powder specimen but for fabrication of photocathodes it is preferred to deposit a thin film from evaporation beads. Alkali metals are evaporated from intermetallic alloys (alvasources \circledR). Typically, the specimen is exposed to alkali metal vapour until the quantum efficiency reaches a plateau which indicates saturation of the compound. Application of excess alkali metal reduces the QE again.

\subsection{Sample Preparation}

A preparation chamber for alkali antimonide photocathodes was recently developed and built at Helmholtz-Zentrum Berlin (HZB) by Dr. Susanne Schubert. In an effort to foster an international cooperation on cathode research and development, the whole system was moved to Brookhaven National Lab (BNL) in Upton, NY, USA, where it was assembled and commissioned.

\begin{figure}[!hb]
\begin{center}
	\mbox{\includegraphics[width=0.65\textwidth]{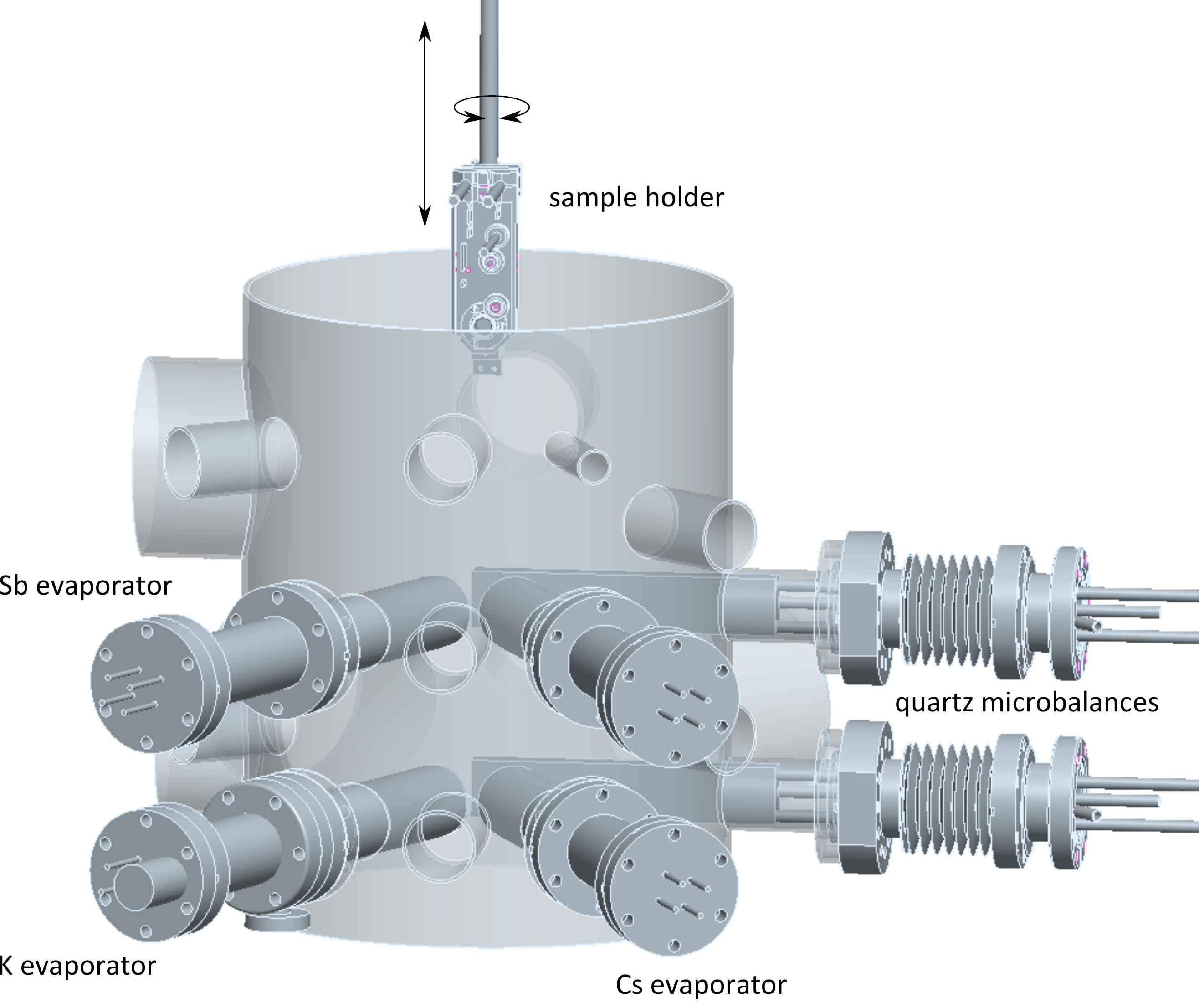}
		  \includegraphics[width=0.35\textwidth]{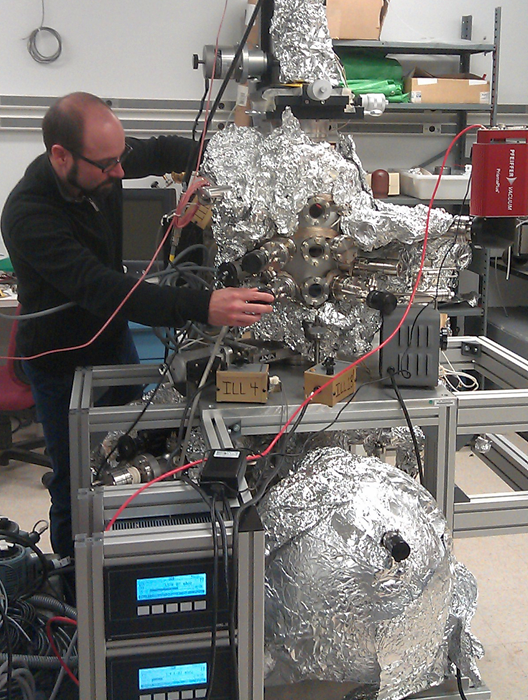}}
\end{center}
\caption[CAD drawing and photograph of the preparation chamber]{CAD drawing and photograph of the preparation chamber of the system. The sample holder, evaporation sources, and quartz crystal monitors (QCM) are displayed. Vacuum components are omitted for clarity.}
\label{fig:prepchamber}
\end{figure}

The samples are prepared in a dedicated chamber that allows physical vapour deposition (PVD) of Potassium, Caesium and Antimony from evaporation sources as shown in figure \ref{fig:prepchamber}. The deposition rate is monitored by quartz microbalances.

\begin{figure}[!hb]
\begin{center}
	\includegraphics[width=0.65\textwidth]{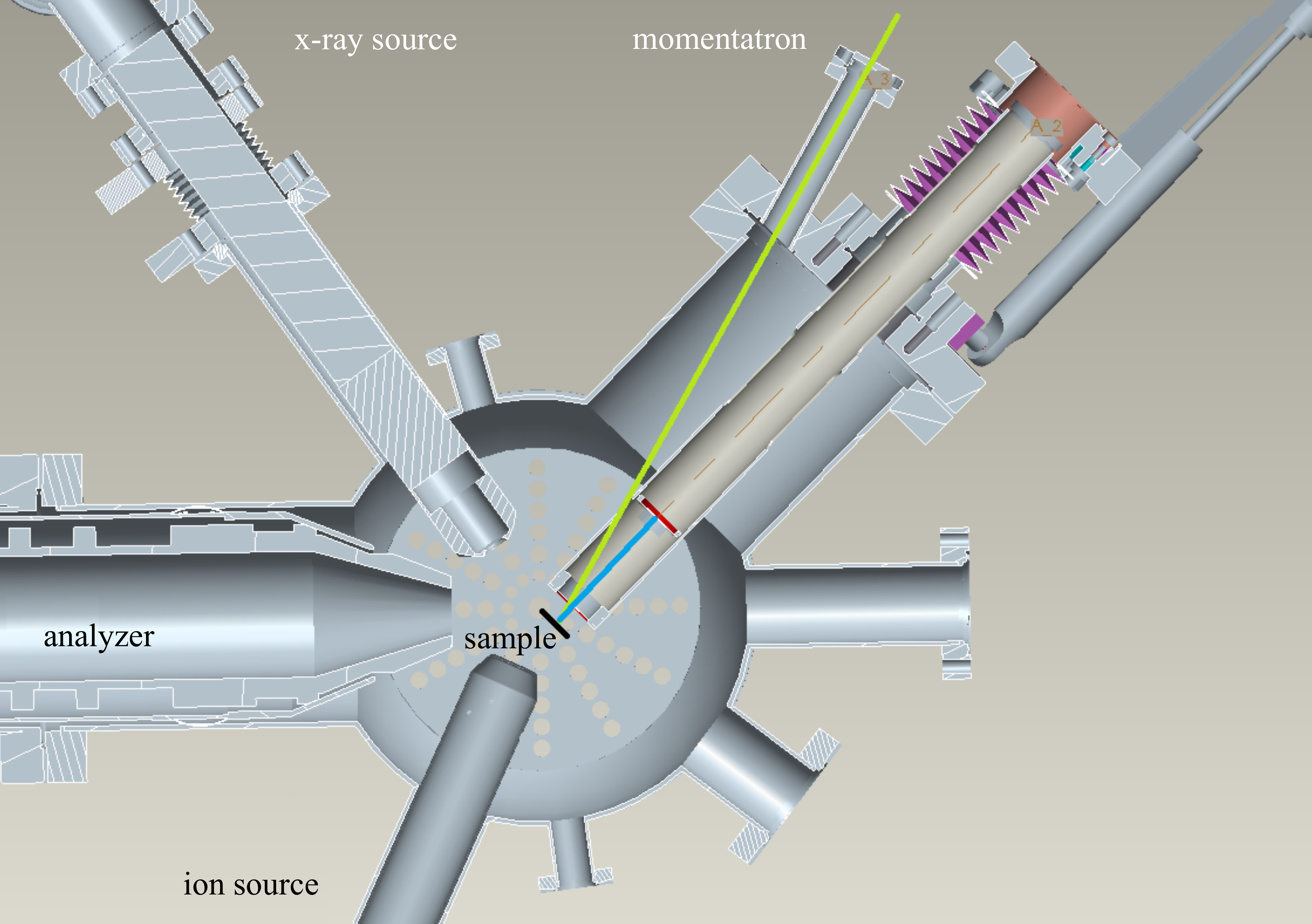}
\end{center}
\caption[CAD drawing of the analysis chamber]{CAD drawing of the momentatron mounted in the analysis chamber mounted in the top right flange. The laser and electron paths are shown as green and blue lines. On the left side an X-ray source, the electron analyser, and an ion source are mounted, from top to bottom.}
\label{fig:momtron-cad}
\end{figure}

Below the preparation chamber and separated by a large valve, an analysis chamber is available and the sample can be transferred without breaking the vacuum. The momentatron was mounted in this chamber. Additionally, it supports an X-Ray source and a hemisphere electron detector for X-ray photoelectron spectroscopy (XPS) and an ion source for low-energy ion scattering experiments (LEIS). A 3-axis manipulator transfers the sample between the two chambers. The sample holder 
is electrically isolated and a bias voltage of up to $\pm 100$V can be applied. Resistive heating is available but the heater's electrical isolation is bad, so it has to be unplugged for any measurements that require a bias applied to the sample.

Due to their sensitivity to poisoning by residual gases, semiconductor photocathodes have to be prepared in an ultra high vacuum (UHV) environment. Too high partial pressures of, especially, oxygen but also carbon monoxide and water will severely reduce the lifetime of the prepared sample or prevent crystallization of the desired compounds in the first place \cite{Sertore2011}. Practically, this requires to achieve partial pressures of oxygen below the sensitivity of the residual gas analyser (ion current below $1\cdot 10^{-14}\,$A or partial pressure below $1\cdot 10^{-11}\,$mbar) and partial pressures of carbon monoxide and water below $2\cdot 10^{-10}\,$mbar. The base pressure of the system will then be about $1\cdot 10^{-9}\,$mbar.

\subsection{Preparation of a Caesium Antimonide Cathode} \label{chap:first-cathode}

A caesium antimonide photocathode was grown in the preparation system described above. Antimony was deposited from three evaporation beads on a polished molybdenum substrate. The deposition rate was calibrated by means of a quartz crystal monitor (QCM) and adjusted to a stable value of 0.1\,\AA/s. 
The quartz balance was moved to the side and the sample placed in the same position for 15 minutes. After the evaporation, the QCM did not monitor any further deposition, possibly because the beads were empty.

Subsequently, caesium was evaporated on the antimony film from an Alvatec S-type evaporation source. A deposition rate of 1.3\,\AA/s was recorded before the sample was moved in place and, again, after 17 minutes of evaporation the quartz balance displayed no further deposition.

Due to bad electrical isolation of the sample heater and thermocouple, heating of the sample had to be disconnected during Cs evaporation in order to measure the photocurrent. The sample temperature was at about 160$^\circ$C before the antimony deposition and 70$^\circ$C before the caesium deposition.

\begin{figure}[!hb]
\label{fig:I-photo}
\begin{center}
	\includegraphics[width=.85\textwidth]{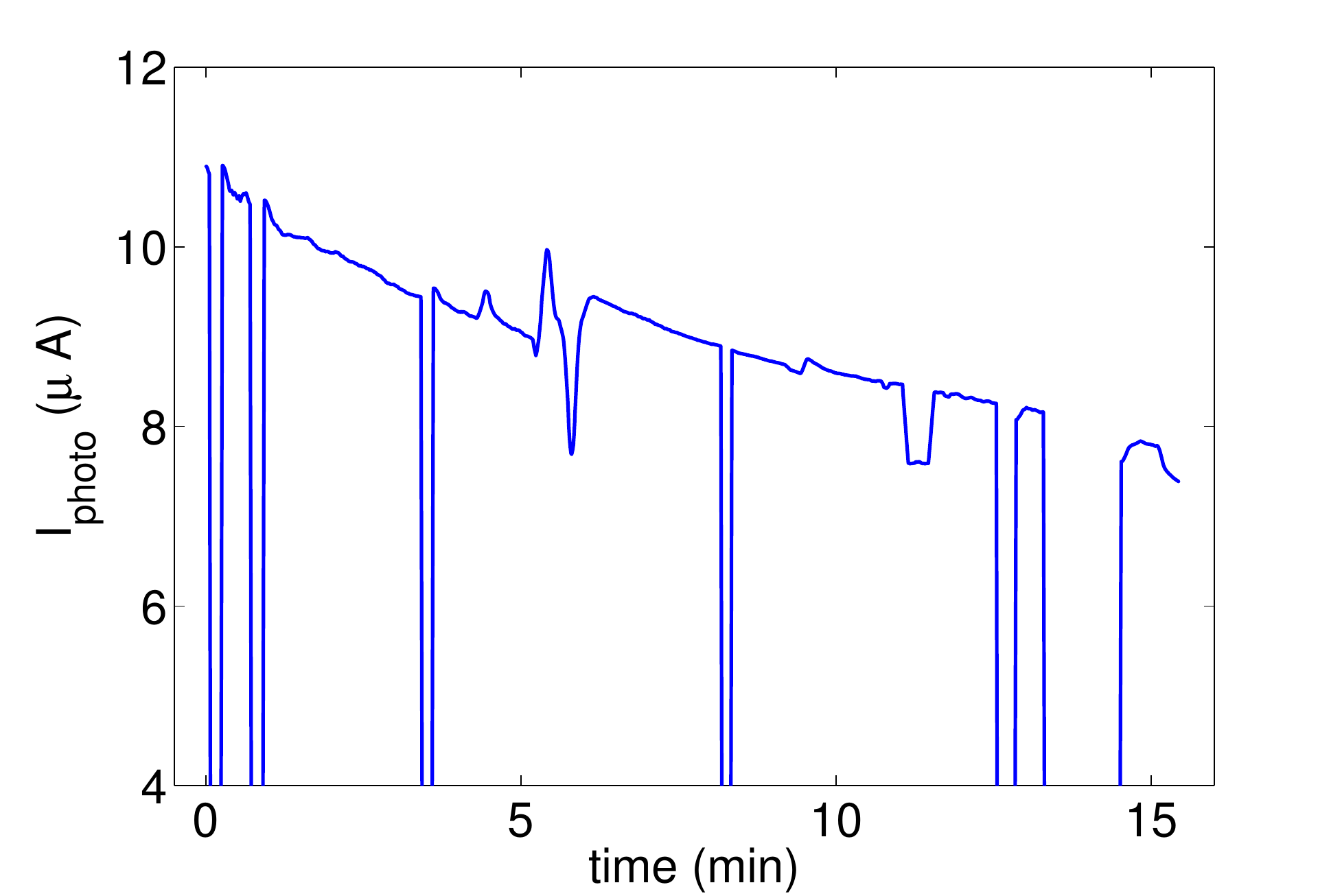}
\end{center}
\caption[Photocurrent from a Cs$_3$Sb photocathode over time.]{Photocurrent from the Cs$_3$Sb photocathode under illumination by a 5\,mW green laser (532nm) over time. The drops in the signal are from manually blocking the laser light. The wiggle at about 5 minutes is due to a movement of the laser spot on the sample surface.}
\end{figure}

During the caesium deposition, a 5\,mW green laser (532nm) was directed at the sample in order to produce photoelectrons and the photocurrent was measured by a picoampere-meter while the sample was at a -10\,V bias. The recorded photocurrent is displayed in figure \ref{fig:I-photo}. The signal drops at several points in time where the drive laser was blocked and the wiggle at about 5\,min is due to a movement of the laser spot on the sample surface.

Significant degassing from the evaporation sources increased the vacuum base pressure during the sample preparation. The base pressure was at $2\cdot10^{-9}$\,mbar in the beginning of the experiment and increased occasionally to $3\cdot10^{-7}$\,mbar during evaporation. Partial pressures of $4.5\cdot10^{-9}$\,mbar and $1.3\cdot10^{-10}$\,mbar were recorded for water and oxygen, respectively, which may have allowed oxidation of the material and thus impeded the formation of large Cs$_3$Sb crystallites. One of the potassium sources degassed strongly which compromised the vacuum pressure before the caesium deposition and prevented the use of this source.

A photocurrent of 10.7\,$\mu$A was initially recorded which reduced to 8\,$\mu$A in 15 minutes. The 1/e lifetime of the cathode is estimated to 50\,minutes. Nominally, the QE of the sample from these values is about 0.5\%. However, this must be considered as a lower boundary as the electron emission may have been space charge limited. The photocurrent decreased by a factor of 107.1 when the laser light was attenuated by a factor of 225.9. This suggests an uncertainty of a factor of 2 in the quantum efficiency. Estimating the space charge limit for electron emission with the \textit{Child-Langmuir}-Law for parallel diodes \cite{Reiser2008}
\begin{equation}
\label{eq:child-langmuir}
I = \frac{4 \pi \epsilon_0}{9} \sqrt{\frac{2e}{m_e}} U^{3/2} \frac{r^2}{d^2}
\end{equation}
yields a maximum current of 0.1\,$\mu$A. Here, a voltage of $U=-10\,V$, an illuminated spot radius of $r=1$\,mm, and a distance between cathode and anode of $d=50$\,mm were assumed. The considerably larger measured current may be possible due to a much larger anode surface area, which is assumed to be of the same size as the cathode spot in the derivation of equation \ref{eq:child-langmuir}.

The vacuum system consists of three ion getter pumps (one 500\,l/s pump below the analysis chamber and two 75\,l/s pumps at the deposition chamber) and one 80\,l/s turbomolecular pump which is backed by a dry scroll pump. The performance of the vacuum system may be limited by the small conductance of the piping between the turbo pump and the chamber. Additionally, a differential pumping connection between the manipulator and the scroll pump might compromise the foreline vacuum for the turbo pump. For a second measurement campaign, it is planned to install a second turbo pump directly to a 40\,mm flange at the preparation chamber. It is expected, that this will decrease the bakeout time and base pressure and allow faster pumping of gas load from the evaporation sources.

\chapter{In-Situ diagnostics of Initial Transverse Momentum} \label{chap:momentatron}

The momentatron is a device that allows measurement of the intrinsic transverse momentum distribution of an electron beam emitted from a photocathode. It was mounted to the photocathode preparation and analysis chamber where it will be used to perform in-situ diagnostics of the prepared cathodes without exposure to ambient conditions. 
The goal is to find correlations between cathode preparation parameters and the transverse momentum distribution. Information from the momentatron can provide insight on the influence of structural properties of the cathode like crystal orientations, crystallite sizes, surface roughness, and stoichiometry on the emittance as a figure of merit for accelerator physics.

\section{Working Principle}

\begin{figure}[!ht]
\centering
\includegraphics[width=.85\textwidth]{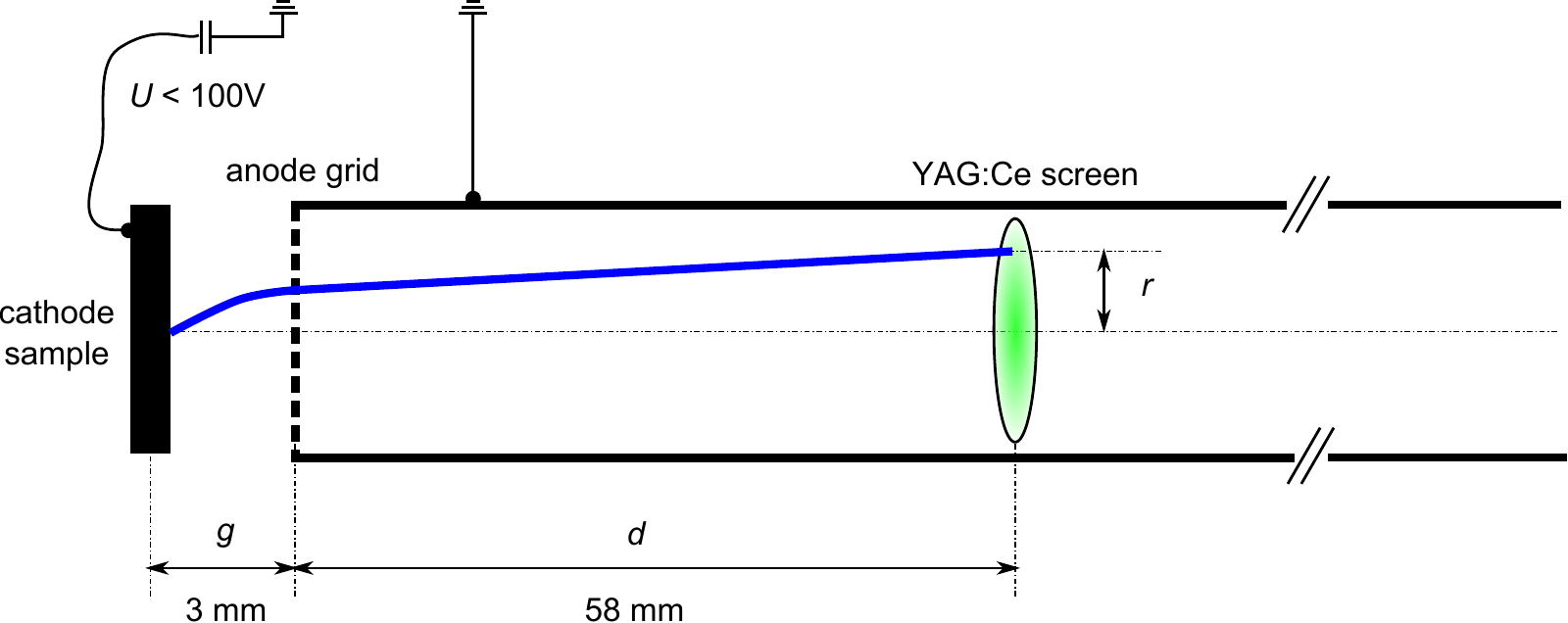}
\label{fig:momtron-scheme-dist}
\caption[Scheme and parameters of the momentatron]{Scheme and parameters of the momentatron.}
\end{figure}

\begin{figure}[ht]
\centering
\mbox{\includegraphics[width=.45\textwidth]{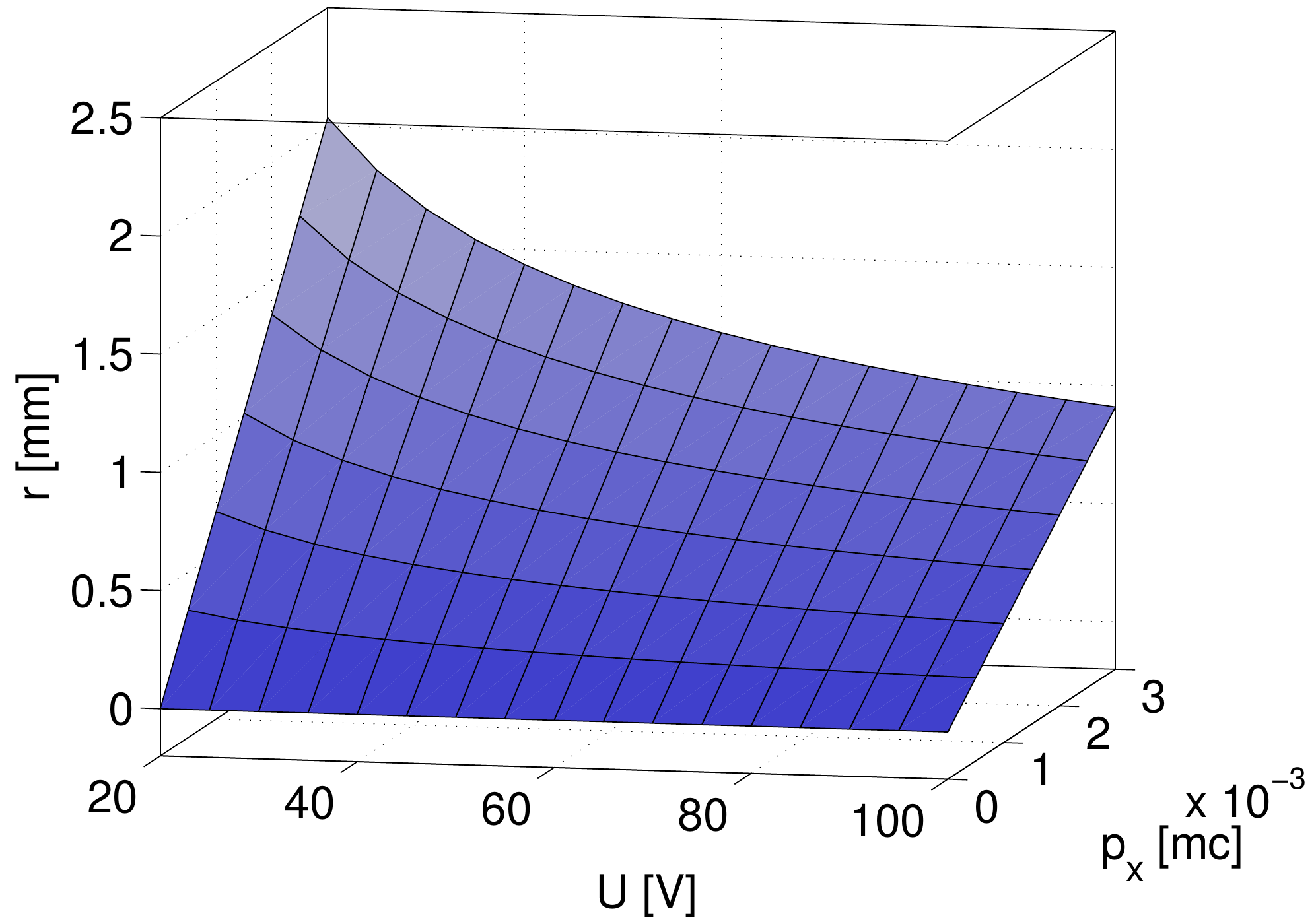}
\includegraphics[width=.425\textwidth]{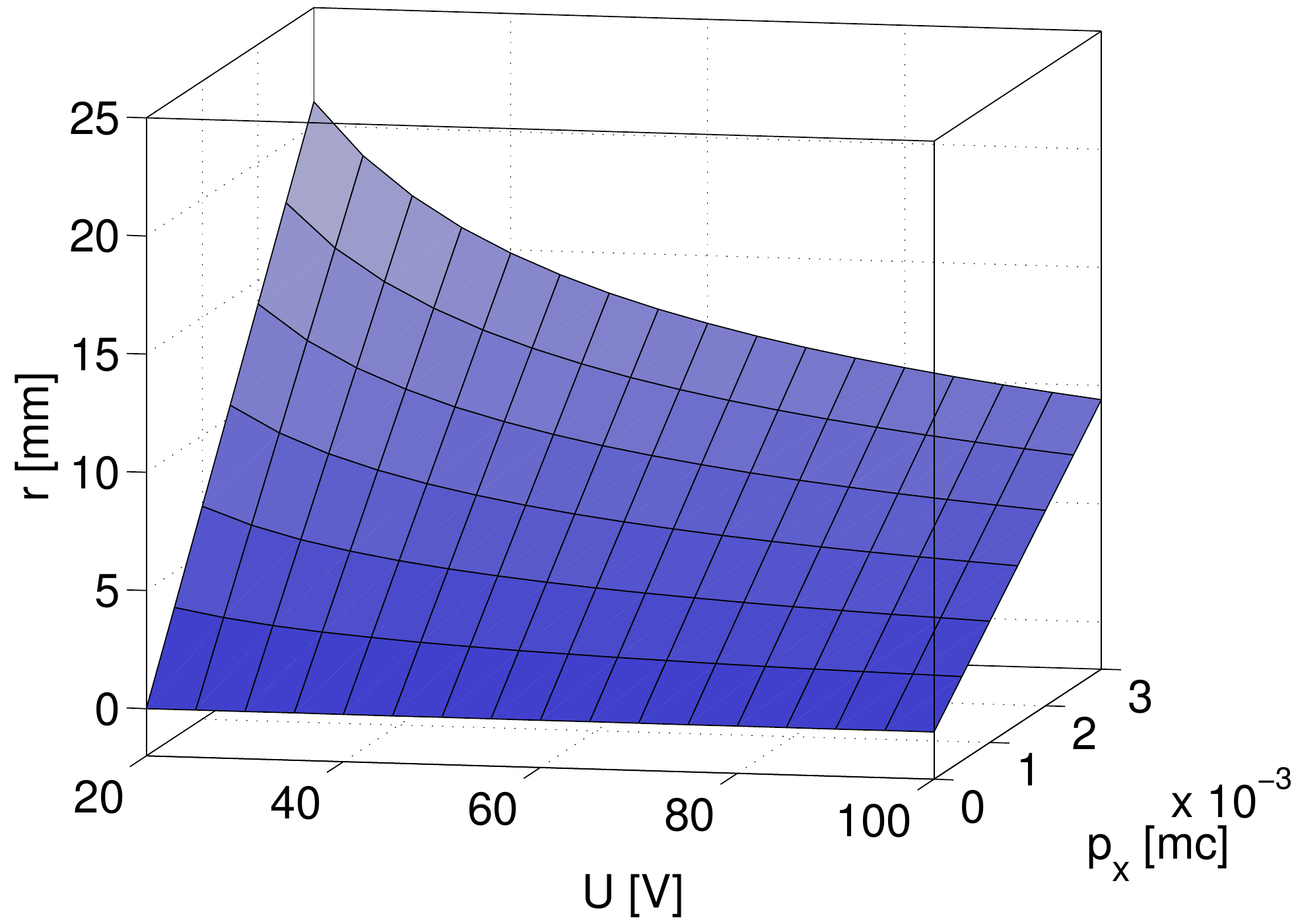}}
\caption[Beam Sizes in the Momentatron]{Expected radial coordinates for particles with different transverse momentum $p_x$ at various accelerating potentials taken at the position of the grid (left) and the screen (right).}
\label{fig:beamsize}
\end{figure}

The electrons are emitted from a small laser-illuminated spot on the cathode sample.
After acceleration in a short gap $g$ between the sample and a mesh, they are allowed to drift in a field free region $d$ towards a screen.
See figure \ref{fig:momtron-scheme-dist} for an illustration.
The radial coordinate $r$ at the position of the screen depends linearly on the intrinsic transverse momentum $p_x$
\begin{equation}
\frac{p_x}{mc} = \frac{r}{2g+d} \sqrt{\frac{2eU}{mc^2}} \, ,
\label{eq:beamsize}
\end{equation}
and a radial distribution of $p_x$ can be obtained from the radial intensity distribution on the screen.
Here, $p_x$ is expressed in units of $mc$ (electron mass times speed of light), $e$ is the elementary charge and $U$ is the potential difference between sample and grid.
\begin{figure}[!ht]
\centering
\includegraphics[width=.85\textwidth]{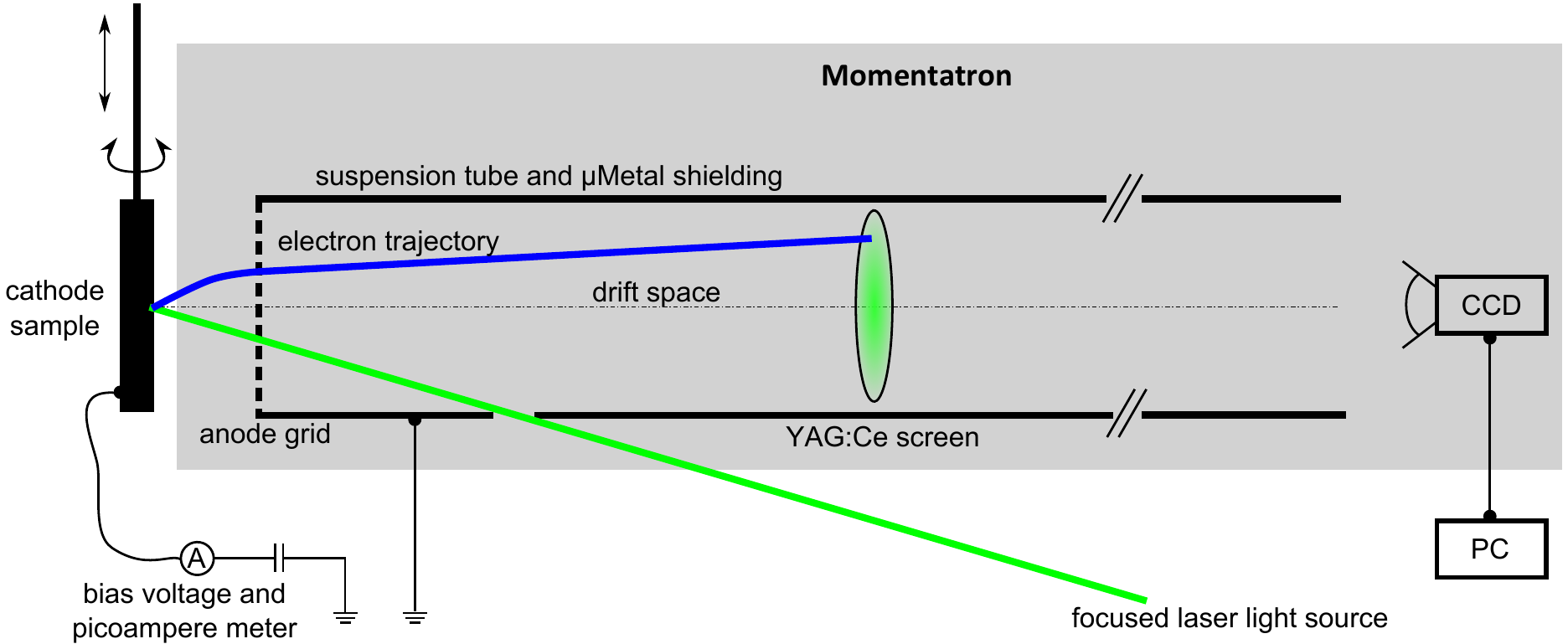}\\
\vspace{0.5cm}
\includegraphics[width=.85\textwidth]{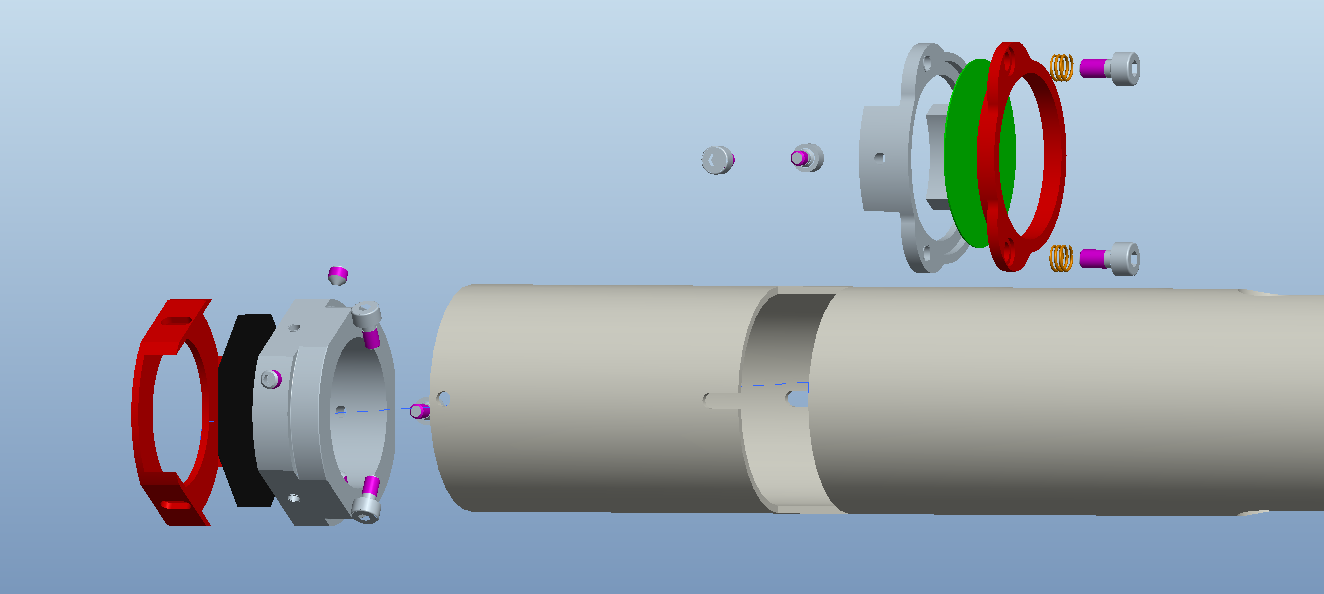}\\
\vspace{0.5cm}
\includegraphics[width=.85\textwidth]{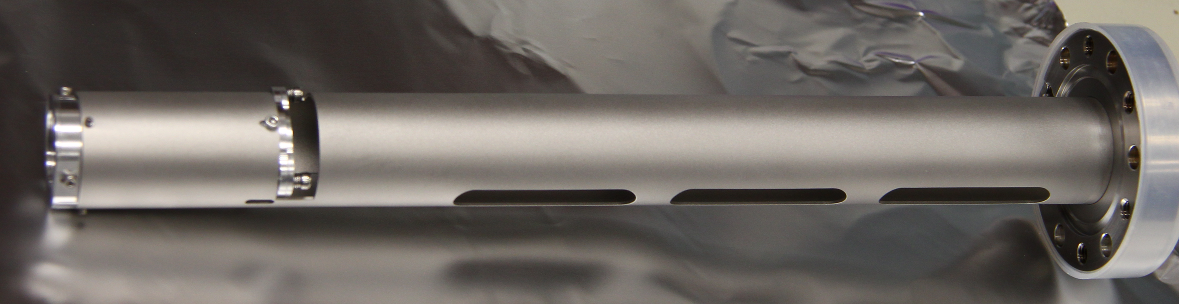}
\label{fig:momtron-scheme}
\caption[Working principle of the Momentatron and exploded view]{Schematic illustration of the working principle of the momentatron (top), exploded view (middle) and photograph of the device (bottom).}
\end{figure}
The beam size on the screen depends on the drift and gap distances $d$ and $g$ and on the applied voltage $U$.
At fixed gap and drift lengths, the maximum transverse momentum that can be imaged on the screen can be tuned by the choice of the accelerating potential, see figure \ref{fig:beamsize}.
$g$ and $d$ were chosen to yield a beam size of 15\,mm diameter on the screen at an accelerating voltage of 50-100\,V. This offers a reasonable resolution when imaging the screen and leaves room for measurements when emittance dilution due to misalignments of the system or rough cathode surfaces is present.
The normalized emittance $\varepsilon_n$ of a K$_2$CsSb cathode was previously measured as 0.37\,mm\,mrad per mm rms laser spot size \cite{Vecchione2011} and a theoretical estimate is 0.4\cite{Dowell2010}.
Considering equation \ref{eq:eps-from-px2}, the transverse momentum distribution is thus expected to fall off within 0.0015\,mc.
At $U=50$\,V and with distances $g=3$\,mm and $d=56$\,mm, the maximum transverse momentum that can be measured is $0.0022$\,mc corresponding to to 9.75\,mm radius on the screen.

\begin{equation}
\frac{\varepsilon_n}{\sigma_x} = \frac{<p_x^2>^{\frac{1}{2}}}{mc}
\label{eq:eps-from-px2}
\end{equation}

%



\section{Experimental Setup}

A grid acts as an anode during the acceleration but lets most of the electron pass, so they can be detected after travelling through the drift region.
A 1000 mesh grid with $18\,\mu$m rectangular holes will be employed which is commercially available.
The grid foil is about $5\,\mu$m thick and has an open area of 50\%.

\subsection{Imaging}

The detection of the electrons will be performed by a cerium doped YAG screen that is monitored by a CCD camera (Basler scout sc640-74gm).
Light emission from the screen reaches from 500 to 700\,nm with a maximum at 550\,nm and the sensitivity of the camera covers a range from 400 to 800\,nm with a maximum at 500\,nm, so the camera is well suited to capture images from YAG:Ce screens.
The light output (emitted photons per incident electron) for low energy electrons is unknown but expected to be about 50\,\% from extrapolation of high energy data \cite{myYAGSheet}. At 10\,$\mu$A photocurrent, on average about $2\cdot 10^5$ photons reach a single pixel per second through the lens aperture. Assuming a 50\,\% quantum efficiency of the CCD chip, the exposure time should be about 100\,ms to reach the saturation capacity of $2.5\cdot 10^{4}$ electrons per pixel \cite{CamSpec}.

A green (515\,nm) solid state laser with 1\,mW output power or a blue (405\,nm) diode with 5\,mW power will be used to illuminate the cathode. When using the blue diode laser, stray light from the grid can effectively be blocked with a long pass filter that is transparent to the green scintillation light.
Stray light from the green laser could be suppressed with a notch filter that is transparent to visible wavelengths except a narrow band around the laser wavelength. The problem is however, that the scintillator emits at very similar wavelengths and a large fraction of the signal intensity will be lost too.
The beam is expanded in diameter from one to three millimetres and focused on the cathode to an rms spot size of $40\,\mu$m.
Due to refraction at the grid, the spot size will be larger, see section \ref{sec:errors}. 

\begin{figure}[ht]
\centering
\label{fig:laser-optics}
\includegraphics[width=.85\textwidth]{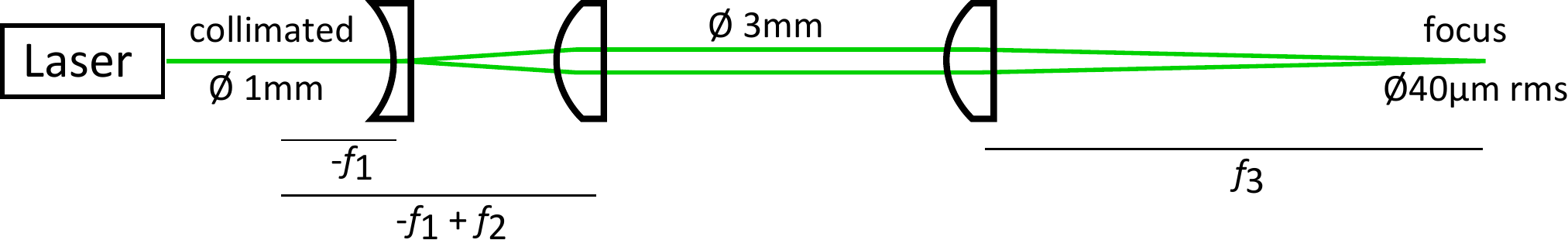}
\caption[Sketch of the laser optics]{Sketch of the laser optics.}
\end{figure}

Looking for camera optics, it is desirable to image the 20\,mm diameter of the screen on 4\,mm sensor area with a distance between lens and screen of about 320\,mm. The camera chip's dimensions are about 4.9 by 6.6\,mm. From basic optics one has
\begin{eqnarray}
B &=& \frac{fG}{g-f}\\
f &=& \frac{g}{\frac{G}{B}+1} = 53.33\,\mathrm{mm} \, .
\end{eqnarray}
Here, G is the size of the screen (object size), B is the size of the image on the sensor, f is the focal length, and g is the object distance. Using a 50\,mm lens with 60\,mm image distance results in an object distance of 300\,mm.

This yields an image scale between screen and camera of 5 to 1. The scale from transverse momentum to the radial coordinate on the screen is 0.00025\,mc to 1\,mm. Through the whole system the scale is thus 0.001\,mc per mm on the camera chip, where pixels have a size of $10\,\mu$m, so the resolution will be $1\cdot 10^{-5}$\,mc per pixel. The resolution has to be compared with an expected maximum transverse momentum value of $1.5\cdot 10^{-3}$\,mc which results in a relative resolution of about 1\,\%. This is also the uncertainty of the rest of the system, as discussed below.

\subsection{Data Acquisition}

In order to acquire images from the camera, either MATLAB\cite{myMATLAB} or a custom program is used. The camera manufacturer provides driver software and a C++ library, the Pylon API, that allows to capture single frames or a continuous stream of images. These functions were wrapped in a C++ class that can be accessed via short C functions in MATLAB's MEX interface. A wrapper class is necessary to make the Pylon library and the state of the camera connection persistent in MATLAB memory. A graphical user interface allows convenient monitoring of the camera data and displays histograms and Gaussian fits of the data.

The wrapper class is also employed in a C++ implementation of the same user interface using the Qt\cite{myQt} Framework and Qwt extension for displaying graphics and plots. Generally, the C++ implementation is faster 
and might thus be used for setup and calibration of the optics where direct visual feedback is required. On the other hand, the MATLAB implementation proved to be convenient for final data analysis and archival which is also performed in MATLAB.
\begin{figure}[ht]
\centering
\includegraphics[width=.85\textwidth]{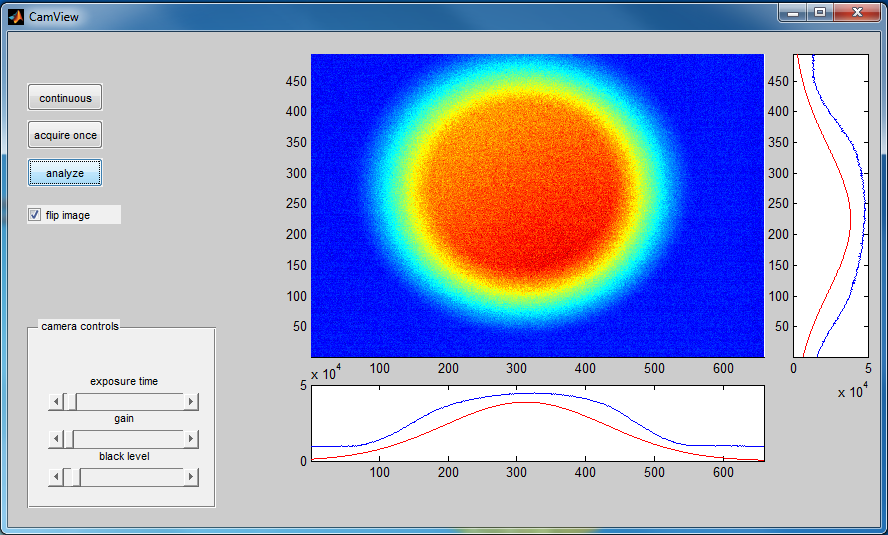}
\caption[Image acquisition tool]{Screenshot of the image acquisition tool implemented in MATLAB. The image is displayed in 6bit false color, but the analysis and storage use the 12bit greyscale data provided by the camera.}
\label{fig:camview-matlab}
\end{figure}
%
%
The intensity distribution of electrons at the screen position is represented as a 2D image from the CCD camera.
The radial intensity distribution is reconstructed by azimuthally integrating the acquired data.

First, the center of gravity and width of the distributions is estimated from a Gaussian fit.
This information is used to assign the intensity value of each pixel to $N$ bins that are formed by concentric circles of equal distance around the center of gravity.
The bins are normalized to their area and then yield the radial distribution in arbitrary units with respect to the distance to the center of gravity.

Background noise from the CCD chip is efficiently suppressed by subtracting a dark image and averaging over the circle (bin) area.
A Gaussian test distribution can well be recovered at signal to noise amplitude ratios down to -4\,dB.

\begin{figure}[!ht]
\centering
\includegraphics[width=.85\textwidth]{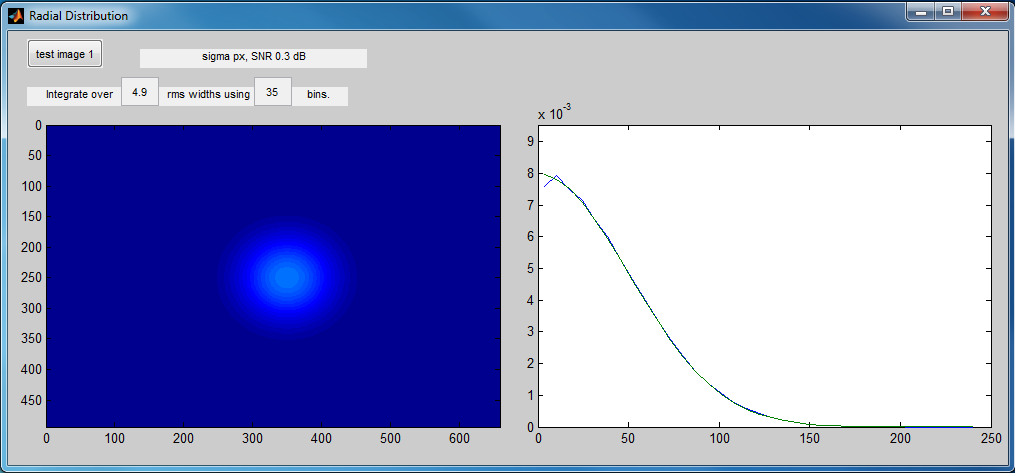}
\caption[Tool for azimuthal integration]{Screenshot of the azimuthal integration implemented in MATLAB. The current image is displayed on the left, the plot on the right shows the obtained radial distribution and a normal distribution.}
\label{fig:radial-matlab}
\end{figure}

The distribution obtained from the azimuthal integration is a convolution of the transverse momentum distribution and the light intensity profile on the cathode. In order to obtain the second order moment of the distribution, the addition theorem of statistics allows to simply subtract the rms values of the distributions. To recover the momentum distribution a Wiener filter approach will be used to deconvolve the measured distribution, approximating the light profile with a Gaussian point spread function.

\clearpage
\subsection{Detection of the Photocurrent}
The electrical isolation of the sample holder allows to measure the photocurrent. A Keithley 6517B picoampere-meter is used both as a bias voltage source and to detect the photocurrent. The picoampere-meter is connected in series between a 10\,k$\Omega$ resistor and the sample in a circuit driven by the integrated voltage source. For the QE monitoring during deposition the preparation chamber acts as an anode and during the momentatron experiment the grid is the anode.
%
%

\section{Estimate of Errors} \label{sec:errors}

\subsection*{Electric Fields}

\begin{figure}[!ht]
\centering
\includegraphics[width=.85\textwidth]{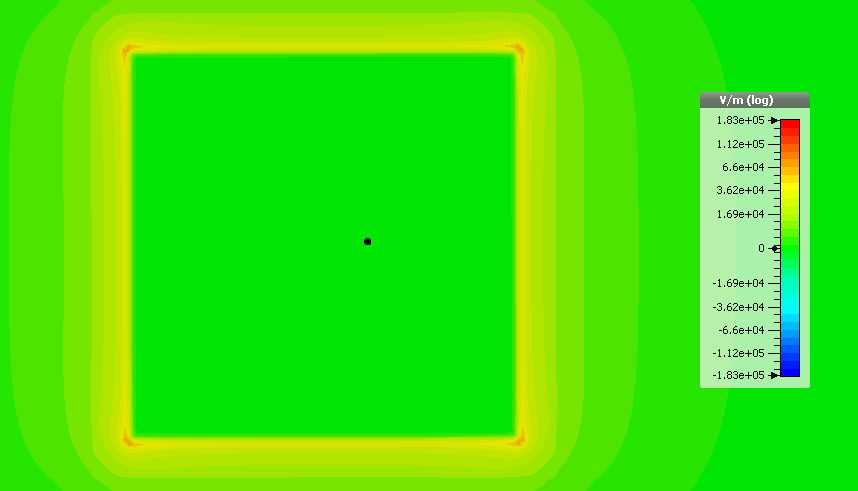}
\caption[Transverse field components in the accelerating gap]{Tangential field components in the accelerating gap. The field has strong transverse components at the edges of the 10x10\,mm sample holder but is homogeneous in the center of the gap.}
\label{fig:E-tan}
\end{figure}

The electric field should ideally be the single source of action on the electrons. In order to receive accurate results, the electric field has to be homogeneous in the volume populated by electrons between the cathode and the anode and it has to be efficiently suppressed in the drift region. As is shown in figure \ref{fig:E-tan}, the tangential field components are very low in the center of the accelerating gap. The field integral of one transverse component along a straight line through the gap in 1\,mm distance from the center is 0.017\,V at 50\,V accelerating potential, so the error is negligible for any realistic beam size in the gap. Figure \ref{fig:E-tan} shows the position of the line integral as a black dot. Details of the electrostatic simulation can be found in appendix \ref{chap:sim}.

Due to different field gradients $G_1$ and $G_2=0$ in front and behind the grid, the electron beam will be defocused at the grid apertures. The focal length for round holes is \cite{Davisson1932}

\begin{equation}
f=\frac{4U}{G_1-G_2}=\frac{4U}{\frac{U}{g}}=4g \, .
\end{equation}		

Thus, following Shipley \cite{Shipley1960}, the transverse resolution of the system is

\begin{equation}
F=\frac{Dd}{4g}=\frac{D\cdot56\,mm}{4\cdot3\,mm}=4.67D
\end{equation}

where $D$ is the hole size. This yields $F=233\,\mu$m resolution at $D=50\,\mu$m.
A finer grid with D=$18\,\mu$m and F=$84\,\mu$m is favoured and was used in the setup. Assuming a typical total beam radius at the screen of 7.5\,mm, the lens effect of the grid introduces a relative uncertainty of 1\,\%.

This is independent of the accelerating voltage although at higher voltages the drift length will be increased. Thus, at a lower voltage a higher resolution will be achieved.

\subsection*{Magnetic Fields}

The earth's magnetic field may introduce dispersive steering to an electron beam motion. The absolute field strengths are about 50\,$\mu$T (0.5 Gauss) in Brookhaven and Berlin \cite{Padmore2012, geomag}, and the vacuum pumps may also create stray fields in the chamber. An electron beam in an uniform magnetic field travels on a circle with radius
\begin{equation}
r = \frac{1}{B} \sqrt{\frac{2Um}{e}} \, .
\end{equation}
This is prohibitive as the radius is 0.67\,m for an energy of 100\,eV, similar to the length of the drift region and far greater than the diameter of the screen. The drift tube and analysis chamber are thus made of high-$\mu$ material which suppresses external magnetic fields in the drift path.
The residual magnetic field is expected to have negligible influence on the trajectories.

\subsection*{Angle between Cathode and Anode}

\begin{figure}[!ht]
\centering
\includegraphics[width=.85\textwidth]{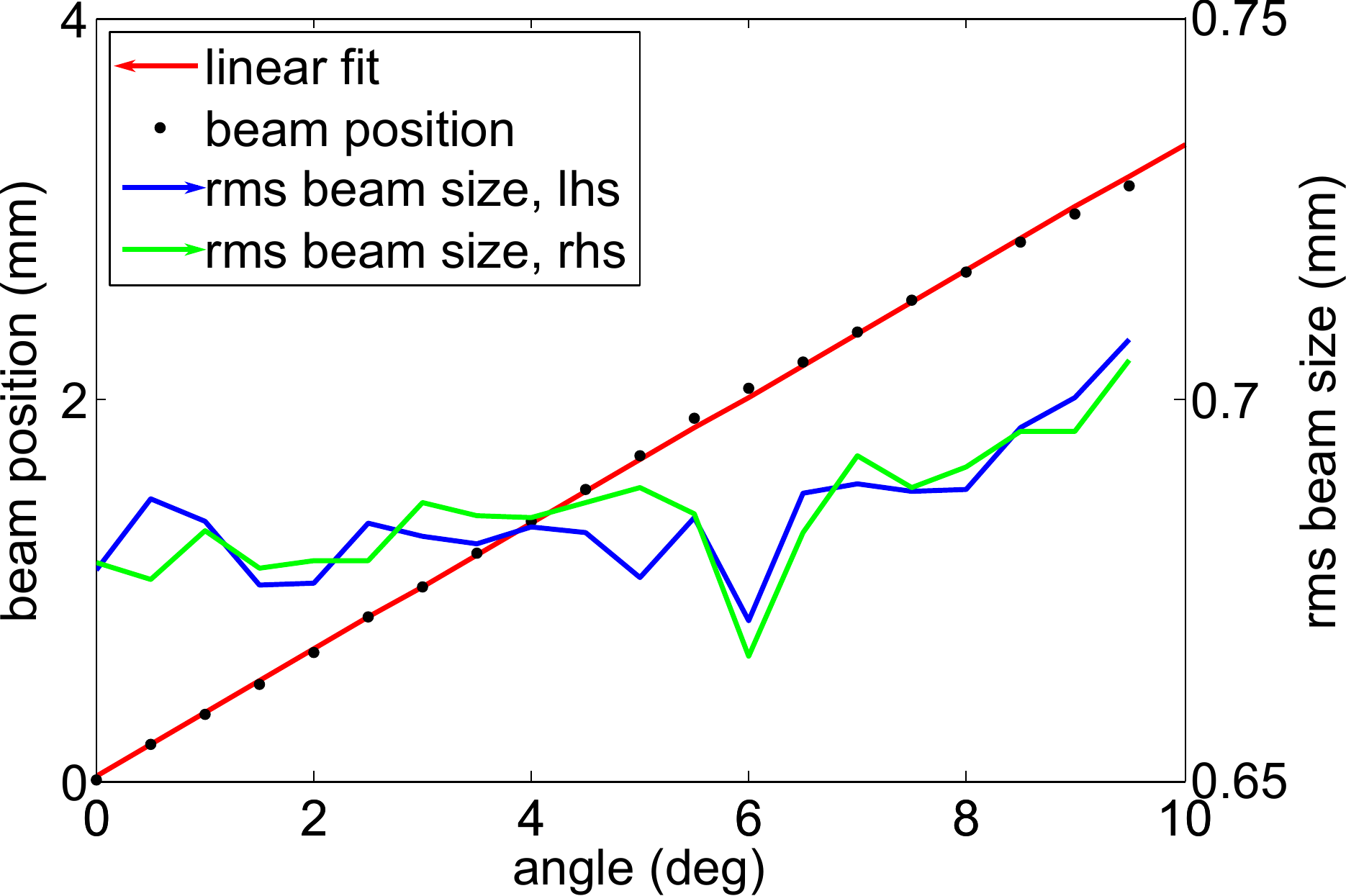}
\caption[Influence of an angle between anode and cathode]{Influence of an angle between anode and cathode. For small angles, the beam position is steered linearly and the rms width on the left and right hand side stay constant. The slope of the steering is 330\,$\mu$m per degree angle.}
\label{fig:pos-vs-angle}
\end{figure}

\begin{figure}[!ht]
\centering
\includegraphics[width=.85\textwidth]{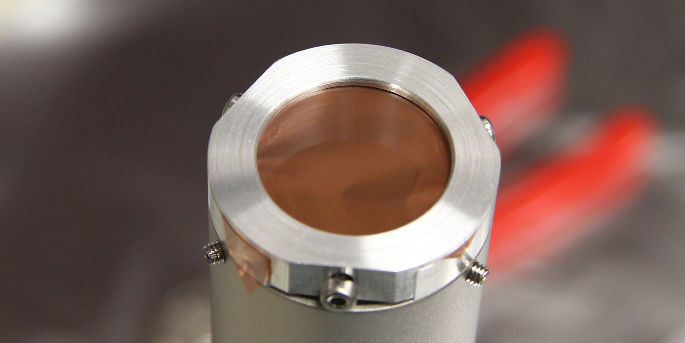}
\caption[Grid mesh mounted to the drift tube]{Photo of the grid mesh mounted to the drift tube. Soft folds of the mesh are discernible.}
\label{fig:photo-grid}
\end{figure}

The sample holder can be rotated around the vertical axis by a stepper motor which defines the angle between cathode and anode in this plane. Due to spatial constraints, no feedback system is implemented that could provide information on the angle from direct measurement.
The influence of such an angle was studied using an FEM simulation of the electric field and particle tracking in CST Studio Suite \cite{CST}. For angles below $10^\circ$ only linear beam steering of 330\,$\mu$m per degree was observed, as can be seen in figure \ref{fig:pos-vs-angle}. The total and also the left- and right-hand side rms values of the distribution stayed constant. These results are independent from the applied voltage.
Using the beam position as an indicator for the angle might be possible, thus allowing beam based alignment.
This information can also be used to estimate the influence of a non-planar grid on the measurement. From visual inspection, the grid in its current mount is expected to be  planar within 100\,$\mu$m, see figure \ref{fig:photo-grid}. If the folds are 100\,$\mu$m deep over a distance of 10\,mm, the local mesh angle is 0.5$^\circ$ which results in a deviation of the position on the screen of 145\,$\mu$m. Assuming a typical total beam radius at the screen of 7.5\,mm, the folds of the grid introduce a relative uncertainty of 2\,\%.
\pagebreak
\subsection*{Uncertainties of Distances}
The derivative of equation \ref{eq:beamsize} with respect to $g$ is
\begin{equation}
\frac{d}{dg} [ \frac{r}{2g+d} \sqrt{\frac{2eU}{mc^2}}]=\frac{2r}{(2g+d)^2} \sqrt{\frac{2eU}{mc^2}} \, .
\end{equation}
For  $r$=7.5\,mm, $g$=3\,mm, $d$=58\,mm, $U$=50\,V and a misalignment of $\Delta g$ = 0.1\,mm one has an uncertainty of $\Delta p_x/mc=6.83/\mathrm{m}^2 * \Delta g*r = 6.83/\mathrm{m}^2 * 0.1$\,mm* 7.5\,mm = $5.1\cdot 10^{-6}$ which is 0.5\,\% of the expected maximum transverse momentum.

\subsection*{Scintillator and Imaging}
Light from scintillation sites in the screen is emitted isotropically which blurs the image of the beam. In this case multiple scattering events can be ignored due to the low energy of the electrons. Light will only be generated in a very thin surface layer of the scintillator. The imaging system has a very small angle of acceptance of $\vartheta_2<3.8^\circ$ due to the small aperture of the viewscreen and lens. YAG has an index of refraction of 1.8, so light from a scintillation site in the crystal is refracted away from the surface normal when leaving the crystal. The maximum angle inside the crystal at which light will pass the aperture is $\vartheta_1 = \vartheta_2 / 1.8 < 2.12^\circ$. One can estimate the uncertainty due to this effect as $\Delta r = d_{YAG} \tan \vartheta_1 = 7.4\,\mu$m, where $d_{YAG}=200\,\mu$m is the thickness of the screen. Given the large influence of the grid this effect can be neglected, but it should be kept in mind when designing scintillation based detection systems for high energy beams or when larger apertures are used. The angle of total reflectance is $33^\circ$, so without aperture limitations, the radius of a point source would be $d_{YAG} \tan(33^\circ) = 130\,\mu$m.


\subsection*{Wavelength Spread}
\begin{figure}[!ht]
\centering
\includegraphics[width=.85\textwidth]{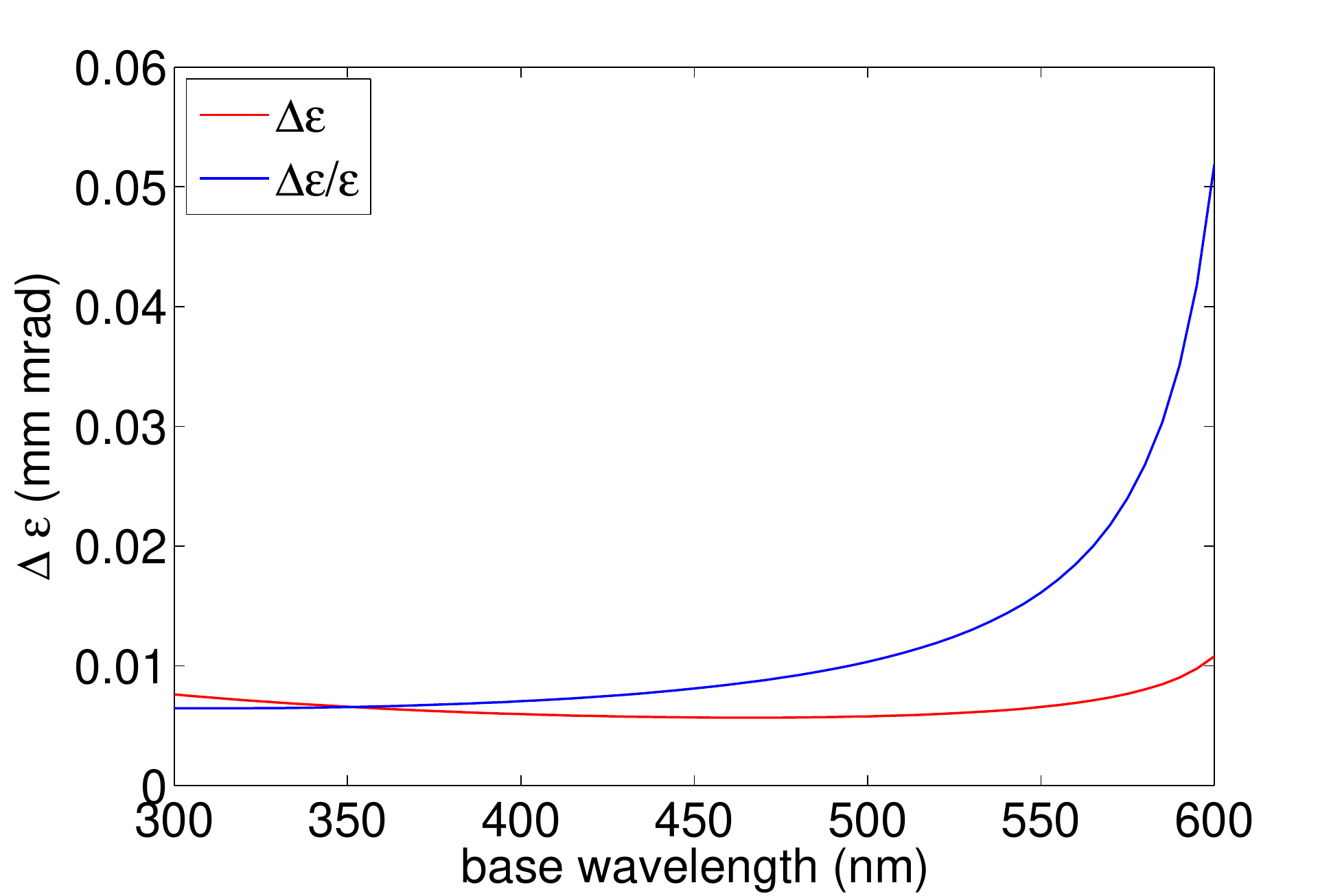}
\caption[Emittance dilution and relative error due to a wavelength spread]{Estimated emittance dilution and relative error due to a wavelength spread of 2\,nm of the light source.}
\label{fig:energy-spread}
\end{figure}

Presently, the sample will be illuminated by a solid state laser diode at 532\,nm wavelength which is temperature stabilized and has negligible wavelength spread. However, spectrally resolved measurements of QE and emittance 
are envisaged and a tunable light source will have some wavelength spread. A typical monochromator for a tungsten halogen lamp with 125\,mm arm length, 600\,l/mm grating and a 120\,$\mu$m exit slit has a wavelength resolution of 2\,nm. Using equation \ref{eq:sc-emittance-E-xcs} one may estimate the uncertainty in transverse energy spread introduced by a wavelength spread $\Delta \hbar \omega$ (or energy spread $\Delta E$) as
\begin{eqnarray}
\Delta \sigma_{x'} &=& \Delta \left( \sqrt{\frac{<p_x^2>}{mc}} \right) = \Delta E \frac{1}{6mc^2} \sqrt{\frac{3mc^2}{E_0-E_G-E_A}} \\
 &=& \Delta \lambda \frac{hc}{\lambda_0 \cdot 6mc^2} \sqrt{\frac{E_0-E_G-E_A}{3mc^2}} \label{eq:eps-dilution} \, .
\end{eqnarray} 
The relative error is estimated by dividing equation \ref{eq:eps-dilution} by \ref{eq:eps-from-px2}.
As shown in figure \ref{fig:energy-spread} for an example where the work function is 2\,eV, the expected emittance dilution due to the wavelength spread of such a system varies with the base wavelength between 0.006 and 0.0075\,mm\,mrad which amounts to an relative error of less than one percent below 500\,nm.
The estimated error rises quickly above 500\,nm as it approaches a singularity at the threshold wavelength.
\pagebreak
\subsection*{Summary}

The total rms uncertainty of the radial coordinate on the screen that originates from the folds (145\,$\mu$m) and lens effect (84\,$\mu$m) of the grid and the imaging of the scintillator (7.4\,$\mu$m) amounts to 167.7\,$\mu$m. Thus, one has an uncertainty of the transverse momentum of
\begin{equation}
\frac{\Delta p_x}{mc} = \frac{\Delta r}{2g+d} \sqrt{\frac{2eU}{mc^2}}= 3.7\cdot 10^{-5}
\end{equation}
which is 2.5\,\% of the expected maximum transverse momentum and a contribution of $5.1\cdot 10^{-6}$ from the uncertainty of the sample position which is 0.5\,\% of the expected maximum transverse momentum.

The evaluation of sources of uncertainties above shows, that the contributions from the lens effect and local mesh curvature both rise with $\sqrt{U}$ and the contribution from the gap distance falls with $1/\sqrt{U}$. At the scintillation screen, scattering effects will introduce additional uncertainties at higher energies. The development of a high voltage version of the momentatron thus makes sense only to investigate the effect of higher gradients on the cathode but not to reduce the errors.

\begin{table}[!ht]
\label{tab:errors}
\centering{
\begin{tabular}{L{2.5cm} C{3.5cm} C{2.2cm} C{2.6cm} C{2.8cm} }
 &                 resolution limit  & voltage dependence  & magnitude                   & rel. error on $p_x$ \\
 \midrule
electric field   & $\dfrac{D d}{4 g}$   & $d \propto \sqrt{U}$    & 84\,$\mu$m                     & 1\%                 \\
angle \& mesh    & $\tan (0.5^\circ) d$ & $\sqrt{U}$    & 145\,$\mu$m                    & 2\%                 \\
$\Delta d$ (gap) & $\dfrac{\Delta p_x}{mc} \approx \dfrac{2r}{d^2} \sqrt{\dfrac{2eU}{mc^2}}$
                                        & 1/$\sqrt{U}$  & $\dfrac{\Delta p_x}{mc} \sim 5\cdot 10^{-6}$  & 1\%  \\
imaging          & aperture             &         -     & 8\,$\mu$m                      & 0.1\%
\end{tabular}   
\caption{Sources of error in the transverse momentum measurement. }}
\end{table}

\section{Status of the Device}

Currently, the preparation and analysis chamber is at Brookhaven National Lab. The momentatron is installed in the analysis chamber and awaits testing. A copper grid mesh with 1000 lines per inch was obtained from SPI and mounted as anode. From visual inspection, the flatness of the mesh is good enough to support the assumptions made for local mesh curvature in section \ref{sec:errors}. A cerium doped YAG screen without coating from CRYTUR is used as scintillator. The screen has a thickness of 200\,$\mu$m and proved to be fairly robust in handling. When the screen is mounted to its holder using the screws and springs that were intended it does not fit into the slit left for the holder, it is now mounted without springs and with some offset from the intended position. The drift length is 58.06\,mm.
The laser optics were set up using the green 1\,mW laser module and a circular beam profile with an rms radius of about 40\,$\mu$m was obtained.

During the first measurement campaign, commissioning of the preparation system was started and a first Cs_3Sb cathode was grown. Testing of the momentatron was intended, but the sample holder could not be moved in front of the grid due to space constraints (the nozzle of an ion gun was in the way). In a second campaign, the ion gun will be removed and improvements on the vacuum system are planned, allowing testing and commissioning of the momentatron.
\chapter{Conclusions and Outlook}

SRF photoinjectors for high brightness beams are an enabling technology for next-generation accelerator projects like bERLinPro. The beam parameters of a photoinjector prototype (gun 0.2) were characterized and a spectrometer for the initial transverse momentum distribution of photoelectrons was developed.

\section*{Performance of the Gun}

Gun 0.2 demonstrated production of a low current electron beam at energies between 1 and 2.5\,MeV. The normalized emittance was measured to be 1.9\,mm\,mrad per mm rms laser spot size by the slit mask technique and 1.8\,mm\,mrad\,/\,mm rms (horizontal) and 2.5 \,mm\,mrad\,/\,mm rms (vertical) by the solenoid scanning technique. Overall, the performance has much improved over the previous setup, where the normalized emittance was between 5.2 and 5.7\,mm\,mrad / mm rms. This is probably due to better control of the lead deposition process on the cathode plug. Visually, the beam was less structured than on images from the previous setup.
The QE for cavity 0.1 is higher than for cavity 0.2 by a factor of 10 because laser cleaning with an excimer laser of the Pb cathode film was only performed with cavity 0.1 \cite{BardayIPAC2013}.
Lower available average drive laser power further reduced the average current generated from the cathode by a factor of 3.
Generally, the plug-gun concept offers advantages over the direct coating of the inner back wall. Lower dark current and lower emittance offer high performance characteristics for the hybrid Nb/Pb gun.

\section*{Consequences for Future Injector Setups}

There are several features of the data collected from the injector test setup that require further investigation in order to identify the causes.
Currently, the data does not allow to quantify the influence of cavity and emission dynamics and the influence of dynamic range differences or distinguish between them. Thus, it was not possible to clearly attribute the discrepancy between the results of slit mask and solenoid scan measurements to one of them. The effect of solenoid aberrations on both measurements could be mitigated by weak focusing, which implies that the back screen should be used for solenoid scans and a larger screen is required to image all beamlets when a large beam diameter is scanned over the slit aperture for phase space characterization.
The emittance increase with the launch phase can  be attributed to the chromatic aberration of the solenoid.

Regarding the measurements themselves, future ones should be conducted with greater care for comparability. Bunch charges should be kept equal when changing the laser diameter. Emission phases and the field gradient in the cavity should be equal to allow comparison of different measurement techniques.
The back screen is currently in a position where a sharp image of the cathode's emission surface is created when the beam is focused on the slit mask.
A vertical slit in front of the dipole is required to make meaningful measurements of the momentum spread.

A setup to directly measure the phase space in short time (about 1~min) using a double slit design has been developed at Cornell \cite{Bazarov2008}. The entirely electric measurement delivers a higher and more reliable dynamic range. Additionally, faster emittance measurement would allow to better correlate gun parameters and beam emittance.

\section*{Momentatron}

In order to aid with the development of \PCA~cathodes, a compact spectrometer for the transverse momentum distribution of photoemitted electrons was designed, built and set-up.
The initial transverse momentum of the electrons emitted from the cathode sets a limit for the beam brightness in the accelerator.

The system was made compatible to the existing photocathode preparation and analysis system and will allow in-situ characterization of cathodes, without breaking vacuum after preparation.
It consists of a short accelerating gap between the sample and a grid anode and a drift region. After passing the drift region, electrons are detected by a scintillating screen which is imaged by a CCD camera. The radial coordinate at the screen position scales linearly with the initial transverse momentum of the electron.
Design parameters of the system are the gap and drift distances and the applied bias voltage, as well as type and material of the grid and scintillation screen.
The drift distance was optimized to obtain maximum resolution of the electron beam on the screen, while the accelerating voltage was limited to -100\,V by the electrical isolation of the available sample holder.

To minimize influence of the earth magnetic field and possible stray field from the vacuum pumps, the drift region is shielded by high-$\mu$ material.
The flatness of the accelerating field depends on the angle between the cathode sample and the anode grid, which is ideally zero but cannot be measured directly in the experiment. Finite-element simulations predict that the electron beam is only steered but the transverse distribution is not distorted when the angle is below 10$^\circ$. Possibly the sample can be positioned by beam based alignment.
A systematic error is introduced by the spot size of the laser light, which can be estimated by reproducing the optical path and corrected for.
Additional uncertainties arise from the lens effect of the grid, a local curvature (folds) in the grid mesh, the uncertainty of the gap length, and the point spread function of the scintillator. The folds of the grid introduce the largest error of 2\,\% at the expected transverse energies. Adding all errors results in a total relative uncertainty of 3\,\% for the transverse momentum.

Currently, the spectrometer is set up and will be tested in a second measurement campaign. During the first campaign, the cathode preparation system was commissioned and a first caesium antimonide cathode was prepared. A photocurrent of 10\,$\mu$A was obtained from the photocathode. The photocurrent measurement was space charge limited, so only a lower limit of the QE can be reported which is 0.5\%. One of the potassium sources was degassing strongly which compromised the vacuum pressure and prevented the use of this source. For the second campaign improvements of the vacuum system are planned which should shorten the bakeout time and help achieve base pressures below $10^{-9}$\,mbar.

As explained in section \ref{chap:momentatron}, increasing the accelerating voltage will have a negative effect on the accuracy of the measurement.
For future measurements, if the signal from the scintillation screen is weak, one could think of a similar setup with a microchannelplate as a signal amplifier to increase the dynamic range of the measurement.
With a retarding field analyser, one could measure the longitudinal energy distribution and, using adiabatic transverse expansion of the electron beam in a longitudinal magnetic field, resolve the transverse energy distribution. Such measurements were performed in Heidelberg \cite{Orlov2001, Pastuszka1997} and allow 2D resolution of the longitudinal and transverse momentum distributions.
For a spectrally resolved measurement one could use different laser wavelengths or a grating monochromator and light from a tungsten or laser plasma source. This would allow to measure the work function of the samples and investigate effects of the light wavelength on the momentum distribution.

In the framework of the Photocathodes for High Brightness Electron Beams (PCHB) cooperation a vacuum suitcase and compatible plug design are under development to allow exchange of photocathodes between the partner institutes at Mainz University and Helmholtz Zentrum Dresden Rossendorf (HZDR). At HZB and HZDR cathodes will be tested in the SRF gun environments of GunLab and the ELBE injector while in Mainz measurements of the response time of the cathodes are possible\cite{Kirsch2014}.

\appendix

\chapter{Details of the Electrostatic Simulation} \label{chap:sim}

The effect of an angle between the cathode and grid anode was studied in an electrostatic simulation using the finite elements approach implemented in CST Studio Suite \cite{CST}. Particle tracking was done by using the Runge-Kutta tracking algorithms in CST and in Astra \cite{Astra} with field maps exported from CST. It proved problematic to export field maps from CST with the required mesh density due to large file sizes, so the final results are from the CST simulation.

\begin{figure}[ht]
\centering
\label{fig:cst-sim}
\includegraphics[width=.85\textwidth]{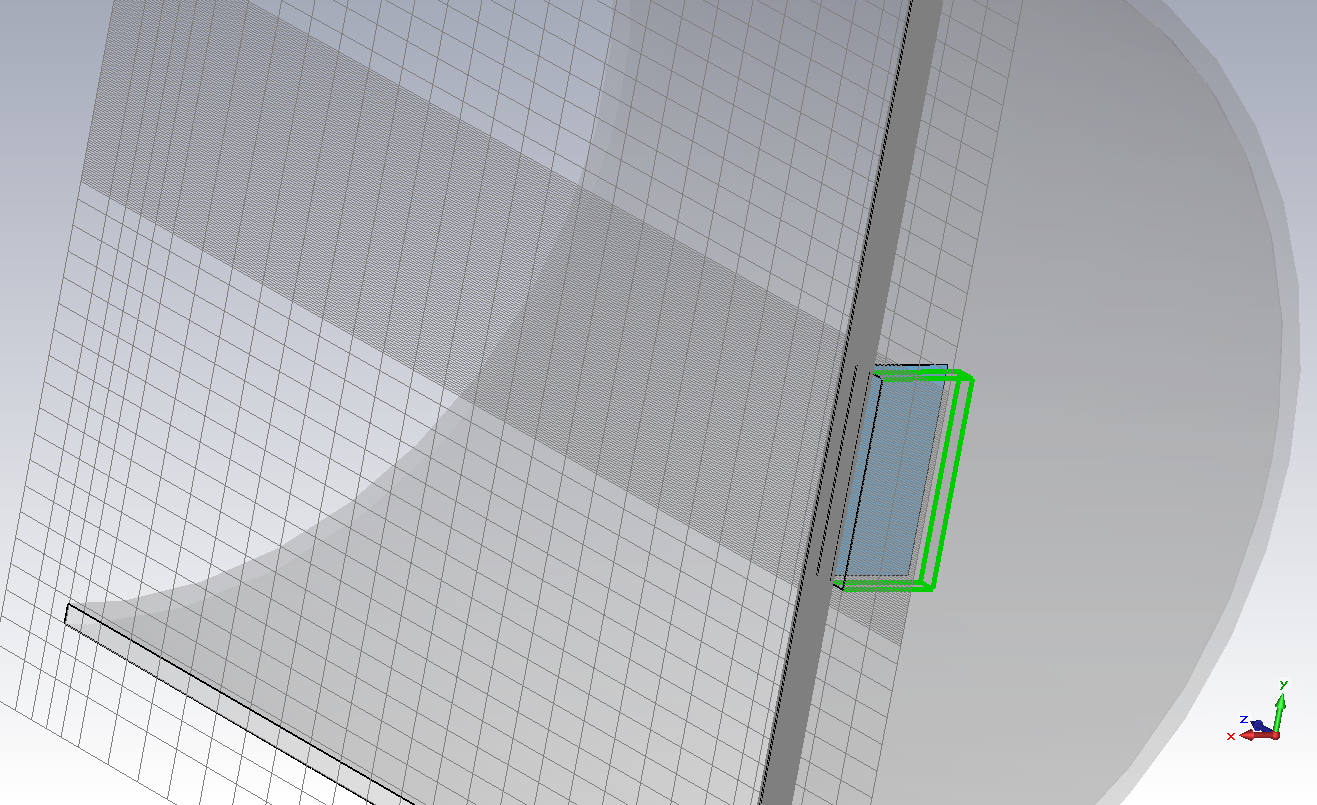}
\caption[Geometry of the FEM simulation]{Geometry and mesh definition of the FEM simulation. The sample holder is represented by the green brick at z=0. The drift tube and anode grid are shown as grey primitives. A vacuum brick (blue layer) between cathode and anode defines the region for refined mesh properties. Particles start at the cathode surface and travel in positive z direction.}
\end{figure}

The geometry of the simulation is shown in figure \ref{fig:cst-sim}. Material with infinite conductivity and magnetic permeability was used to model the solid components and the grid anode was modelled as a conductive sheet with full transparency to the particles. The acceleration gap is 3\,mm wide and the particles are allowed to drift for 58\,mm in the tube.
A hexahedral mesh was used and the results converged at a mesh step size of 0.1\,mm (transverse) and 0.02\,mm (longitudinal) in the gap region. The mesh step size in the drift region was relaxed to 0.1\,mm in all directions.

\chapter{Derivation of the cathode emittance formula for photoemission}

We solve the integrals for the rms transverse momentum of electrons from photoemission (following Jensen \cite{Jensen2006}). Emittance is defined through second moments of the transverse momentum distribution. Moments are obtained by 3D integration over all possible momentum vectors.

\begin{eqnarray}
<k_r^2> &=& \frac{M_2}{M_0} \\
M_n &=& \int k_r^n f(E(\vec{k})) d\vec{k}
\end{eqnarray}

where $f(E(\vec{k}))$ is the probability distribution of the momentum (assuming it only depends on the energy). The integral is transformed to spherical coordinates:

\begin{eqnarray}
M_n &=& \int k_r^n f(E(\vec{k})) |\vec{k}|^2 \sin \vartheta \Dd \phi \dd \vartheta \dd |\vec{k}| \, , 
\end{eqnarray}

and from momentum to energy integration:


\begin{eqnarray}
\frac{d|\vec{k}|}{dE} &=& \frac{1}{2} \sqrt{\frac{2m}{\hbar^2}} (E+\hbar\omega)^{-1/2} \\
M_n &=& \frac{1}{4\pi} \left(\frac{2m}{\hbar^2}\right)^{3/2}     \int k_r^n \, (E+\hbar\omega)^{1/2} \sin \vartheta \, f(E) \Dd \vartheta \dd E \\
    &=& \frac{1}{4\pi} \left(\frac{2m}{\hbar^2}\right)^{3/2}     \int\limits_0^\infty \dd E \int\limits_0^\pi \Dd \vartheta \left(\frac{2m}{\hbar^2} (E+\hbar\omega) \right)^{n/2} (E+\hbar\omega)^{1/2} (\sin \vartheta)^{n+1} f(E)\\
    &=& \frac{1}{4\pi} \left(\frac{2m}{\hbar^2}\right)^{(n+3)/2} \int\limits_0^\infty \dd E \int\limits_0^\pi \Dd \vartheta (E+\hbar\omega)^{(n+1)/2} (\sin \vartheta)^{n+1} f(E)
\end{eqnarray}

The probability distribution consists of a term for the transmission probability T, that accounts for the fact that electrons need sufficient longitudinal momentum to overcome the potential barrier. It it essentially a step function at the maximum escape angle (defined later). The two latter terms account for the population probability of the initial and final states, respectively. In the $0\,$K limit, the Fermi-Dirac distribution is also well represented by a step function. 
\begin{equation}
f(E) = T \left[ (E+\hbar\omega) \cos^2 \vartheta \right] \, f_\lambda (\cos \vartheta, E+\hbar\omega) \, f_{FD}(E) \, \left[ 1-f_{FD}(E+\hbar\omega)\right]
\end{equation}
$f_\lambda$ accounts for scattering and is given by \cite{Jensen2007}
\begin{equation}
f_\lambda(\cos \vartheta, E) = \frac{\cos \vartheta}{\cos \vartheta + p(E)}
\end{equation}
the asymptotic emittance is independent of p(E), so the energy dependence of scattering is neglected here.
\begin{eqnarray}
M_n &=& \frac{1}{4\pi} (\frac{2m}{\hbar^2})^{(n+3)/2} \int\limits_{\mu-\hbar\omega+\phi}^\mu \dd E \, (E+\hbar\omega)^{(n+1)/2} \int\limits_0^{\cos^{-1} \vartheta_{max}} \dd \vartheta \, (\sin \vartheta)^{n+1} f_\lambda (\cos \vartheta, E+\hbar\omega)\\
    &=& \frac{1}{4\pi} (\frac{2m}{\hbar^2})^{(n+3)/2} \int\limits_{\mu-\hbar\omega+\phi}^\mu \dd E \, (E+\hbar\omega)^{(n+1)/2} \int\limits_0^{\cos^{-1} \vartheta_{max}} \dd \vartheta \, (\sin \vartheta)^{n+1} \frac{\cos \vartheta}{\cos \vartheta+p(E)}\\
    &=& \frac{1}{4\pi} (\frac{2m}{\hbar^2})^{(n+3)/2} \int\limits_{\mu-\hbar\omega+\phi}^\mu \dd E \, (E+\hbar\omega)^{(n+1)/2} \int\limits_{\vartheta_{max}}^1 \dd (\cos \vartheta) \, (\sin \vartheta)^{n} \frac{\cos \vartheta}{\cos \vartheta+p(E)}\\
    &=& \frac{1}{4\pi} (\frac{2m}{\hbar^2})^{(n+3)/2} \int\limits_{\mu-\hbar\omega+\phi}^\mu \dd E \, (E+\hbar\omega)^{(n+1)/2} \int\limits_{\vartheta_{max}}^1 \dd x  \, \frac{x(1-x)^{n/2}}{x+p(E)}
\end{eqnarray}

Jensen \cite{Jensen2006} offers an approximation for the x ( originally $\vartheta$ ) integral :

\begin{equation}
\label{eq:jensen-defineG}
G(p,b,s) = \int\limits_b^1 \frac{x(1-x)^{s}}{x+p} \dd x\approx \frac{(1-b^2)^{s+1}}{2(s+1)(1+p)}
\end{equation}

which is valid if $(1-b^2) \ll 1$. (That means $(\mu+\phi) \approx (E+\hbar\omega)$ which is usually given.)

The interval of the energy integral is mapped to $y \in [0,1]$
\begin{eqnarray}
E &\in& [\mu-\hbar\omega+\phi, \mu] \\
E &=&   \mu-\hbar\omega+\phi+y(\hbar\omega-\phi) \\
\frac{\dd E}{\dd y} &=& \hbar\omega-\phi
\end{eqnarray}
\begin{equation}
M_n = \frac{1}{4\pi} \left(\frac{2m}{\hbar^2} (\hbar\omega-\phi) \right)^{(n+3)/2} \int\limits_0^1 \dd y \, \left[ \frac{\mu+\phi}{\hbar\omega-\phi} +y\right]^{(n+1)/2} \, G\left[ p(y), \left( \frac{1}{1+\Delta y}\right)^{1/2}, \frac{n}{2} \right]
\end{equation}

After neglecting y compared to $(\mu+\phi)/(\hbar\omega-\phi)$ and applying \ref{eq:jensen-defineG} one arrives at

\begin{eqnarray}
M_0 &\propto& \int\limits_0^1 1-\frac{1}{1+\Delta y} \Dd y = \int\limits_0^1 \frac{\Delta y}{1+\Delta y} dy \\
    &=&        \frac{\Delta-\mathrm{log}(1+\Delta)}{\Delta}= \frac{\Delta}{2} + O(\Delta^3) \\
M_2 &\propto& \int\limits_0^1 (\frac{\Delta y}{1+\Delta y})^2 \Dd y \\
    &=&        1+\frac{1}{1+\Delta}-\frac{2 \mathrm{log}(1+\Delta)}{\Delta}= \frac{\Delta^2}{1+\Delta} + O(\Delta^3)
\end{eqnarray}

where the logarithms were approximated by a series to second order, $\Delta = (\hbar\omega-\phi)/(\mu+\phi)$. Finally, the moments of the transverse momentum distribution are

\begin{equation}
M_n \approx \frac{1}{4\pi^2}\left(\frac{2m}{\hbar^2 }(\hbar \omega -\phi )\right)^{\frac{n+3}{2}}
\begin{cases}
\dfrac{\sqrt{(\hbar\omega-\phi)(\mu+\phi)}}{\mu+\hbar\omega} \dfrac{1}{12(1+p)} & (n=2) \\
 \sqrt{\dfrac{\hbar\omega-\phi}{\mu+\phi}}\dfrac{1}{4(1+p)}  & (n=0)
\end{cases}
\end{equation}
The fraction of the two moments gives the rms transverse momentum, so the emittance is
\begin{equation}
\varepsilon \approx \, <x> \sqrt{ \dfrac{\hbar\omega-\phi}{3mc^2} \dfrac{\mu+\phi}{\mu+\hbar\omega} }
\end{equation}
in the threshold limit, where $\hbar\omega \approx \phi$, the result is the same as the one deduced by Dowell \textit{et al.} \cite{Dowell2009}
\chapter{Derivation of the QE expression for metals} \label{appendix:QE}

$\vartheta_{max}$ is defined by the requirement, that the momentum component $p_z$ normal to the surface needs to be greater than the surface barrier
\begin{eqnarray}
p_z = \sqrt{2m(E+\hbar\omega)} cos \vartheta \geq \sqrt{2m(E_F+\Phi)} \, ,\\
\cos \vartheta_{max} = \sqrt{\frac{E_F + \Phi}{E+\hbar\omega}} \, .
\end{eqnarray}
The fraction of excited electrons that gets excited above the vacuum level and has a momentum vector in the escape cone is calculated.
\begin{eqnarray}
\frac{\alpha_{PE}}{\alpha} P_{dir} &=& \frac{ \int\limits_{E_F +\Phi-\hbar\omega}^{E_F} \int\limits_0^{\vartheta_{max}} \Dd \vartheta \Dd E}
										   { \int\limits_{E_F-\hbar\omega}^{E_F}       \int\limits_0^\pi \Dd \vartheta \Dd E} \\
  								  &=& \frac{ \int\limits_{E_F +\Phi-\hbar\omega}^{E_F} \int\limits_{\sqrt{(E_F+\Phi)/(E+\hbar\omega)}}^1 \Dd (\cos \vartheta) \Dd E}
										   { \int\limits_{E_F-\hbar\omega}^{E_F}       \int\limits_{-1}^1 \Dd (\cos\vartheta) \Dd E}\\
								  &=& \frac{\hbar\omega - \Phi + 2(E_F+\Phi) - 2 \sqrt{\frac{E_F+\hbar\omega}{E_F+\Phi}}(E_F+\Phi)}{2\hbar\omega}\\
								  &=& \frac{E_F+\hbar\omega}{2\hbar\omega} \left[ 1- \sqrt{\frac{E_F+\Phi}{E_F+\hbar\omega}} \right]^2
\end{eqnarray}

\printbibliography

\chapter*{Acknowledgements}

I am indebted to Prof. Dr. Andreas Jankowiak for giving me the opportunity to work on this topic and for his support, as well as Prof. Dr. Kurt Aulenbacher for valuable discussions and encouragements.

Great thanks goes to Thorsten Kamps, who continually inspired me during this work and provided friendly assistance and support.

The commissioning and operation of a photoinjector test stand in the HoBiCaT cryomodule is a great collaborative effort and I would like to thank the whole team that made it possible and allowed me to take data there.

I greatly appreciate the help and friendly, encouraging support of Susanne Schubert and Miguel Ruiz-Osés during my stay at Brookhaven National Lab, as well as stimulating discussions with John Smedley, Erik Muller and Triveni Rao. 

I would also like to thank my colleagues in the PCHB collaboration Monika Dehn, Rong Xiang, and Jochen Teichert who helped me with numerous ideas and remarks during the regular meetings.

Additionally, this work would not have been possible without the valuable support of Daniel Böhlick who carried out the actual construction of the momentatron and John Walsh at BNL who provided expert technical assistance.

Jens, Julia, Eva, Sebastian, Stephi - the type of colleague who proof-reads a thesis in the week before its due date and shares a jacket on the way home - thank you for making HZB a fun place to work!

\selectlanguage{ngerman}
Zu guter Letzt möchte ich meinen Eltern, meinem Bruder und meiner Freundin danken auf deren volle Unterstützung ich mein ganzes Studium lang vertrauen konnte, die nicht müde wurden, mich zu ermuntern und die in jeder Situation für mich da waren.

\selectlanguage{ngerman}
\LARGE{\textbf{Selbstständigkeitserklärung}}
\vspace{0.5cm}
\normalsize\\

Hiermit versichere ich, dass ich die vorliegende Arbeit selbstständig verfasst, keine anderen als die angegebenen Quellen und Hilfsmittel verwendet habe und erstmalig eine Masterarbeit einreiche. Die Bearbeitung erfolgte unter Beachtung der gültigen Prüfungsordung. \\
\vspace{1cm}\\
Berlin, den \dcdatesubmitted

%
\selectlanguage{english}





\end{document}